\documentclass[preprint]{aastex}
\usepackage{graphics,graphicx}
\usepackage{subfigure}
\usepackage{epsfig}
\usepackage{lscape}
\usepackage{longtable}
\usepackage{url}
\usepackage{bigstrut,bigdelim,multirow}

\shorttitle{Spectral Properties of \textit{Fermi} LAT/GBM GRBs}
\shortauthors{Zhang et al.}
\begin{document}
\title{A Comprehensive Analysis of {\em Fermi} Gamma-Ray Burst Data.
 I.  Spectral Components and Their Possible Physical Origins of LAT/GBM GRBs}
\author
{Bin-Bin Zhang\altaffilmark{1}, Bing Zhang\altaffilmark{1},  En-Wei
Liang\altaffilmark{1,2}, Yi-Zhong Fan\altaffilmark{3,1},
Xue-Feng Wu\altaffilmark{1,3}, Asaf Pe'er\altaffilmark{4},
Amanda Maxham\altaffilmark{1}, He Gao\altaffilmark{1},Yun-Ming Dong\altaffilmark{1,5}}
\altaffiltext{1}{Department of Physics and Astronomy,
University of Nevada Las Vegas, Las Vegas, NV 89154, USA.}
\altaffiltext{2}{Department of Physics, Guangxi University, Guangxi 530004,
China.} \altaffiltext{3}{Purple Mountain Observatory, Chinese Academy of
Sciences, Nanjing 210008, China} \altaffiltext{4}{Space
Telescope Science Institute, 3700 San Martin Drive, Baltimore, MD 21218, USA ;
Riccardo Giacconi fellow}\altaffiltext{5}{Department of Physics,
Sichuan Normal University, Sichuan 530004, China.}

\begin{abstract}

We present a systematic analysis of the spectral and temporal properties
of 17 gamma-ray bursts (GRBs) co-detected by Gamma-Ray Monitor (GBM) and
Large Area Telescope (LAT) on board the {\em Fermi} satellite by May 2010.
We performed a time-resolved spectral analysis of all the bursts
with the finest temporal resolution allowed by statistics, in order
to {reduce} temporal smearing of different spectral components.
We found that the time-resolved spectra of 14 out of 17 GRBs are best 
modeled with the classical ``Band'' function over the entire Fermi 
spectral range, which may suggest a common origin for emissions 
detected by LAT and GBM. GRB 090902B and GRB 090510 
require the superposition of a MeV component and an extra power law component, with the former
having a sharp cutoff above $E_p$. For GRB 
090902B, this MeV component becomes progressively narrower as 
the time bin gets smaller, and can be fit with a Planck function as 
the time bin becomes small enough. In general, we speculate
that phenomenologically there may be three elemental spectral
components that shape the time-resolved GRB spectra: a Band-function
component (e.g. in GRB 080916C) that extends in a wide energy range and 
does not narrow with decreasing time bins, which may be of non-thermal 
origin; a quasi-thermal component (e.g. in GRB 090902B) with the spectra 
progressively narrowing with reducing time bins; 
and another non-thermal power law component extending to high energies. 
The spectra of different bursts may be decomposed into one or more
of these elemental components. 
We compare this sample with the BATSE sample and investigate some
correlations among spectral parameters.
We discuss the physical implications of the data analysis results 
for GRB prompt emission, including jet composition (matter-dominated
vs. Poynting-flux-dominated outflow), emission sites
(internal shock, external shock or photosphere), as well as radiation
mechanisms (synchrotron, synchrotron self-Compton, or thermal Compton
upscattering).

\end{abstract}

\keywords{gamma-ray burst: general}
\slugcomment{2011 ApJ, in press}

\section{Introduction}

Although observationally accessed much earlier, GRB prompt emission
is stil less understood than afterglow. The fundamental uncertainties
lie in the following three poorly known important properties of GRBs
(e.g. Zhang \& M\'esz\'aros 2004 for a review): (1) Ejecta composition:
Are the ejecta mostly composed of baryonic matter or a Poynting flux?
(2) Energy dissipation site: Is the emission from internal 
shocks (Rees \& M\'esz\'aros 1994; Kobayashi et al. 1997),
the photosphere (Pacz\'ynski 1986; Goodman
1986; M\'esz\'aros \& Rees 2000; M\'esz\'aros et al. 2002; Pe'er 2008), 
some magnetic dissipation regions
(Lyutikov \& Blandford 2003; Zhang \& Yan 2011), or the external shock 
(Rees \& M\'esz\'aros 1992; M\'esz\'aros \& Rees 1993; Dermer \& Mitman 
1999)? (3) Is the radiation mechanism synchrotron/jitter radiation
(M\'esz\'aros et al. 1994; Medvedev 2000), synchrotron self-Compton
(Panaitescu \& M\'esz\'aros 2000; Kumar \& McMahon 2008), or
Comptonization of thermal photons (e.g. Thompson 1994; Pe'er et al. 
2005, 2006; Beloborodov 2010; Lazzati \& Begelman 2009)?

Before {\em Fermi}, understanding of GRB prompt emission has progressed slowly. 
Observations of early X-ray afterglows by
{\em Swift} revealed a steep decay phase that is smoothly connected
to prompt emission (Tagliaferri et al. 2005; Barthelmy et al. 2005),
which suggests that the prompt emission region is detached from the
afterglow emission region, and that the prompt emission site is
``internal'' (Zhang et al. 2006). Other than this, the properties
of prompt emission were poorly constrained. The main factor that
hampers progress has been the narrow energy band of the gamma-ray
detectors of previous missions. Theoretical models usually predict rich
features in the prompt spectra (e.g. Pe'er et al. 2006, see Zhang 2007;
Fan \& Piran 2008 for reviews on high energy emission processes). However,
within the narrow observational spectral window, these features
cannot be fully displayed. Instead, most previous spectral analyses
revealed an empirical ``Band''-function (Band et al. 1993),
which is a smoothly-joint broken power law, whose physical
origin is not identified. For the bright BATSE GRB sample,
the typical low and high energy photon indices are distributed around
$\alpha \sim -1$ and $\beta \sim -2.2$, respectively, while the
spectral peak energy $E_p$ is distributed around 200-300 keV
(Preece et al. 2000). Later observations suggested that the
distribution of $E_p$ can be much wider, extending to a few
keV in the soft regime for X-ray flashes (Sakamoto et al. 2005)
and to greater than 100 MeV in the hard regime
(e.g. $\gtrsim$ 170 MeV for GRB 930506, Kaneko et al. 2008).
Some BATSE GRBs were also detected by EGRET in the GeV range
(Kaneko et al. 2008). For example,
It was found that the GeV emission can last much longer than
the MeV emission (e.g. GRB 940217, Hurley et al. 1994), and
that it can form a distinct spectral component (e.g. GRB 941017,
Gonz\'alez et al. 2003). In the softer regime, an X-ray excess
component with respect to the Band function was discovered
in some BATSE GRBs (Preece et al. 1996). However, the previous
data were not adequate to place meaningful constraints on the
three main questions discussed above.

The {\em Fermi} satellite ushered in a new era of studying GRB prompt
emission. The two instruments on board {\em Fermi}, the Gamma-ray
Burst Monitor (GBM; Meegan et al. 2009) and the Large Area Telescope
(LAT; Atwood et al. 2009), provide an unprecedented spectral
coverage for 7 orders of magnitude in energy (from $\sim$8 keV to
$\sim$300 GeV). Since the beginning of GBM/LAT science operation in
August 2008 and as of the writing of this paper (May 2010), there
have been 17 GRBs co-detected by LAT and GBM, with a detection rate
comparable to the expectation assuming that the LAT-band emission
is the simple extrapolation of the Band spectrum to the GeV range
(Abdo et al. 2008; L\"u et al. 2010). As will be shown below, the
Band-function fits apply to most LAT GRBs, although some outliers
do exist. Broad band spectral analyses have been published by the
{\em Fermi} team for several individual GRBs, e.g. GRB 080916C (Abdo
et al. 2009a), GRB 090510 (Abdo et al. 2009b, Ackermann et al. 2010), 
GRB 090902B (Abdo et al. 2009c, Ryde et al. 2010), GRB 080825C 
(Abdo et al. 2009d), and GRB 081024B (Abdo et al. 2010a), which
revealed several interesting features, such as the nearly featureless
Band spectra covering 6 orders of magnitude in all epochs for GRB
080916C, the existence of an extra power law component extending to
high energies in GRB 090510 and GRB 090902B, the existence of a
quasi-thermal emission component in GRB 090902B, the delayed onset
of the LAT-band emission with respect to the GBM-band emission,
as well as an extended rapidly decaying GeV afterglow for most GRBs.

These discoveries have triggered a burst of theoretical investigations
of GRB prompt emission. Zhang \& Pe'er (2009) argued that
the lack of a thermal component in the nearly featureless spectra of
GRB 080916C suggests a Poynting flux dominated flow for this
burst. The conclusion was strengthened by a follow up study of Fan 
(2010, see also Gao et al. 2009). On the other hand, the quasi-thermal
component in GRB 090902B (Ryde et al. 2010) is well-consistent with
the photosphere emission of a hot fireball (Pe'er et al. 2010, Mizuta
et al. 2010), suggesting that the burst is not highly magnetized.
The possibility that the entire Band function spectrum is photosphere 
emission was discussed by several authors (Fan 2009; Toma et al. 
2010; Beloborodov 2010; Lazzati \& Begelman 2010; Ioka 2010).
These models have specific predictions that can be tested by the
available data. In the high energy regime, Kumar \& Barniol Duran
(2009, 2010), Ghisellini et al. (2010) and Wang et al. (2010)
suggested that the GeV afterglow is of external shock origin,
which requires some unconventional parameters (Li 2010a; Piran \& Nakar 2010).
On the other hand, the fact that LAT emission is the natural spectral
extension of GBM emission in some GRBs suggests that the GeV emission
may be of an internal origin similar to MeV emission (Zhang \& Pe'er
2009). Finally, the delayed onset of the GeV emission has been
interpreted as emergence of the upscattered cocoon emission (Toma
et al. 2009), synchrotron emission from shock accelerated protons
(Razzaque et al. 2009), as well as delayed residual internal shock
emission (Li 2010b). Again these models have specific predictions
that may be tested by a detailed analysis of the data.

Our goal is to systematically analyze the GRB data collected by the 
{\em Fermi} mission, aiming at addressing some of the above mentioned
problems in prompt GRB emission physics. Here we report the first paper 
in the series, which focuses on a comprehensive analysis of the GRBs that 
were co-detected by LAT and GBM. This sample has a much broader spectral
coverage than the GBM-only GRBs, and therefore carries much more
information about GRB prompt emission. The plan of the paper is the
following. In Section 2 we describe the details of our sample selection
and data analysis method. The data analysis results are presented in
Section 3, with emphases on the unique features of some GRBs. We 
also present spectral parameter distributions and some possible 
correlations. In Section 4, we summarize the results and speculate 
on the existence of at least three elemental spectral components, and
discuss their possible physical origins and possible combinations.
In Section 5, we present the comparison between the emissions
detected in the GBM band and that detected in the LAT band and discuss
their physical connections. 
Our conclusions are summarized in Section 6 with some discussion.

\section{Sample and Data Reduction}

As of May 2010, 17 GRBs have been co-detected by
{\em Fermi} LAT and GBM. Our sample includes all 17 GRBs (Table 1).
We downloaded the GBM and LAT data for these GRBs
from the public science support center at the official {\em Fermi} web
site http://fermi.gsfc.nasa.gov/ssc/data/. An IDL code was developed
to extract the energy-dependent lightcurves and time-dependent spectra
for each GRB. This code was based on the {\em Fermi} RMFIT
package (V3.3), the {\em Fermi} Science Tools (v9r15p2) and the
HEASOFT tools, which allows a computer to extract
lightcurves and spectra automatically. The human
involvement is introduced later to refine the analysis when needed. The
code automatically performs the following tasks.

\begin{enumerate}
\item Extract the background spectrum and lightcurve of the GBM
data. {\em Fermi} records GBM data in several formats.  For background
reduction we use the CSPEC format data because it has a wider temporal
coverage than the event data (time-tagged event, TTE, format). The 
background spectrum and lightcurve are extracted from some appropriate
time intervals  before and after the burst\footnote{{ An 
appropriate background time 
interval is typically when the lightcurve is ``flat'' with Poisson 
noise photons. For each burst, we select background time intervals as 
[-$t_{b,1}$,-$t_{b,1}$] before the burst and [$t_{b,3}$,$t_{b,4}$] 
after the burst, where $t_{b}$'s are typically in the order of tens to 
hundreds of seconds. The exact values vary for different bursts due to 
their different brightnesses and the corresponding orbit slewing 
phases.}}, and the energy-dependent background lightcurves are modeled
with a polynomial function $B(E_{\rm ch},t)$, where $E_{\rm ch}$ is a 
specified energy band.

\item Extract the source spectrum and lightcurve of the GBM data.
This is done with the event (TTE) data. GBM has 12 NaI detectors
(8 keV--1 MeV) and 2 BGO detectors (200 keV--40 MeV). The overall 
signal-to-noise ratio (SNR) and peak count rate are calculated for
each detector. The brightest NaI and BGO detectors are
usually used for the analyses.
If several detectors have comparable brightnesses, all of
them (usually 2-4 detectors) are taken for the analyses. By subtracting
the background spectrum and lightcurve obtained in the previous
step, the time-dependent spectra
and energy-dependent lightcurves of the source in the GBM band are then obtained.

\item Estimate the LAT-band background. Since only a small number of
photons are detected by LAT for most GRBs, the background estimation
should be performed cautiously. It is not straightforward to
estimate an accurate LAT background using off-source regions around
the trigger time. In our analyses, the LAT background is extracted using 
on-source region data long after the GBM trigger when the photon counts 
merge into a Poisson noise.

\item Extract the LAT-band spectrum and lightcurve. Both
``diffuse'' and ``transient'' photons (level 0-3) are included. Since the 
LAT point spread function (PSF) strongly depends on the incident energy
and the convention point of the tracker (Ohno et al. 2010), the photons 
are grouped into FRONT and BACK classes and their spectra are 
extracted separately based on different detector response
files. The region of interest (ROI) that contains significant counts
of LAT photons is further refined when
necessary (Atwood et al. 2009; Abdo et al. 2009d).

\item
Extract the background-subtracted GBM and LAT lightcurves for
different energy bands. In our analysis, the lightcurves are extracted
in the following energy bands: 8--150 keV, 150--300 keV, 300 keV--MeV, 
1--30 MeV, and the LAT band (above 100 MeV).

\item Make dynamically time-dependent spectral fits. Initially, the
burst duration is divided in an arbitrary number of slices. The code
then automatically refines the number of slices and the time interval
for each slice, so that the photon counts in each bin 
{(typically minimum 20 counts for GBM spectra)} give adequate
statistics for spectral fitting (the reduced $\chi^2$ is typically
in the range of 0.75 - 1.5, a special case is GRB 090510, see
Sect.\ref{sec:090510}). The time slices are defined to be
be as small as possible as long as the extracted spectra satisfy these
statistical criteria. The GBM spectra of the selected NaI
and BGO detectors and the LAT ``FRONT'' and ``BACK'' type spectra are
all extracted for each slice. These spectra, together with the
corresponding response files (using the same one as the CSPEC data
for LAT, or generated using gtrsp for GBM) are input into XSPEC
(V 12.5.1) simultaneously to perform spectral fitting. The following
spectral functions are considered (in order of increasing free
parameters): single power law (PL), blackbody (BB, Planck function), 
power-law with exponential cutoff (CPL), and Band function. The models 
are tested based on the following principles:
(1) If a one-component model can adequately describe the data
(giving reasonable reduced $\chi^2$, say, between 0.75 and 1.5), 
two-component models are not 
considered; (2) for one-component models, if a function with
less free parameters can describe the data adequately, it is
favored over the models with more parameters. {(3) 
In addition, the Akaike's Information Criterion\footnote{AIC is defined 
by  $ AIC = n \ln{ \left( \frac{\chi^{2}}{n} \right) } + 2 k$,
where $n$ is the number of data points, $k$ is the number of free 
parameters of a particular model, and $\chi^{2}$ is the residual sum 
of squares from the estimated model (e.g. Shirasaki et al. 2008).}
(AIC, Akaike 1974) is calculated to evaluate each model by considering 
both the fitting goodness ($\chi^2$) and the complexity of the model. 
We confirmed that the model with minimal AIC is the preferred model we 
choose based on the first two criteria. } 
Nonetheless, since most GRBs have a Band-function spectra (see 
below), we also apply the Band function to those time bins that
do not demand it in order to compare the fitting results between
the Band function and other functions with less parameters 
(e.g. power law, blackbody, or power law with exponential cutoff).
\end{enumerate}

To assess the quality of a spectral fit, we use the traditional
$\chi^2$ statistics. Due to the low count rate of LAT photons, we use
the Gehrels (1986) weighting method in the high energy regime. We
also employed the C-stat method (as used by the {\em Fermi} team),
and found that the two methods usually give consistent results. We chose 
the $\chi^2$ method since it gives more reliable error estimates.
All the model fitting parameters and $\chi^2$ statistics are presented
in Table 2. For each burst, we present the time-dependent
spectral parameters in the designated time bins defined by the statistics
of spectral fitting, as well as the time-integrated spectral fit
during the entire burst in the last row.

\section{Data Analysis Results}

The data analysis results are presented in Figs. \ref{080825C}-\ref{100414A}. 
Each figure corresponds to one burst, and contains 10-11 panels. In the left
panels, the lightcurves in 5 energy bands (8--150 keV, 150--300 keV,
300 keV--1 MeV, 1--30 MeV, and $>$ 100 MeV) are presented in linear
scale, together with the temporal evolution
of the spectral parameters ($\alpha$, $\beta$, $E_p$ for Band
function, $kT$ for blackbody function, and $\Gamma$ for single
power law photon index). The top right panel is an
example photon spectrum with model fitting, typically taken at
the brightest time bin. The time-dependent model spectra are presented
in the mid-right panel. The time-slices for the time-resolved spectral
fitting are marked with vertical lines in the left panel lightcurves.
In the bottom right panel, the GBM and
LAT lightcurves are presented and compared in logarithmic scale.

In the following, we discuss the results of several individual bright GRBs
(Sect.\ref{sec:080916C}-\ref{sec:090926A}),
and then discuss other GRBs in general (Sect.\ref{sec:otherGRBs}).
We then present statistics of spectral parameters (Sect.\ref{sec:distributions})
and some possible correlations (Sect.\ref{sec:correlations}).

\subsection{GRB 080916C}\label{sec:080916C}

As shown in Fig.\ref{080916C}, GRB 080916C is a long GRB with a duration
$\sim 66$ s. The entire lightcurve can be divided into 6 segments. The
smallest time bins during the brightest epochs (first two) are 3.7 s and
5.4 s, respectively. This corresponds to a rest-frame time interval
$\leq 1$ s (given its redshift 4.35, Greiner et al. 2009). In all the
time intervals, we found that the Band-function gives excellent fits
to the data, consistent with Abdo et al. (2009a). Initially there is
a spectral evolution where the spectra ``widen'' with time ($\alpha$
hardening and $\beta$ softening), but later the spectral parameters
essentially do not evolve any more. We note that the steep $\beta$ 
in the first time bin is mostly because of the non-detection in the 
LAT band. The tight upper limit above 100 MeV constrains the range
of $\beta$ not to be too hard. On the other hand, with GBM data alone,
the data of in first time bin can be still fit as a Band function,
with $\beta \sim -2.12^{+0.158}_{-0.107}$ similar to the values at 
later epochs. This suggests an alternative interpretation to the 
data: The high energy spectral index may be similar throughout the 
burst. The delayed onset of LAT-band emission may be because initially 
there is a spectral cutoff around 100 MeV, which later moves to much 
higher energies (e.g. above 13.2 GeV in the second time bin).

It is interesting to note that the time integrated spectrum of GRB 
080916C throughout the burst is also well fit with a Band function, 
where the spectral indices do not vary with time resolution. As an 
example, we present in Fig.\ref{080916C-090902B} the $\nu F_\nu$ 
spectra of GRB 080916C for three time bins with varying time 
resolution. Remarkably, the parameters do not vary significantly:
$\alpha \sim -1.12$, $\beta \sim -2.25$ for 3.5-8 s; 
$\alpha \sim -1.0$, $\beta \sim -2.29$ for 2-10 s;
$\alpha \sim -1.0$, $\beta \sim -2.27$ for 0-20 s.
This is in stark contrast with GRB 090902B discussed below.

\subsection{GRB 090510}\label{sec:090510}

The short GRB 090510 was triggered with a precursor 0.5 s prior to 
the main burst. Two LAT photons were detected before the main burst.
During the first time slice (0.45-0.5 s), no LAT band emission is
detected, and the GBM spectrum can be well fit with a PL with an
exponential cutoff (CPL hereafter). In the subsequent time slices,
an additional PL component shows up, and the
time-resolved spectra are best fit by the CPL + PL model.
If one uses a Band + PL model to
fit the data, the high energy spectral index $\beta$ of the Band
component cannot be constrained. If one fixes $\beta$ to a particular
value, it must be steeper than -3.5 in order to be consistent with the
data. The CPL invoked in these fits has a low energy photon index
$\Gamma_{\rm CPL} \sim -(0.6-0.8)$, which is very different from the case
of a BB (where $\Gamma_{\rm CPL} \sim +1$). On the other hand, the 
high-energy regime (exponential cutoff) is very similar to the behavior 
of a BB.

Since this is a short GRB, we do not have enough photons to perform
very detailed time-resolved spectral analysis. However,
in order to investigate spectral evolution and the interplay between
the MeV component and the extra PL component, we nonetheless make
4 time bins (see also Ackermann et al. 2010). As a result, the
reduced $\chi^2$ of each segment is outside the range of
$0.75 \leq\chi^2/{\rm dof} \leq 1.5$  as is required for other GRBs.
Our reduction results are generally consistent with
those of the Fermi team (Abdo et al. 2009b; Ackermann et al. 2010).

\subsection{GRB 090902B}

The spectrum of GRB 090902B is peculiar. Abdo et al. (2009c)
reported that both the time-integrated and time-resolved spectra of
this GRB can be fit with the Band+PL model. Ryde et al. (2010) found
that the time-resolved spectra can be fit with a PL plus a multi-color
blackbody model. This raises the interesting possibility that
a blackbody-like emission component is a fundamental emission unit shaping 
the observed GRB spectra.

In order to test this possibility, we carried out a series of 
time-resolved spectral analysis on the data (Fig.\ref{080916C-090902B}). 
We first fit the time-integrated data within the time interval 0-20 s, 
and found that it can be fit with a model invoking a Band function 
and a power law, but with a poor $\chi^2/{\rm dof} \sim 
3.52$. Compared with the Band component of other GRBs, this
Band component is very narrow, with $\alpha \sim -0.58$,
$\beta \sim -3.32$. A CPL + PL model can give comparable
fit, with $\Gamma_{\rm CPL} \sim -0.59$. Next we zoom into the 
time interval 8.5 - 11.5 s, and perform
spectral fits. The Band+PL and CPL+PL models can now both 
give acceptable fits, with parameters suggesting a narrower
spectrum. For the Band+PL model, one has 
$\alpha \sim -0.07$, $\beta \sim -3.69$ with $\chi^2/{\rm dof}
=1.26$. For the CPL+PL model, one has $\Gamma_{\rm CPL} \sim -0.08$
with $\chi^2/{\rm dof} =1.30$. Finally we room in the smallest
time bin (9.5 - 10 s) in which the photon counts are just enough 
to perform adequate spectral fits. We find that the Band + PL
model can no longer constrain $\beta$. 
The spectrum becomes even narrower, with $\alpha \sim 0.07$
and $\beta < -5$. The CPL+PL model can fit the data with a range
of allowed $\Gamma_{\rm CPL}$. In particular,
if one fixes $\Gamma_{\rm CPL} \sim +1$ (the Rayleigh-Jeans slope
of a blackbody), one gets a reasonable fit with  
$\chi^2/{\rm dof} =0.92$. This encourages us to suspect that 
a blackbody (BB) + PL model can also fit the data. We test it and
indeed found that the model can fit the data with 
 $\chi^2/{\rm dof} =1.11$. These different models require
different $\Gamma_{\rm PL}$ for the extra PL component, but
given the low photon count rate at high energies, all these
models are statistically allowed. Since the BB + PL model has
less parameters than the CPL + PL and Band + PL models, we 
take this model as the simplest model for this smallest time
interval.

Next, we tried to divide the lightcurve of GRB 090902B into as
many as time bins as possible so that the photon numbers in each time bin
are large enough for statistically meaningful fits to be 
performed. Thanks to its high flux, we managed to
divide the whole data set (0-30 s) into 62 time bins. 
We find that the data in each time bin can be well
fit by a BB+PL model, and that the BB temperature evolves with time.
The fitting results are presented in Table 2 and Fig.\ref{090902B}.
The time-integrated spectrum, however, cannot be fit with such
a model ($\chi^2/{\rm dof}=14732/276$). A Band+PL model gives a much
improved fit, although the fit is still not statistically
acceptable ($\chi^2/{\rm dof}=2024/275$).  The best fitting parameters are
$\alpha=-0.83$, $\beta=-3.68$, $E_p=847$ keV, and $\Gamma=-1.85$.
Notice that the high energy photon index of the time-integrated
Band spectrum is much steeper/softer than that observed in typical GRBs
(Fig.\ref{distributions}).

In Fig.\ref{090902B-lightcurves}, we display the lightcurves of both the
thermal and the power-law components. It is found that the two
components in general track each other. This suggests that
the physical origins of the two components are related to each
other (see Section 5 for more discussion).

An important inference from the analysis of GRB 090902B is that a Band-like 
spectrum can be a result of temporal superposition of many blackbody-like
components. This raises the interesting possibility of whether all
``Band'' function spectra are superposed thermal spectra. From the
comparison between GRB 090902B and GRB 080916C (Fig.\ref{080916C-090902B}),
we find that such speculation is far-fetched. As discussed above,
GRB 080916C shows no evidence of ``narrowing'' as the time bin becomes 
small ($\sim 1$ s in the rest frame). In the case of GRB 090902B,
a clear ``narrowing'' feature is seen. For the time integrated spectrum,
GRB 080916C has a wide Band function (with $\alpha \sim -1.0$, 
$\beta \sim -2.27$), while GRB 090902B (0-20 s) has a narrow Band 
function (with $\alpha=-0.58$, $\beta=-3.32$) with worse reduced 
$\chi^2$. Another difference between GRB 090902B and GRB 080916C is
that the former has a PL component, which leverages the BB spectrum
on both the low-energy and the high-energy ends to make a BB spectrum
look more similar to a (narrow) Band function. GRB 080916C does not
have such a component, and the Band component covers the entire
{\em Fermi} energy range (GBM \& LAT).
We therefore conclude that GRB 090902B is a special case, whose
spectrum may have a different origin from GRB 080916C (and probably
most other LAT GRBs as well, see Sect.\ref{sec:otherGRBs} below).

\subsection{GRB 090926A}\label{sec:090926A}

This is another bright long GRB with a duration $\sim 20$ s. In our 
analysis, the lightcurve is divided into 9 segments. The Band function
gives an acceptable fit to all the time bins (Fig.\ref{090926A}). 
We however notice a
flattening of $\beta$ after $\sim 11$ s after the trigger. Also the
Band function fit gives a worse reduced $\chi^2$ (although still
acceptable) after this epoch. Since our data analysis strategy is
to go for the simplest models, we do not explore more complicated
models that invoke Band + PL or Band + CPL (as is done by the Fermi
team, Abdo et al. 2010b). In any case, our analysis does not
disfavor the possibility that a new spectral component emerges after
$\sim 11$ s since the trigger (Abdo et al. 2010b).

\subsection{Other GRBs}\label{sec:otherGRBs}

The time resolved spectra of other 13 GRBs are all adequately
described by the Band function, similar to GRB 080916C. The Band-function
spectral parameters are generally similar to GRB 080916C. It is likely
that these GRBs join GRB 080916C forming a ``Band-only'' type GRBs.
In the current sample of 17 GRBs, only GRB 090510, GRB 090902B and 
probably GRB 090926A do not belong to this category and have an
extra PL component extending to high energies. One caveat is that
some GRBs in the sample are not very bright, so that we only 
managed to divide the lightcurves into a small number of time bins
(e.g. 3 bins for GRB 080825C, 1 bin for GRB 081024B, 3 bins for
GRB 090328, 3 bins for GRB 091031, 2 bins for GRB 100225A, and
1 bin for GRB 100325A). So one cannot disfavor the possibility
that the observed spectra are superposition of narrower components
(similar to GRB 090902B). However, at comparable time resolution,
GRB 090902B already shows features that are different from these
GRBs: (1) the Band component is ``narrower'', and (2) there is an
extra PL component. These two features are not present in other
GRBs. We therefore suggest that most LAT/GBM GRBs are similar to
GRB 080916C.

\subsection{Spectral parameter distributions}\label{sec:distributions}

Since the time-resolved spectra of most GRBs in our sample 
can be adequately described as a Band function, we present
the distributions of the Band function parameters in this
section. Since their MeV component may be of a different origin,
GRB 090510 and GRB 090902B are not included in the analysis.

The distributions of the spectral parameters $\alpha$,
$\beta$, and $E_p$
are presented in Figure \ref{distributions}, with a comparison 
with those of the bright BATSE GRB sample (Preece et al. 2000). 
It is found that the distributions peak at $\alpha=-0.9$,
$\beta=-2.6$, and $E_p\sim 781$ keV, respectively. The $\alpha$
and $\beta$ distributions are roughly consistent with those found 
in the bright BATSE GRB sample (Preece et al. 2000). The $E_p$
distribution of the current sample has a slightly higher peak than
the bright BATSE sample (Preece et al. 2000). This is likely
due to a selection effect, namely, a higher $E_p$ would favor 
GeV detections. 

\subsection{Spectral parameter correlations}\label{sec:correlations}

For time-integrated spectra, it was found that
$E_p$ is positively correlated with the isotropic gamma-ray energy
and the isotropic peak gamma-ray luminosity (Amati et al. 2002; Wei
\& Gao 2003; Yonetoku et al. 2004). For time resolved spectra,
$E_p$ was also found to be generally correlated with flux (and therefore
luminosity, Liang et al. 2004), although in individual pulses,
both a decreasing $E_p$ pattern and a $E_p$-tracking-flux pattern
have been identified (Ford et al. 1995; Liang \& Kargatis 1996;
Kaneko et al. 2006; Lu et al. 2010).

In Fig.\ref{L-Ep}, we present the $E_p$-luminosity relations. 
Fig.\ref{L-Ep}a is for the global $E_p-L_{\gamma,\rm iso}^p$ correlation. 
Seven GRBs in our sample that have redshift information (and hence, the
peak luminosity) are plotted against previous GRBs (a sample
presented in Zhang et al. 2009). Since the correlation has a large
scatter, all the GBM/LAT GRBs follow the same correlation trend.
In particular, GRB 090902B, whose $E_p$ is defined by the BB
component, also follows a similar trend.
This suggests that even if there may be two different physical
mechanisms to define a GRB's $E_p$, both mechanisms seem to
lead to a broad $E_p-L_{\gamma,\rm iso}^p$ relation. It is interesting
to note that the short GRB 090510 (the top yellow point), even
located at the upper boundary of the correlation, is still not
an outlier. This is consistent with the finding (Zhang et al. 2009;
Ghirlanda et al. 2009) that long/short GRBs are not clearly
distinguished in the $L_{\gamma,\rm iso}^p - E_p$ domain.

In Fig.\ref{L-Ep}b, we present the internal $E_p-L_{\gamma,\rm iso}$ 
correlation. It is interesting to note that although with scatter, 
the general positive correlation between $E_p$ and $L_{\gamma,\rm iso}$ 
as discovered by Liang et al. (2004) clearly stands. More interestingly, 
the BB-defined $E_p$ (in GRB 090902B) follows a similar trend to
the Band-defined $E_p$ (e.g. in GRB 080916C and GRB 090926A),
although different bursts occupy a different space region in
the $E_p-L_{\gamma,\rm iso}$ plane.

In Fig.\ref{correlations}, we present various pairs of spectral parameters 
in an effort to search for possible new correlations. The GRBs with
redshift measurements are marked in colors, while those without
redshifts are marked in gray with an assumed redshift $z=1$.
In order to show the trend of evolution, points for same
burst are connected, with the beginning of evolution marked as
a circle.

No clear correlation pattern is seen in the $E_p -\alpha$ and 
$E_p -\beta$ plots. Interestingly, a preliminary trend of
correlation is found in the following two domains.

\begin{itemize}
\item An $\alpha-\beta$ anti-correlation: Fig.\ref{correlations}a shows 
a rough anti-correlation between $\alpha$ and $\beta$ in individual GRBs. 
This suggests that a harder $\alpha$ corresponds to a softer $\beta$, 
suggesting a narrower Band function. In the time domain, there is evidence 
in some GRBs (e.g. GRB 080916C, GRB 090926A, and GRB 100414A, see 
Figs.\ref{080916C},\ref{090926A},\ref{100414A}) that the 
Band function ``opens up'' as time goes by, but the opposite trend is also 
seen in some GRBs (e.g. GRB 091031, Fig.\ref{091031}). {
The linear Pearson correlation coefficients for individual bursts are 
insert in Fig.\ref{correlations}a }

\item A flux-$\alpha$ correlation: Fig.\ref{correlations}b shows a rough 
correlation between flux and $\alpha$. Within the same burst,
there is rough trend that as the flux increases, $\alpha$ becomes
harder. The linear Pearson correlation coefficients for 
individual bursts are presented in Fig.\ref{correlations}b inset. 
One possible observational bias is that when flux is higher, one tends 
to get a smaller time slice based on the minimum spectral analysis 
criterion. If the time smearing effect can broaden the spectrum, then
a smaller time slice tends to give a narrower spectrum, and hence, a
harder $\alpha$. This would be relevant to bursts similar to GRB 090902B,
but not bursts similar to GRB 080916C (which does not show spectral
evolution as the time resolution becomes finer). More detailed analyses 
of bright GRBs can confirm whether such a correlation is intrinsic or
due to the time resolution effect discussed above.

\end{itemize}

Several caveats should be noted for these preliminary correlations:
First, some bursts do not obey these correlations, so the correlations,
if any, are not universal; Second, the currently chosen time bins are based
on the requirement for adequate spectral analyses, so the time resolution 
varies in different bursts. For some bright bursts, a burst pulse can be 
divided into several time bins, while in some faint others, a time bin 
corresponds to the entire pulse; Third, the current sample is still too 
small. A time-resolved spectral analysis for more bright GBM GRBs may 
confirm or dispute these correlations.

\section{Elemental Spectral Components and their physical origins}

\subsection{Three phenomenologically identified elemental spectral components}

The goal of our time-resolved spectral analysis is to look for
``elemental'' emission units that shape the observed GRB prompt
gamma-ray emission. In the past it has been known that time-integrated
GRB spectra are mostly fit by the Band function (Band et al. 1993).
However, whether this function is an elemental unit in the time-resolved 
spectra is not known. One speculation is that this function is the 
superposition of many simpler emission units. If such a superposition 
relies on adding the emission from many time slices, then these more 
elemental units should show up as the time bins become small enough.

One interesting finding of our time-resolved spectral analyses is that
the ``Band''-like spectral component seen in GRB 090902B is different
from that seen in GRB 080916C and some other Band-only GRBs. While the 
Band spectral indices of GRB 080916C essentially do not change as the
time bins become progressively smaller, that of GRB 090902B indeed
show the trend of ``narrowing'' as the time bin becomes progressively
smaller. With the finest spectral resolution, GRB 090902B spectra can
be fit by the superposition of a PL component and a CPL function,
including a Planck function. 
Even for the time-integrated spectrum, the ``Band''-like component in 
GRB 090902B appears ``narrower'' than that of GRB 080916C. All these 
suggest that the ``Band''-like component of GRB 090902B is
fundamentally different from that detected in GRB 080916C and probably also 
other Band-only GRBs\footnote{Our finest time interval is
around 1s in the rest frame of the burst. Theoretically, how time-integrated
spectra broaden with increasing time bins is subject further study. Our
statement is therefore relevant for time resolution longer than 1s.}. 
Similarly, the time-resolved spectra of the short 
GRB 090510 can be well fit by the superposition of a PL component and a 
CPL spectrum (although not a Planck function). The PL component extends
to high energies with a positive slope in $\nu F_\nu$. The CPL component
may be modeled as a multi-color blackbody spectrum. We therefore 
speculate that the MeV component of GRB 090510 is analogous 
to that of GRB 090902B.

Phenomenologically, the power law component detected in GRB 090902B and 
GRB 090510 is an extra component besides the Band-like component. Such
a component may have been also detected in the BATSE-EGRET burst
GRB 941017 (Gonz\'alez et al. 2003), and may also exist in GRB 090926A
at later epochs.

We therefore speculate that phenomenologically there might be three
elemental spectral components that shape the prompt gamma-ray spectrum.
These include: (I) a {\em Band function component} (``Band'' in 
abbreviation) that covers a wide energy range (e.g. 6-7 orders of magnitude
in GRB 080916C) and persists as the time bins become progressively 
smaller. It shows up in GRB 080916C and 13 other LAT GRBs;
(II) a {\em quasi-thermal component} (``BB'' in 
abbreviation\footnote{Notice that the abbreviation ``BB'' here not
only denotes blackbody, but also includes various modifications to 
the blackbody spectrum such as multi-color blackbody.}) which 
becomes progressively narrower as the time bin becomes smaller, and 
eventually can be represented as a blackbody (or multi-color blackbody)
component as seen in GRB 090902B; (III) a {\em power law component} 
(``PL'' in abbreviation)
that extends to high energy as seen in GRBs 090902B and 090510, 
which has a positive slope in the $\nu F_\nu$ spectrum and should have 
an extra peak energy ($E_p$) at an even higher energy that is not well
constrained by the data.

Figure \ref{Cartoon} is a cartoon picture of 
the $\nu F_\nu$ spectrum that includes
all three phenomenologically identified elemental spectral components. 
The time resolved spectra of the current sample can be understood as
being composed of one or more of these components. 
For example, GRB 080916C and other 13 GRBs have Component I (Band),
GRB 090902B and probably GRB 090510 have Components II (BB) and III
(PL), and GRB 0900926A has Component I initially, and may have 
components I and III at later times.

\subsection{Possible physical origins of the three spectral components}

\subsubsection{Band Component}

The fact that the this component extends through a wide energy range
(e.g. 6-7 orders of magnitude for GRB 080916C) strongly suggests
that a certain non-thermal emission mechanism is in operation.
This demands the existence of a population of power-law-distributed 
relativistic electrons, possibly accelerated in
internal shocks or in regions with significant electron heating,
e.g.  magnetic dissipation.
In the past there have been three model
candidates for prompt GRB emission: synchrotron emission,
synchrotron self-Compton (SSC), and Compton upscattering of a thermal
photon source. In all these models the high energy PL component
corresponds to emission from a PL-distributed electron population.
The spectral peak energy $E_p$ may be related to the minimum 
energy of the injected electron population, 
an electron energy distribution break, or the
peak of the thermal target photons. 

Most prompt emission modeling (M\'esz\'aros et al. 1994; Pilla \& 
Loeb 1998; Pe'er \& Waxman 2004a; Razzaque et al. 2004; Pe'er et al. 
2006; Gupta \& Zhang 2007) suggest that
the overall spectrum is curved, including multiple spectral
components. Usually a synchrotron component is accompanied by
a synchrotron self-Compton (SSC) component. For matter-dominated
fireball models, one would expect the superposition of emissions
from the photosphere and from the internal shock dissipation
regions. As a result, the fact that 14/17 ($\sim 80\%$) of GRBs 
in our sample have a Band-only spectrum is intriguing. 
The three theoretically expected spectral features, i.e. the 
quasi-thermal photosphere emission, the SSC component (if the
MeV component is of synchrotron origin), and a pair-production
cutoff at high energies, are all not observed.
This led to the suggestion that the outflows of these GRBs
are Poynting flux dominated (Zhang \& Pe'er 2009). Within
such a picture, the three missing features can be understood
as the following: (1) Since most
energy is carried in magnetic fields and not in photons, the
photosphere emission (BB component) is greatly suppressed;
(2) Since the magnetic energy density is higher than the
photon energy density, the Compton $Y$ parameter is smaller
than unity, so that the SSC component is naturally suppressed;
(3) A Poynting flux dominated model usually has a larger emission
radius than the internal shock model (Lyutikov \& Blandford 2003
for current instability and Zhang \& Yan 2011 for collision-induced
magnetic reconnection/turbulence). This reduces the two-photon
annihilation opacity and increases the pair cutoff energy. This
allows the Band component extend to very high energy (e.g.
13.2 GeV for GRB 080916C). 

Another possibility, advocated by Beloborodov (2010) and Lazzati
\& Begelman (2010) in view of the {\em Fermi} data {(see also 
discussion by Thompson 1994; Rees \& M\'esz\'aros 2005; Pe'er 
et al. 2006; Giannios \& Spruit 2007; Fan 2009; Toma et al. 2010; 
and Ioka 2010)}, is that the Band component is the emission from 
a dissipative photosphere. This 
model invokes relativistic electrons in the regions where Thomson 
optical depth is around unity, which upscatter photosphere thermal 
photons to high energies to produce a power law tail. This model can 
produce a Band-only spectrum, but has two specific limitations.
First, the high energy power law component cannot extend to
energies higher than GeV in the cosmological rest frame, since for 
effective upscattering, the emission region cannot be too far above the
photosphere. The highest photon energy detected in GRB 080916C
is 13.2 GeV (which has a rest-frame energy $\sim 70$ GeV for its
redshift $z=4.35$). This disfavors the dissipative photosphere
model. {This argument applies if the LAT-band photons are 
from the same emission region as the MeV photons, as suggested by 
the single Band function spectral fits. It has been suggested that 
the LAT emission during the prompt phase originates from a 
different emission region, e.g. the external shock (Kumar \& Barniol 
Duran 2009; Ghisellini et al. 2010). This requires that the two 
distinct emission components
conspire to form a nearly featureless Band spectrum in all temporal
epochs, which is contrived. As will be shown in Sect.\ref{sec:tracking}
later, there is compelling evidence that the LAT emission during the
prompt emission phase is of an internal origin. In particular, the 
peak of the GeV lightcurve of GRB 080916C coincides with the second (the 
brightest) peak of GBM emission, and the 13.2 GeV photon coincides
with another GBM lightcurve peak. All these suggest an internal origin 
of the GeV emission during the prompt phase.}

The second limitation of the dissipative photosphere model is 
that the photon spectral index below $E_p$ is not easy to reproduce. 
The simplest blackbody model predicts a Rayleigh-Jeans spectrum 
$\alpha=+1$. By considering slow heating, this index can be modified
as $\alpha=+0.4$ (Beloborodov 2010).
Both are much harder from the observed $\alpha \sim -1$ value.
In order to overcome this difficulty, one may appeal to the 
superposition effect, i.e. the observed Band spectrum is the
superposition of many fundamental blackbody emission units
(e.g. Blinnikov et al. 1999; Toma et al. 2010; Mizuta et al. 2010;
Pe'er \& Ryde 2010). However, no rigorous 
calculation has been performed to fully reproduce the $\alpha=-1$ 
spectrum. Pe'er \& Ryde (2010) show that when the central engine
energy injection is over and the observed emission is dominated by 
the high-latitude emission, an $\alpha=-1$ can be reproduced with the
flux decaying rapidly with $\propto t^{-2}$. During the phase when the
central engine is still active, the observed emission is always 
dominated by the contribution along the line of sight, which 
should carry the hard low energy spectral index of the blackbody 
function. Observationally, the Band component 
spectral indices are not found to vary when the time bins are reduced
(in stark contrast to the narrow Band-like component identified
in GRB 090902B). This suggests that at least the temporal
superposition of many blackbody radiation units is not the right 
interpretation for this component.

\subsubsection{Quasi-Thermal (BB) Component}

The MeV component in GRB 090902B narrows with reduced time resolution
and eventually turns
into being consistent with a blackbody (or multi-color blackbody)
as the time bin becomes small enough. This suggests a thermal
origin of this component. Within the GRB content, a natural source
is the emission from the photosphere where the photons advected in
the expanding relativistic outflow turn optically thin for Compton
scattering. In fact, the original fireball model 
predicts a quasi-thermal spectrum (Pacz\'ynski 1986; Goodman 1986). 
In the fireball shock model, such a quasi-blackbody component is 
expected to be associated with the non-thermal emission components 
(M\'esz\'aros \& Rees 2000; M\'esz\'aros et al. 2002; Daigne \& Mochkovitch 2002; Pe'er et al.
2006).

Some superposition effects may modify the thermal spectrum to
be different from a pure Planck function. The first is the
temporal smearing effect. If the time bin is large enough,
one samples photosphere emission from many episodes, and
hence, the observed spectrum should be a multi-color blackbody.
This effect can be diminished by reducing the time bin for
time-resolved spectral analyses. GRB 090902B is such an example. 
The second effect is inherited in emission physics of relativistic
objects. At a certain epoch, the observer detects photons coming 
from different latitudes from the line of sight, with different 
Doppler boosting factors. The result is an intrinsic smearing of
the Planck function spectrum. Pe'er \& Ryde (2010) have shown
that after the central engine activity ceases, the high-latitude emission
effect would give an $\alpha \sim -1$ at late times, with a rapidly
decaying flux $F_\nu \propto t^{-2}$. This second superposition
effect is intrinsic, and cannot be removed by reducing the
time bins. 

The case of the thermal component is most evidenced in
GRB 090902B, and probably also in GRB 090510. 
In both bursts, the MeV component can be well fit with a CPL + PL 
spectrum. The exponential cutoff at the high
energy end is consistent with thermal emission with essentially
no extra dissipation. For GRB 090902B, the low energy spectral
index $\Gamma_{\rm CPL}$ is typically $\sim 0$, and can be 
adjusted to $+1$ (blackbody). For GRB 090510, $\Gamma_{\rm CPL}$
is softer ($\sim -0.7$). Since it is
a short GRB, the high-latitude effect may be more important.
The softer low energy spectral index may be a result of the
intrinsic high-latitude superposition effect (Pe'er \& Ryde 2010).

\subsubsection{Power-Law (PL) Component}

This component is detected in GRB 090902B and GRB 090510.
Several noticeable properties of this component are: (1) For our
small sample, this component is always accompanied by a low energy 
MeV component (likely the BB component).
Its origin may be related to this low energy component; (2) It is
demanded in both the low energy end and the high energy end, and
amazingly the same spectral index can accommodate the demanded
excesses in both ends. This suggests that either this PL component
extends for 6-7 orders of magnitude in energy, or that multiple
emission components that contribute to the excesses in both the
low and high energy regimes have to coincide to mimic a single
PL; (3) The spectral slope is positive in the $\nu F_\nu$ space, 
so that the main energy power output { of this component}
is at even higher energies (possibly near or above the upper bound 
of the LAT band). 

Since the non-thermal GRB spectra are expected to be curved
(M\'esz\'aros et al. 1994; Pilla \& Loeb 1998; Pe'er \&
Waxman 2004a; Razzaque et al. 2004; Pe'er et al. 2006;
Gupta \& Zhang 2007; Asano \& Terasawa 2009),
the existence of the PL component is not
straightforwardly expected. It demands coincidences of 
various spectral components to mimic a single PL component 
in the low and high energy ends.
Pe'er et al. (2010) have presented a theoretical model of 
GRB 090902B. According to this model, the apparent PL observed
in this burst is the combination of the synchrotron emission
component (dominant at low energies), the SSC and Comptonization
of the thermal photons (both dominant at high energies). A
similar model was analytically discussed by Gao et al. (2009) 
within the context of GRB 090510.

One interesting question is how Component III (PL) differs from 
Component I (Band). Since both components are non-thermal,
they may not be fundamentally different. They can be two 
different manifestations of some non-thermal emission
mechanisms (e.g. synchrotron and inverse Compton scattering)
under different conditions. On the other hand, since Component III 
seems to be associated with Component II (BB) (e.g. in GRB 090902B
and GRB 090510), its origin may be related to Component II.
One possible scenario is that Component III {(at least 
the part above component II)} is the Compton-upscattered
emission of Component II (e.g. Pe'er \& Waxman 2004b for GRB 941017).
The fact that the lightcurves of the BB component and the PL component
of GRB 090902B roughly track each other (Fig.\ref{090902B-lightcurves})
generally supports such a possibility. { Within this
interpretation, one must attribute the PL part below the thermal
peak as due to a different origin (e.g. synchrotron, see Pe'er
et al. 2010).}
Alternatively, Component I and III may be related to non-thermal 
emission from two different emission sites (e.g. internal vs. 
external or two different internal locations). Indeed, if the late
spectra of GRB 090926A are the superposition of the components
I and III, then both components can coexist, which may
correspond to two different non-thermal emission processes
and/or two different emission sites.

\subsection{Possible Spectral Combinations of GRB Prompt Emission}

Using the combined GBM and LAT data, we have phenomenologically 
identified three elemental spectral components during the prompt 
GRB phase (Fig.\ref{Cartoon}). Physically they may have different 
origins (see above). One may speculate that all the GRB prompt 
emission spectra may be decomposed into one or more of these spectral
components. It is therefore interesting to investigate how many 
combinations are in principle possible, how many have been 
discovered, how many should not exist and why, and how many 
should exist and remain to be discovered. We discuss the following 
possibilities in turn below (see Fig.\ref{spectral-combinations}
for illustrations).

\begin{enumerate}

\item Component I (Band) only:

This is the most common situation, which is observed in 14/17 GRBs 
in our sample exemplified by GRB 080916C. Either the BB and PL 
components do not exist, or they are too faint to be detected above 
the Band component. If the BB component is suppressed, 
these bursts may signify non-thermal emission 
from an Poynting flux dominated flow. 

\item Component II (BB) only:

No such case exists in the current sample. GRB 090902B, and probably
also GRB 090510, have a BB component, but it is accompanied by a PL
component in both GRBs. It remains to be seen whether in the future a 
BB-only GRB will be detected, or whether a BB component is always accompanied
by a PL component. Since the case of GRB 090902B is rare, we
suspect that the BB-only GRBs are even rarer, if they exist at all.

\item Component III (PL) only:

Our PL component stands for the high energy spectral component seen in 
GRB 090902B and GRB 090510, which likely has a high $E_p$ near or above
the boundary of the LAT band.
Observationally, there is no solid evidence for such 
PL-only GRBs\footnote{Most of Swift GRBs can be fit with a PL 
(Sakamoto et al. 2008). However, 
this is due to the narrowness of the energy band of 
the gamma-ray detector BAT on board Swift. The $E_p$ of many Swift
GRBs are expected to be located outside the instrument band. In fact,
using a Band function model and considering the variation of $E_p$
within and outside the BAT band, one can reproduce the apparent
hardness of Swift GRBs, and obtain an effective correlation
between the BAT-band photon index and $E_p$ (e.g. Zhang et al.
2007; Sakamoto et al. 2009). If a GRB is observed in a wider
energy band, the spectrum should be invariably curved.}. In our 
current sample which covers the widest energy band, the PL
component only exists in 2 out of 17 GRBs, and is found to be
associated with the BB component. 
The luminosity of the PL component is found to roughly
track that of the thermal component (Fig.\ref{090902B-lightcurves}).
If the PL component is the Comptonization of a low energy photon
source (e.g. the BB component), then 
PL-only GRBs may not exist in nature.

\item I + II:

Such a case is not found in our sample. If the Band component is the
emission from the internal shocks and the BB component is the emission
from the photosphere, then such a combination should exist and be common
for fireball scenarios.
An identification of such a case would confirm the non-thermal nature
of the Band component (since the thermal component is manifested as
the BB component).  Observationally, an X-ray excess has been observed 
in 12 out of 86 ($\sim 14\%$) bright BATSE GRBs
(Preece et al. 1996). This could be due to the contamination of a
BB component in the X-ray regime. With the excellent spectral
coverage of {\em Fermi}, we expect that such a spectral combination 
may be identified in some GRBs, even if technically it may be 
difficult because there are too many spectral parameters to constrain.

\item I + III:

Such a combination has not been firmly identified in our sample.
Nonetheless, the spectral hardening of GRB 090926A after 11 s may
be understood as the emergence of the PL component on top of
the Band component seen before 11 s. Physically it may be related
to two non-thermal spectral components or non-thermal emission
from two different regions.

\item II + III:

Such a case is definitely identified in GRB 090902B, and likely
in GRB 090510 as well. From the 
current sample, it seems that such a combination is not as common as
the Band-only type, but nonetheless forms a new type of spectrum that
deserves serious theoretical investigations.
Physically, the high-energy PL component is likely the Compton up-scattered 
emission of the BB component, although other non-thermal processes
(e.g. synchrotron and SSC) could also contribute to the observed emission
(Pe'er et al. 2010).

\item I + II + III:

The full combination of all three spectral components (e.g. Fig.\ref{Cartoon})
is not seen from the current sample. In any case, in view of the above
various combinations (including speculative ones), one may assume that 
the full combination of the three spectral components is in principle
possible. Physically this may correspond to one photosphere emission 
component and two more non-thermal components (either two spectral
components or non-thermal emission from two different regions).
Nonetheless, technically there are too many
parameters to constrain, so that identifying such a combination
is difficult. 

\end{enumerate}

\section{LAT-band emission vs. GBM-band emission}

Besides the joint GBM/LAT spectral fits, one may also
use temporal information to investigate the relationship
between the emission detected in the GBM-band and that detected
in the LAT band.  In this section we discuss three topics:
delayed onset of LAT emission, rough tracking behavior between 
GBM and LAT emissions, and long-lasting LAT afterglow.

\subsection{Delayed onset of LAT emission}

The {\em Fermi} team has reported the delayed onset of LAT
emission in several GRBs (GRBs 080825C, 080916C, 090510, 090902B,
Abdo et al. 2009a,b,c,d). Our analysis confirms all these results.
In Table 1, we mark all the GRBs in our sample that show the
onset delay feature.

There have been several interpretations to the delayed onset of GeV
emission discussed in the literature. Toma et al. (2009) suggested
that GeV emission is the upscattered cocoon emission by the internal
shock electrons. Razzaque et al. (2009) interpreted the GeV emission
as the synchrotron emission of protons. Since it takes a longer time
for protons to be accelerated and be cooled to emit GeV photons, 
the high energy emission is delayed.
Li (2010b) interpreted GeV emission as the upscattered prompt
emission photons by the residual internal shocks.

Although it is difficult to test these models using the available
data, our results give some observational constraints to these models.
First, except GRBs 090510 and 090902B whose GeV emission is
a distinct spectral component, other GRBs with onset delay still
have a simple Band-function spectrum after the delayed onset. 
This suggests that
for those models that invoke two different emission components
to interpret the MeV and GeV components, one needs to
interpret the coincidence that the GeV emission appears as
the natural extension of the MeV emission to the high energy regime.

For such delayed onsets whose GeV and MeV emissions form the same
Band component, one may speculate two simpler explanations.
One is that there might be a change in the particle acceleration 
conditions (e.g. magnetic configuration in the particle acceleration 
region). As shown in Sect.\ref{sec:080916C}, the early spectrum during 
the first time bin (before onset of LAT emission) of GRB 080916C
may be simply a consequence of changing the electron
spectral index. One may speculate that early on the particle
acceleration process may not be efficient,
so that the electron energy spectral index is steep. After a while
(the observed delay), the particle acceleration mechanism becomes
more efficient, so that the particle spectral index reaches the
regular value. The second possibility is that
there might be a change in opacity.
The GBM data alone during the first time bin gives a similar 
$\beta$ as later epochs. 
It is possible that there might be a spectral cutoff slightly
above the GBM band early on. A speculated physical picture would 
be that the particle acceleration conditions are similar throughout
the burst duration, but early on the pair production opacity may
be large (probably due to a lower Lorentz factor or a smaller
emission radius), so that the LAT band emission is attenuated.
The opacity later drops (probably due to the increase of Lorentz
factor or the emission radius), so that the LAT band emission
can escape from the GRB. {Within such a scenario, one
would expect to see a gradual increase of maximum photon energy
as a function of time. Figure 25 shows the LAT photon arrival
time distribution of GRB 080916C. Indeed one can see a rough
trend of a gradual increase of the maximum energy with time.}

One last possibility is that the LAT band emission is dominated
by the emission from the external shock, which is delayed with
respect to the GBM-band prompt emission. This possibility
is discussed in more detail below in Sec. \ref{sec:LATafterglow}.

\subsection{Rough tracking behavior}\label{sec:tracking}

Inspecting the multi-band lightcurves (Figs.\ref{080825C}-\ref{100414A} 
left panels), for bright GRBs (e.g. 080916C, 090217, 090323, 090902B)
the LAT emission peaks seem to roughly track some peaks of the
GBM emission (aside from the delayed onset for some of them). 
For example, the peak of the LAT lightcurve of GRB 080916C
coincides with the second GBM peak. This is consistent with the
spectral analysis showing that most time-resolved joint spectra
are consistent with being the same (Band-function) spectral
component. Even for GRB 090902B whose LAT band emission is from
a different emission component from the MeV BB component, the
emissions in the two bands also roughly track each other
(Fig.\ref{090902B-lightcurves}). This suggests that the two physical 
mechanisms that power the two spectral components are related to each 
other.

The rough tracking behavior is evidence against the proposal
that the entire GeV emission is from the external forward shock
(see Sec. \ref{sec:LATafterglow} for more discussion). 
Within the forward shock model, the fluctuation in energy
output from the central engine should be greatly smeared,
since the observed flux change amplitude is related to 
$\Delta E/E \ll 1$ (where $E$ is the total energy already 
in the balstwave, and $\Delta E$ is the newly injected energy
from the central engine), rather than $\Delta E$ itself within
the internal models.

\subsection{Long Term Emission in the LAT Band}\label{sec:LATafterglow}

\begin{figure}
\begin{tabular}{ll}
\multirow{3}{*}{ \includegraphics[angle=0,scale=0.78]{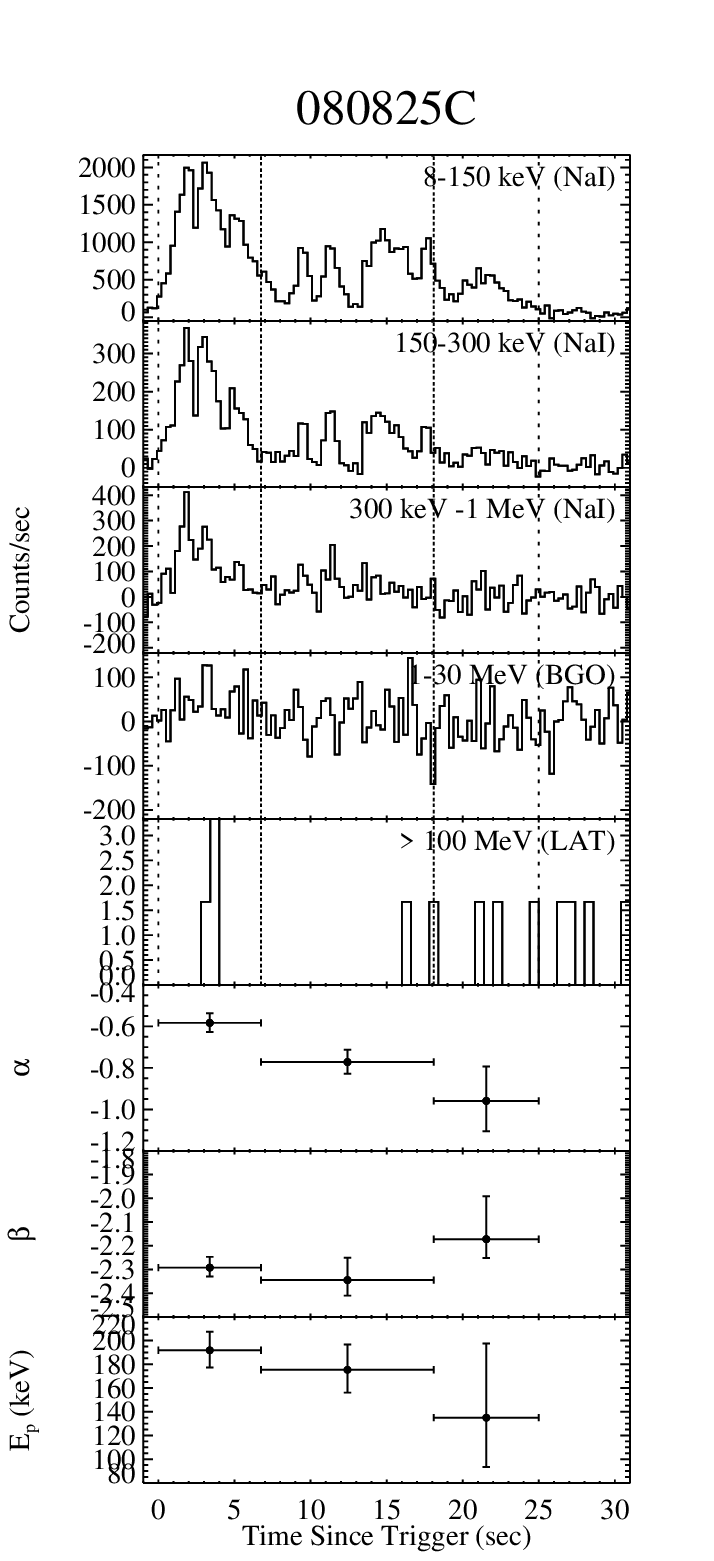}}
 & \includegraphics[angle=270,scale=0.2]{f1b.ps} \\
 & \includegraphics[angle=0,scale=.33]{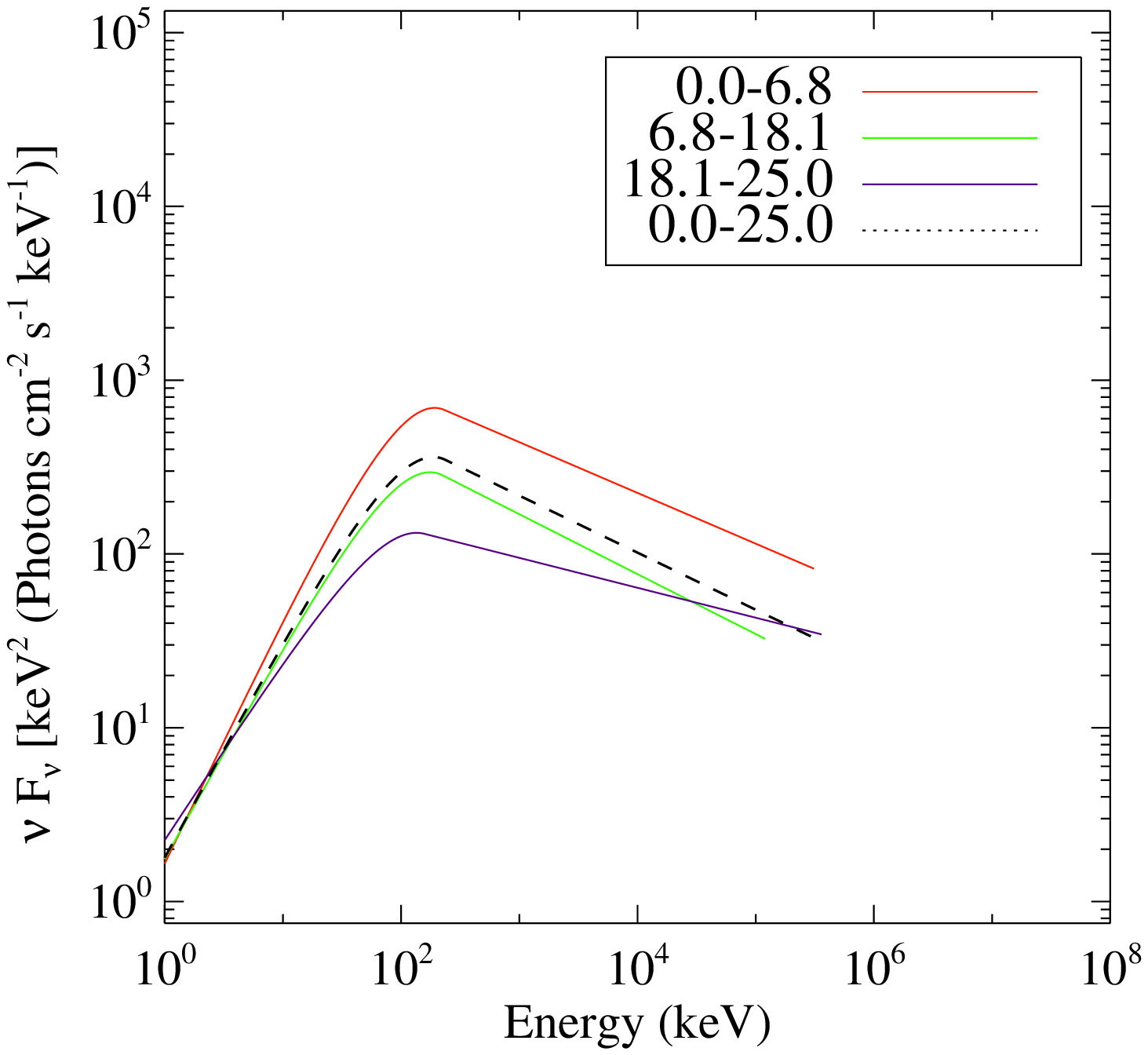} \\
  &  \includegraphics[angle=0,scale=.33]{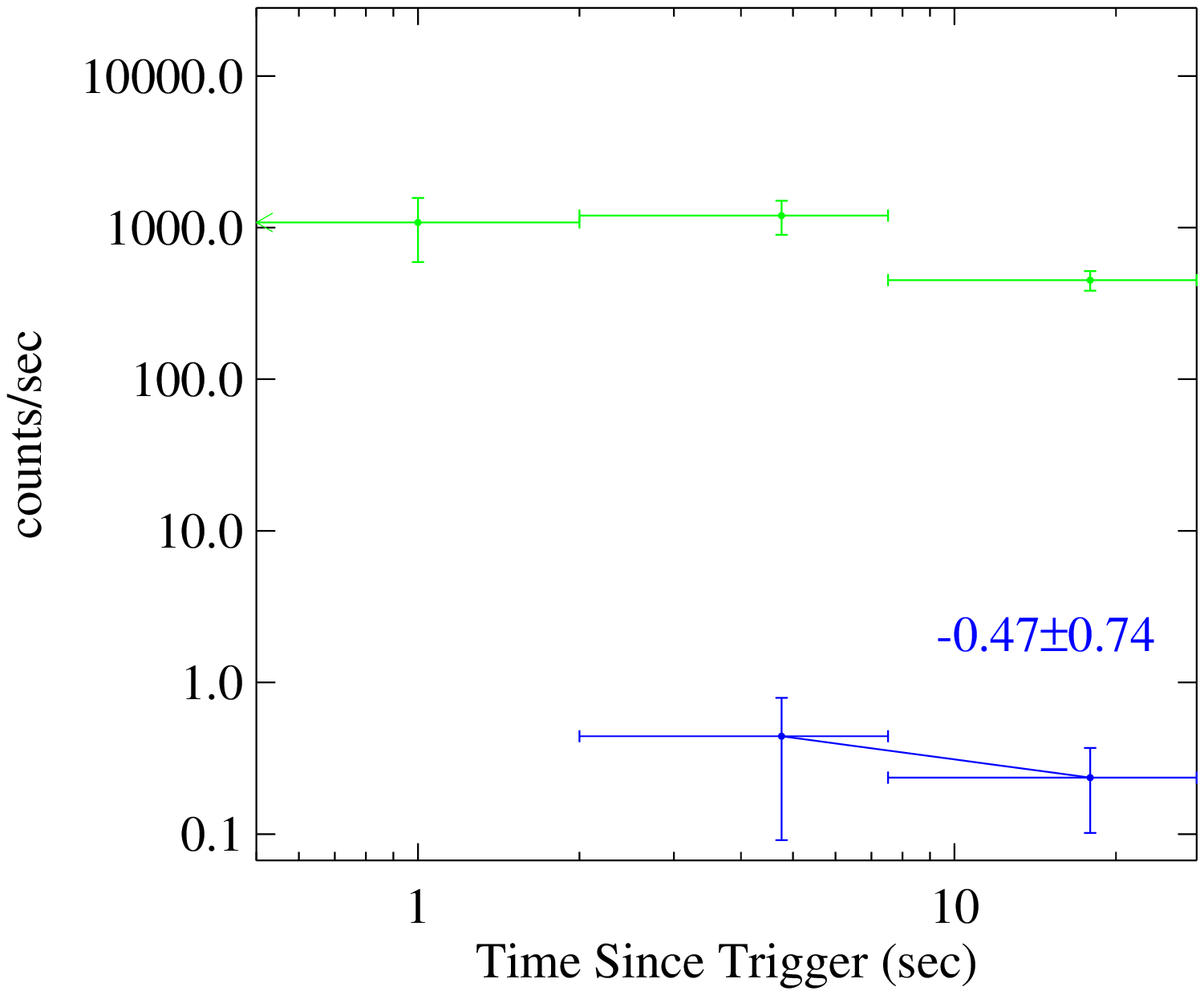}  \\
 \end{tabular}
\caption{Joint temporal and spectral analysis of GBM and LAT data for GRB
080825C. {\em Left panels:} the background-subtracted GBM and LAT lightcurves
(from top: 8-150 keV, 150-300 keV, 300 keV - 1 MeV, 1-30 MeV, $>$100 MeV),
and evolution of spectra parameters ($\alpha$, $\beta$, $E_p$).
{\em Right panels:} an example (the brightest episode) of the observed photon
spectrum as compared with the spectral model ({\em top}), the best fit
$\nu F_\nu$ spectra of all time bins ({\em middle}), and the comparison
between the GBM (green) and LAT (blue) count rate lightcurves in log-scale
({\em bottom}).}
\label{080825C}
\end{figure}

\begin{figure}
\begin{tabular}{ll}
\multirow{3}{*}{ \includegraphics[angle=0,scale=0.78]{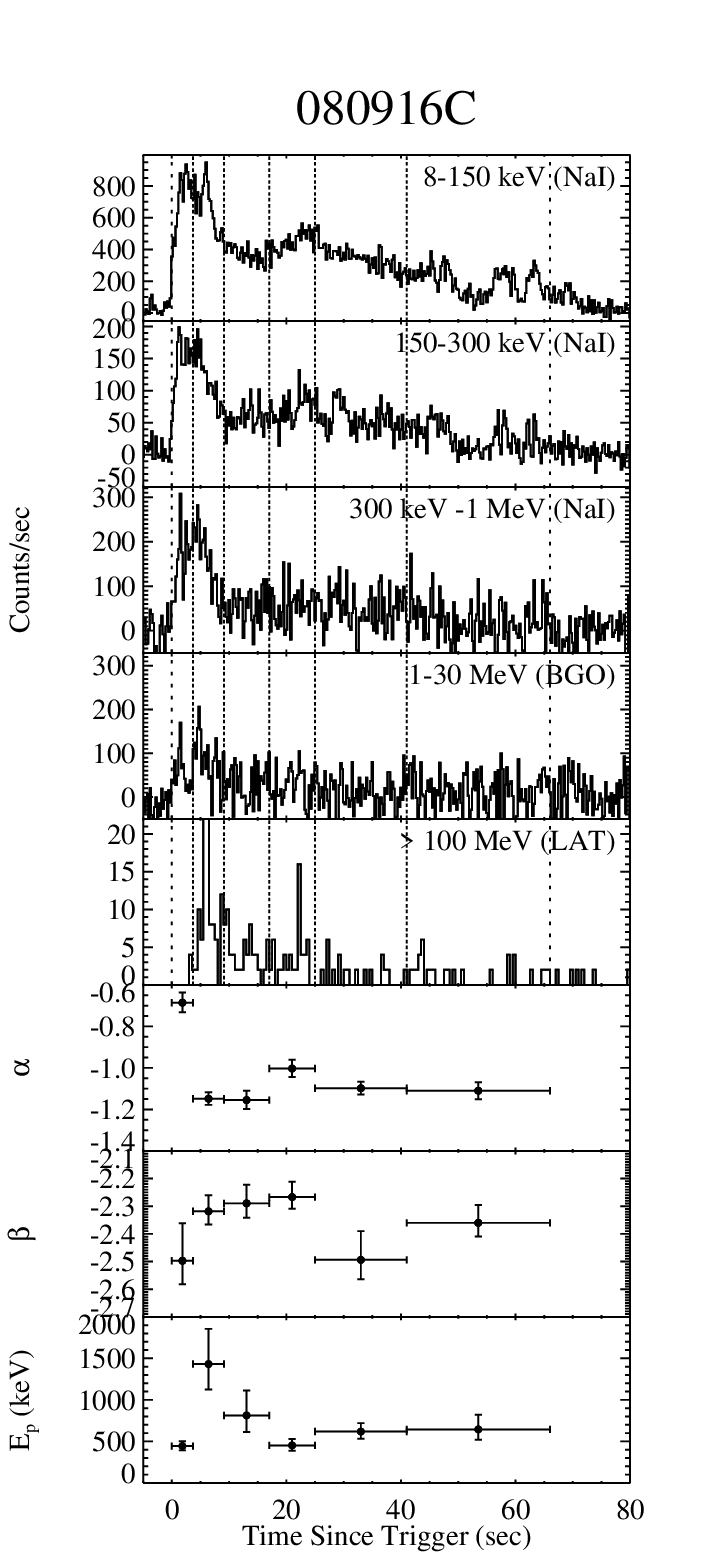}}
 & \includegraphics[angle=270,scale=0.2]{f2b.ps} \\
 & \includegraphics[angle=0,scale=.33]{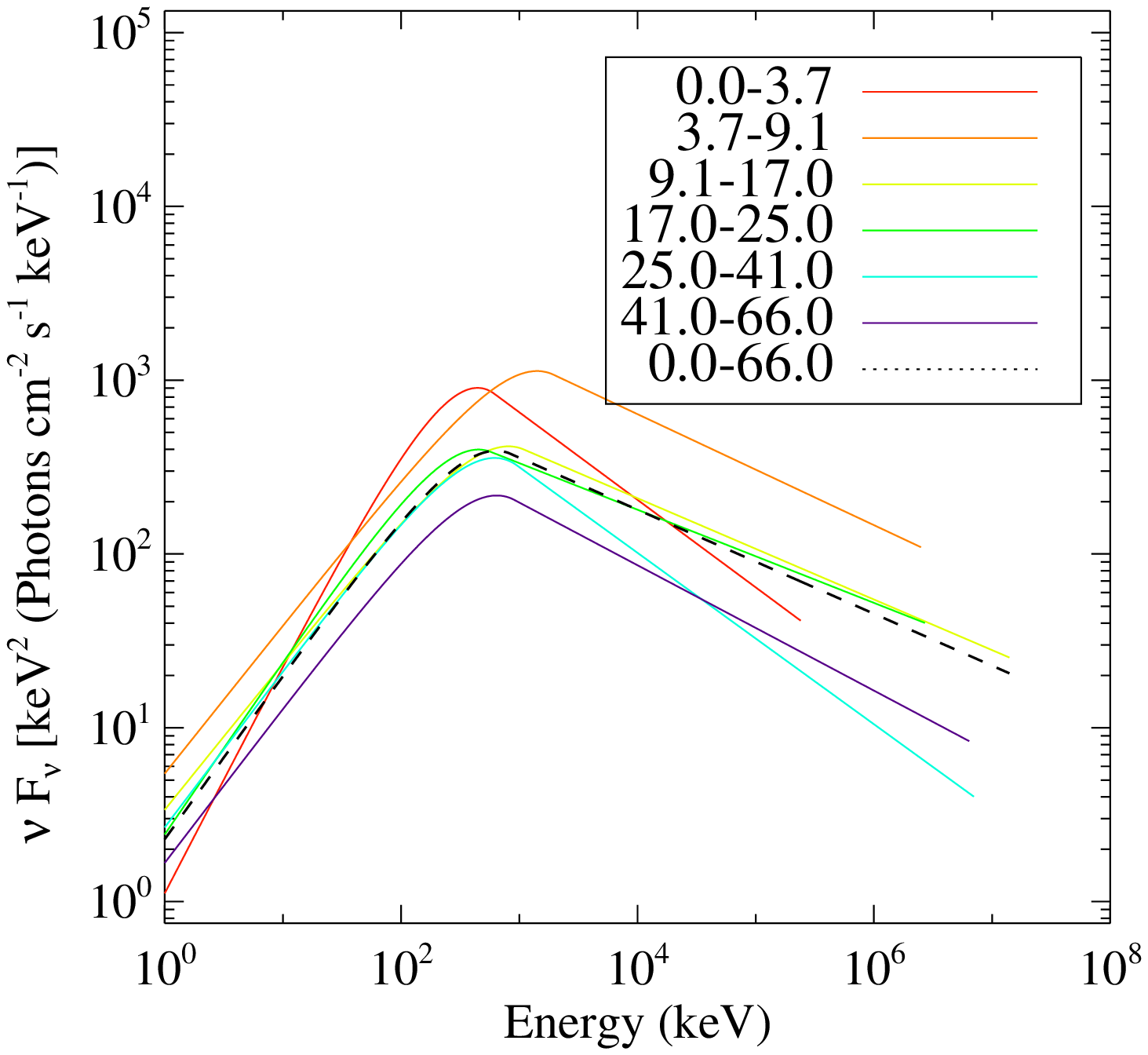} \\
  &  \includegraphics[angle=0,scale=.33]{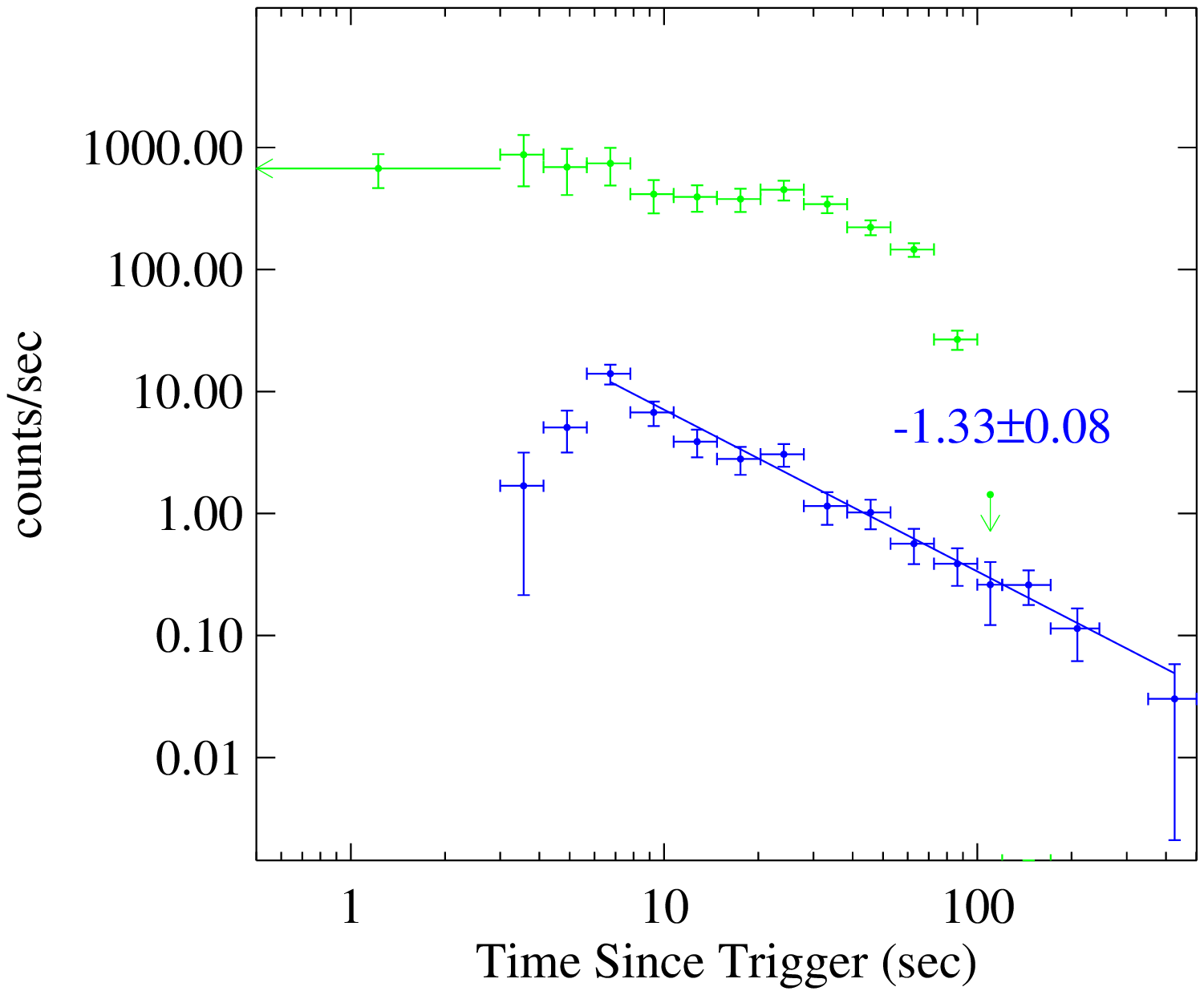}  \\
 \end{tabular}

\caption{Same as Figure 1, but for GRB 080916C.}
\label{080916C}
\end{figure}

\begin{figure}
\begin{tabular}{ll}
\multirow{3}{*}{ \includegraphics[angle=0,scale=0.78]{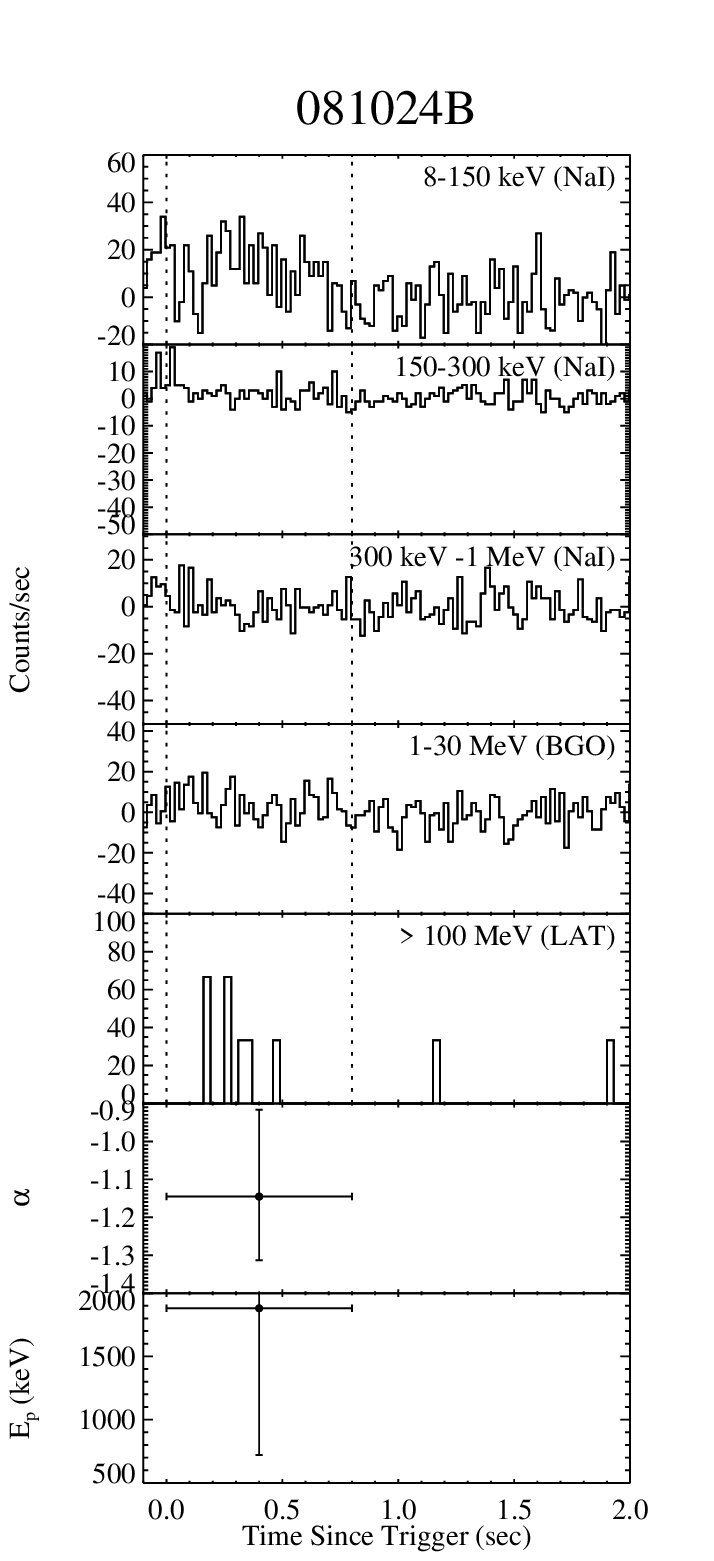}}
 & \includegraphics[angle=270,scale=0.2]{f3b.ps} \\
 & \includegraphics[angle=0,scale=.33]{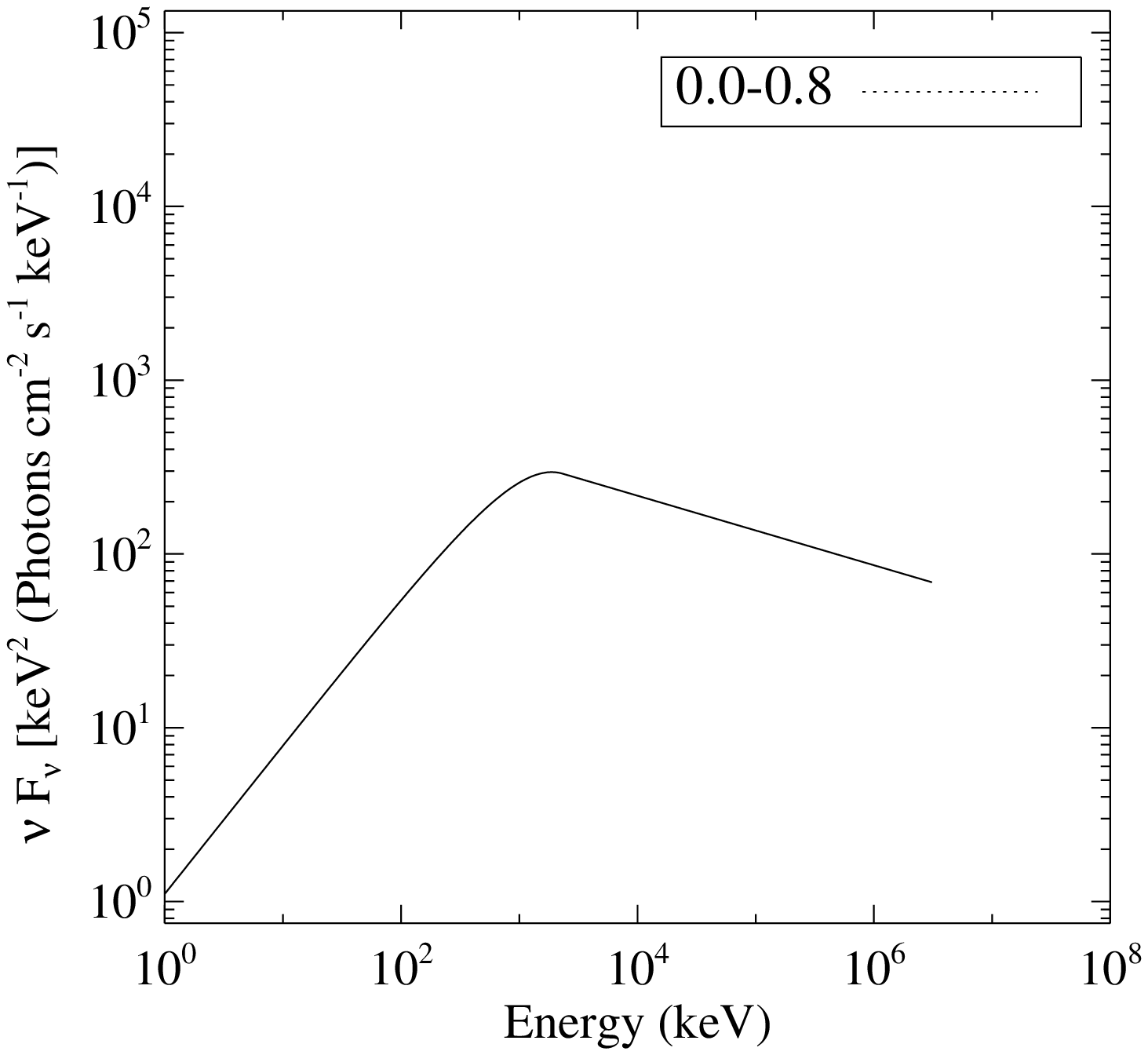} \\
  &  \includegraphics[angle=0,scale=.33]{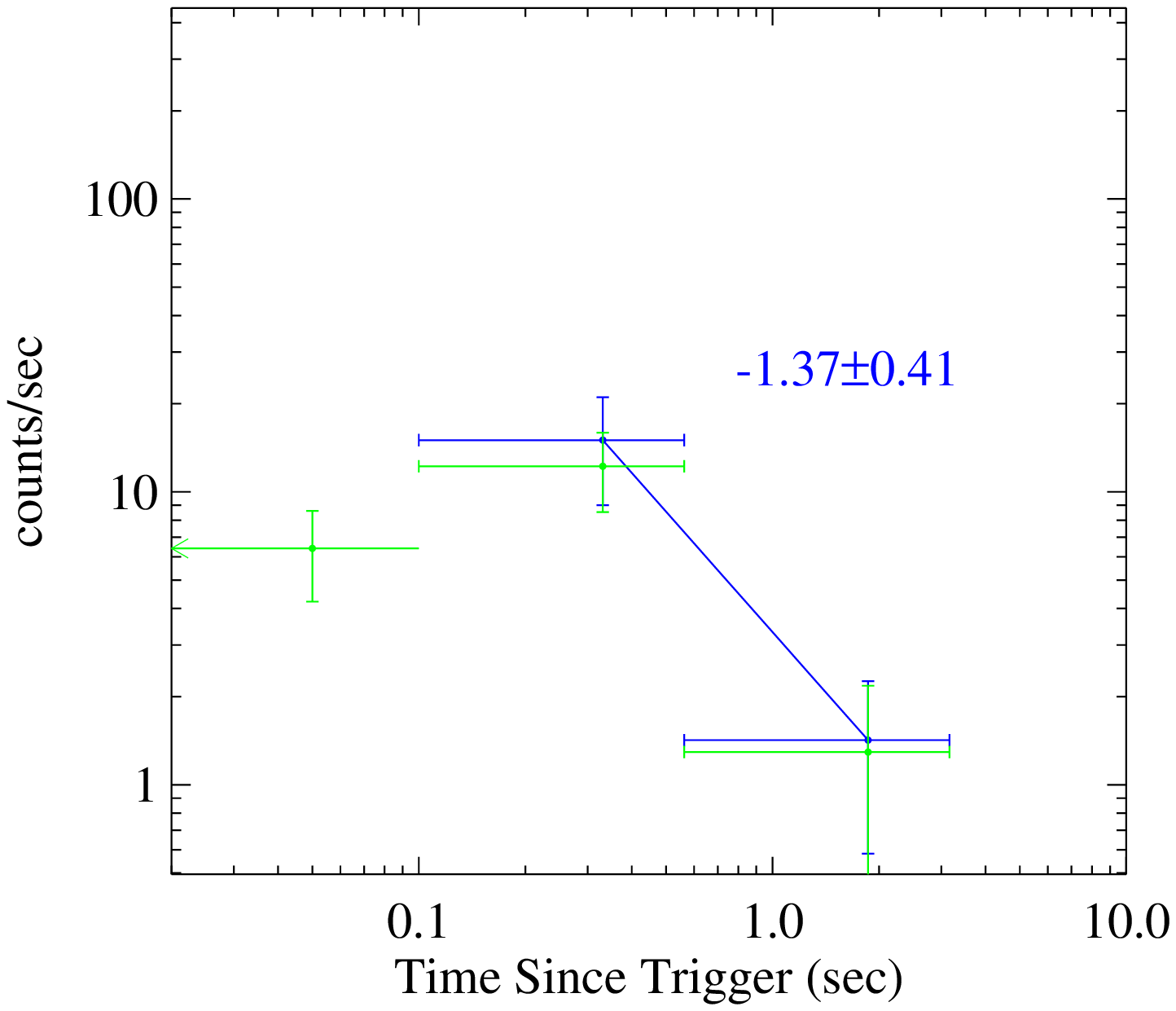}  \\
 \end{tabular}
\caption{Same as Figure 1, but for GRB 081024B.}
\label{081024B}
\end{figure}

\begin{figure}
\begin{tabular}{ll}
\multirow{3}{*}{ \includegraphics[angle=0,scale=0.78]{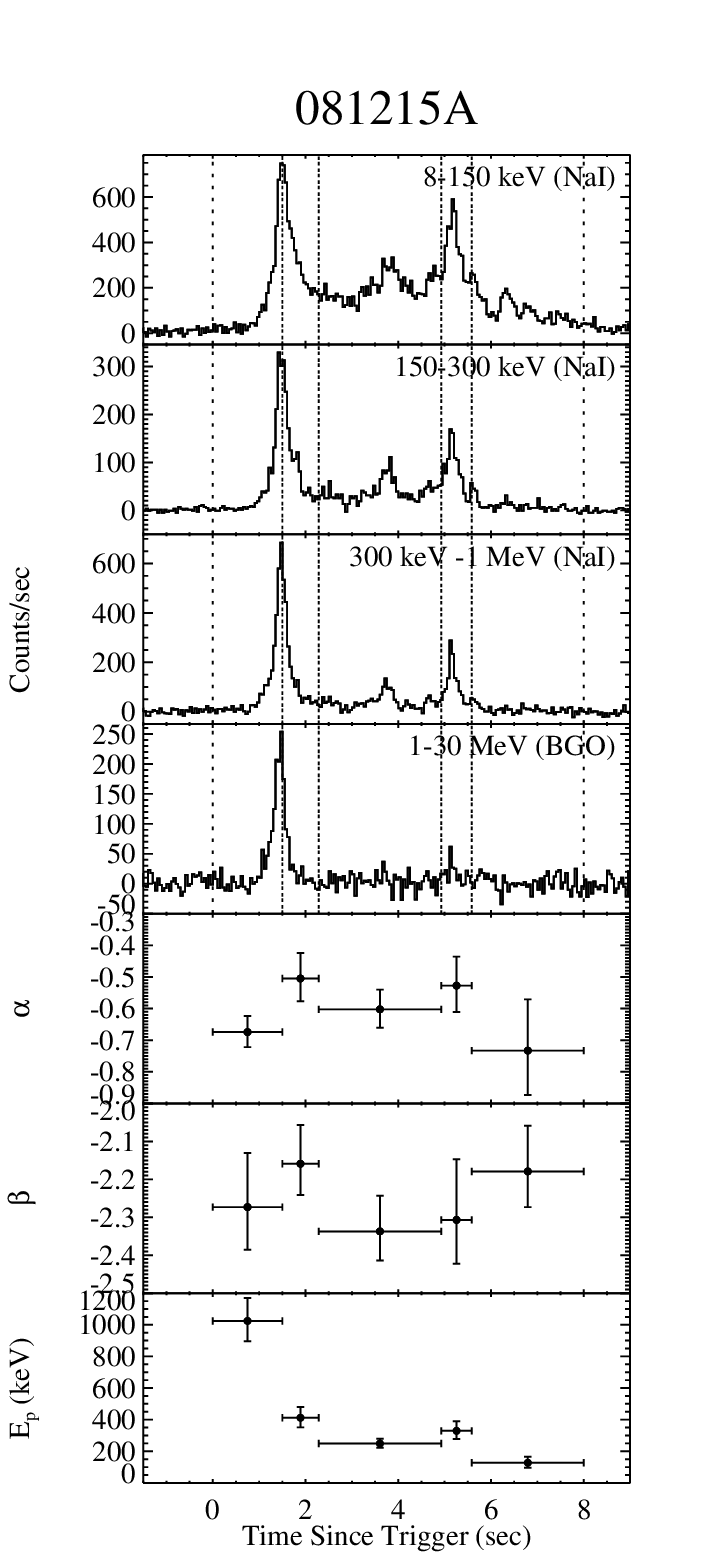}}
 & \includegraphics[angle=270,scale=0.2]{f4b.ps} \\
 & \includegraphics[angle=0,scale=.33]{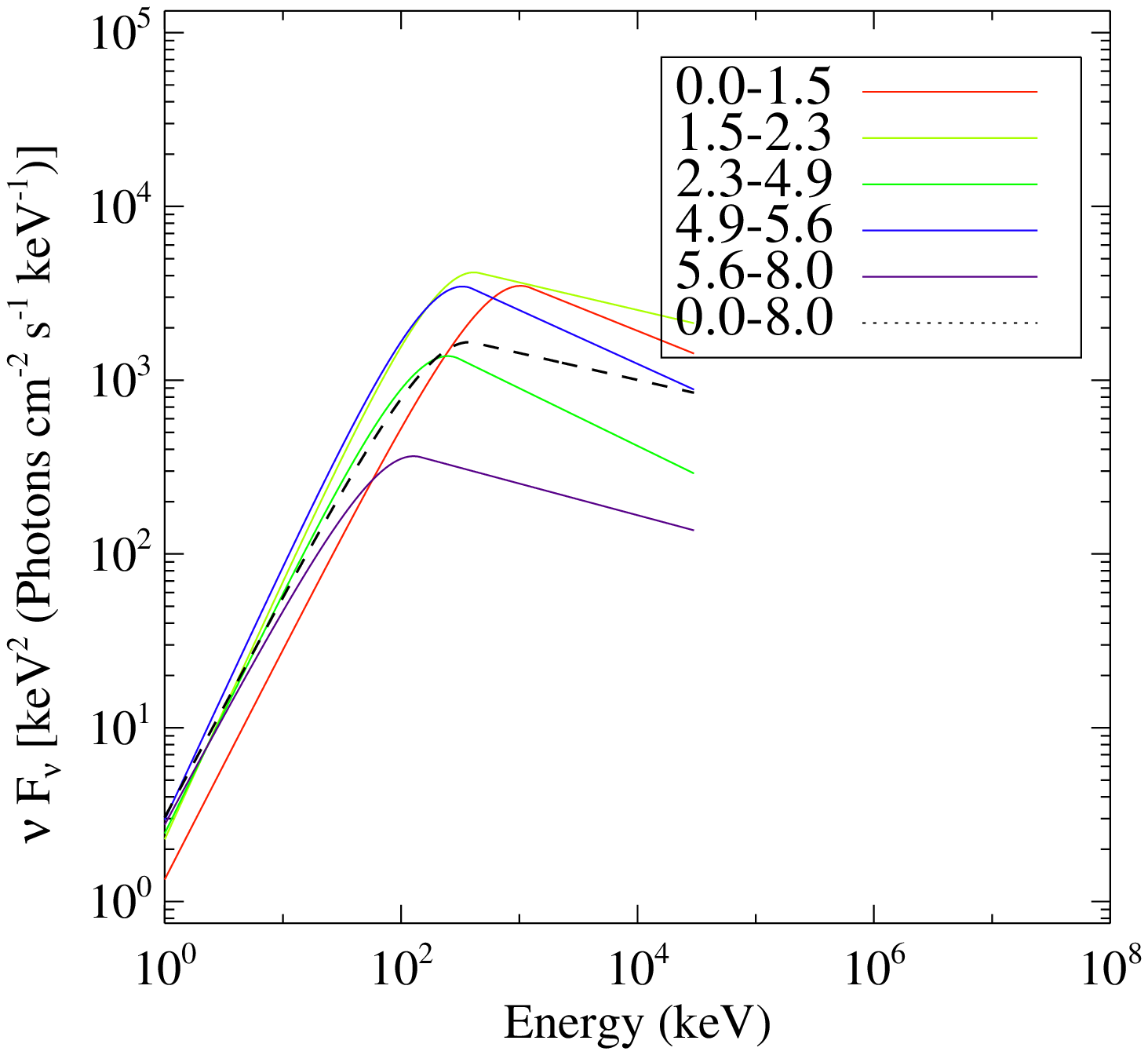} \\
 & \includegraphics[angle=0,scale=.33]{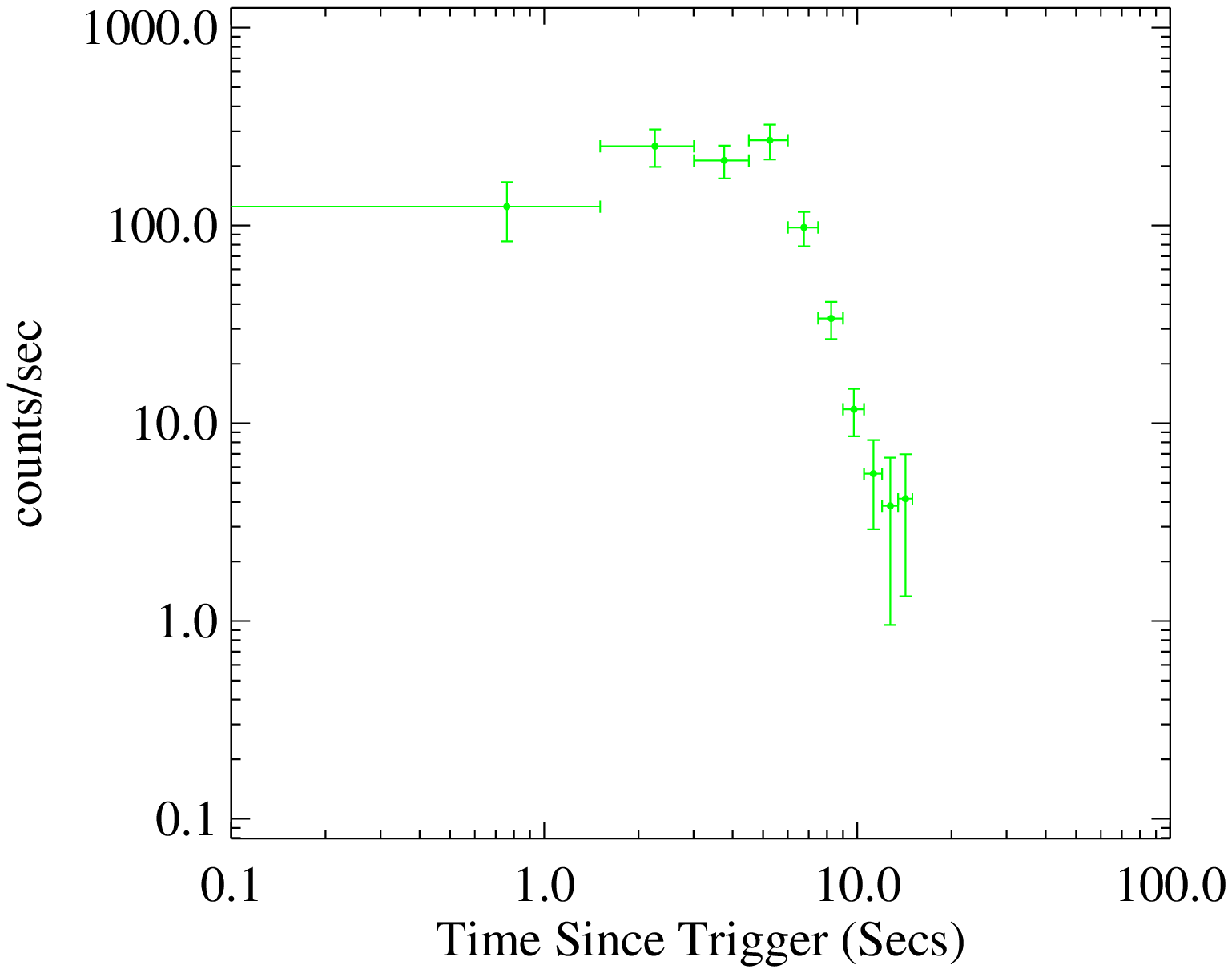}   \\
 \end{tabular}

\caption{Same as Figure 1, but for GRB 081215A. This burst was at an angle
of 86 degrees to the LAT boresight. The data cannot be obtained
with the standard analysis procedures. Using a non-standard data selection,
over 100 counts above background were detected within a 0.5 s interval in
coincidence with the main GBM peak (McEnery et al. 2008). We thus add this
GRB in our sample, but do not add its LAT data in our analysis.}
\label{081215A}
\end{figure}

\begin{figure}
\begin{tabular}{ll}
\multirow{3}{*}{ \includegraphics[angle=0,scale=0.78]{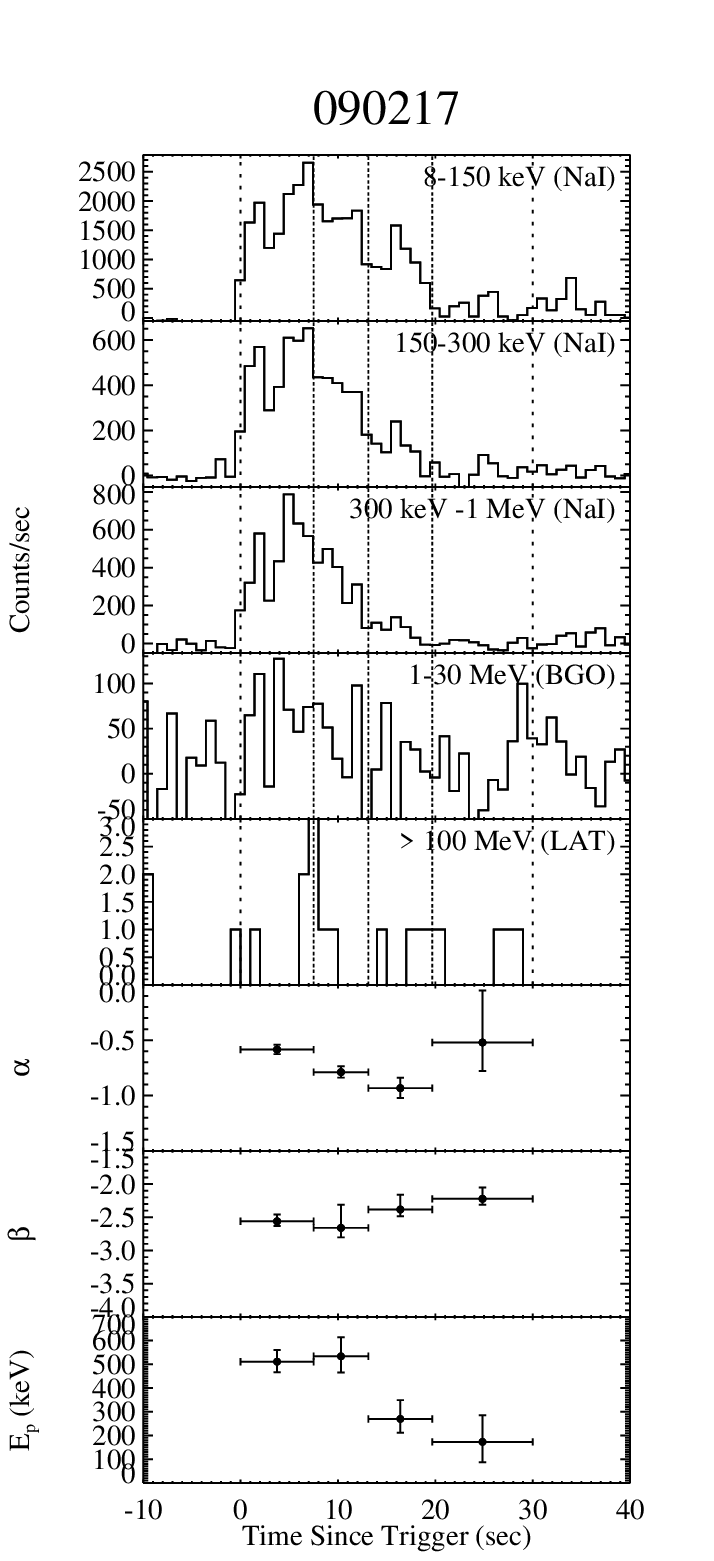}}
 & \includegraphics[angle=270,scale=0.2]{f5b.ps} \\
 & \includegraphics[angle=0,scale=.33]{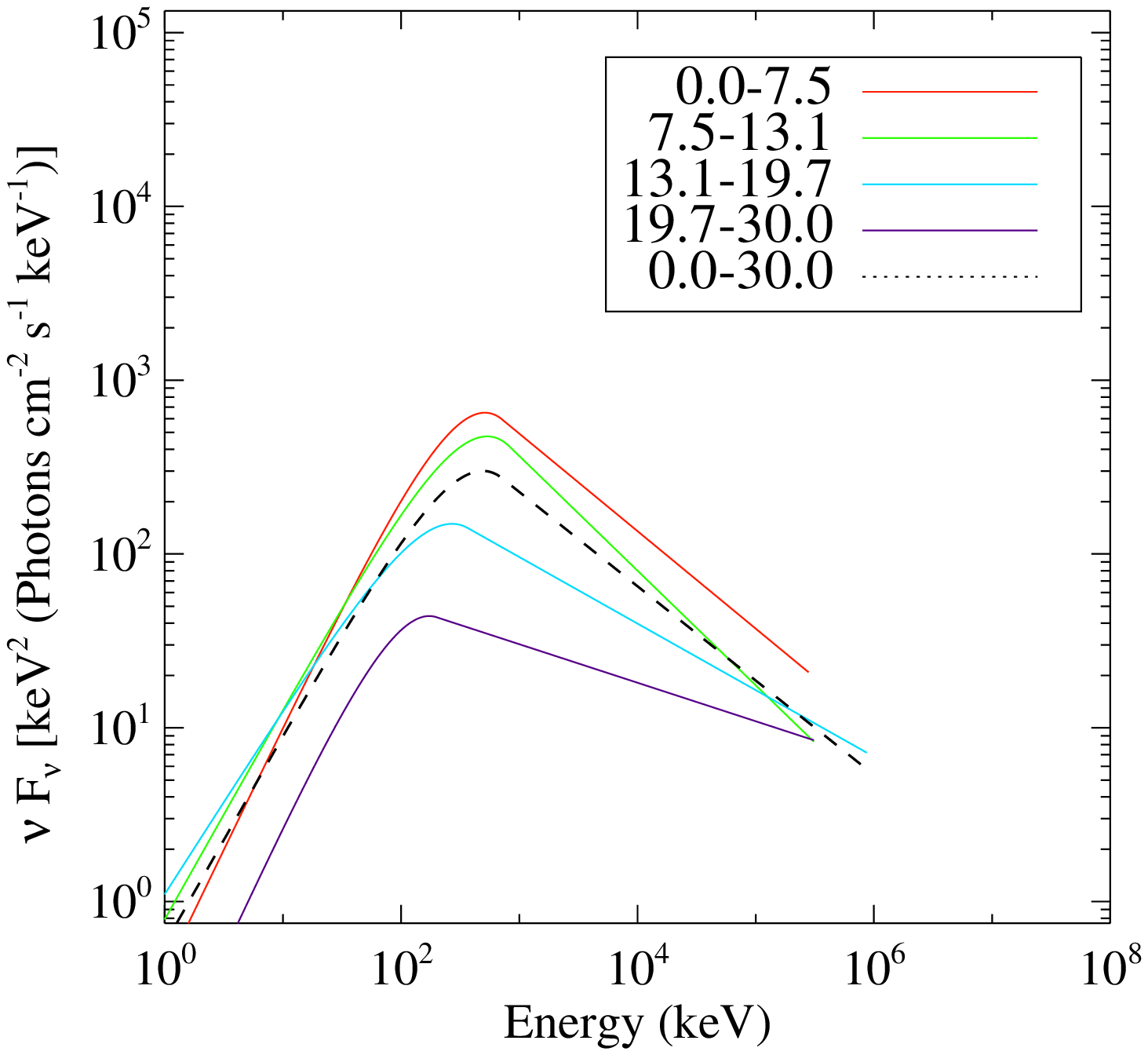} \\
  &  \includegraphics[angle=0,scale=.33]{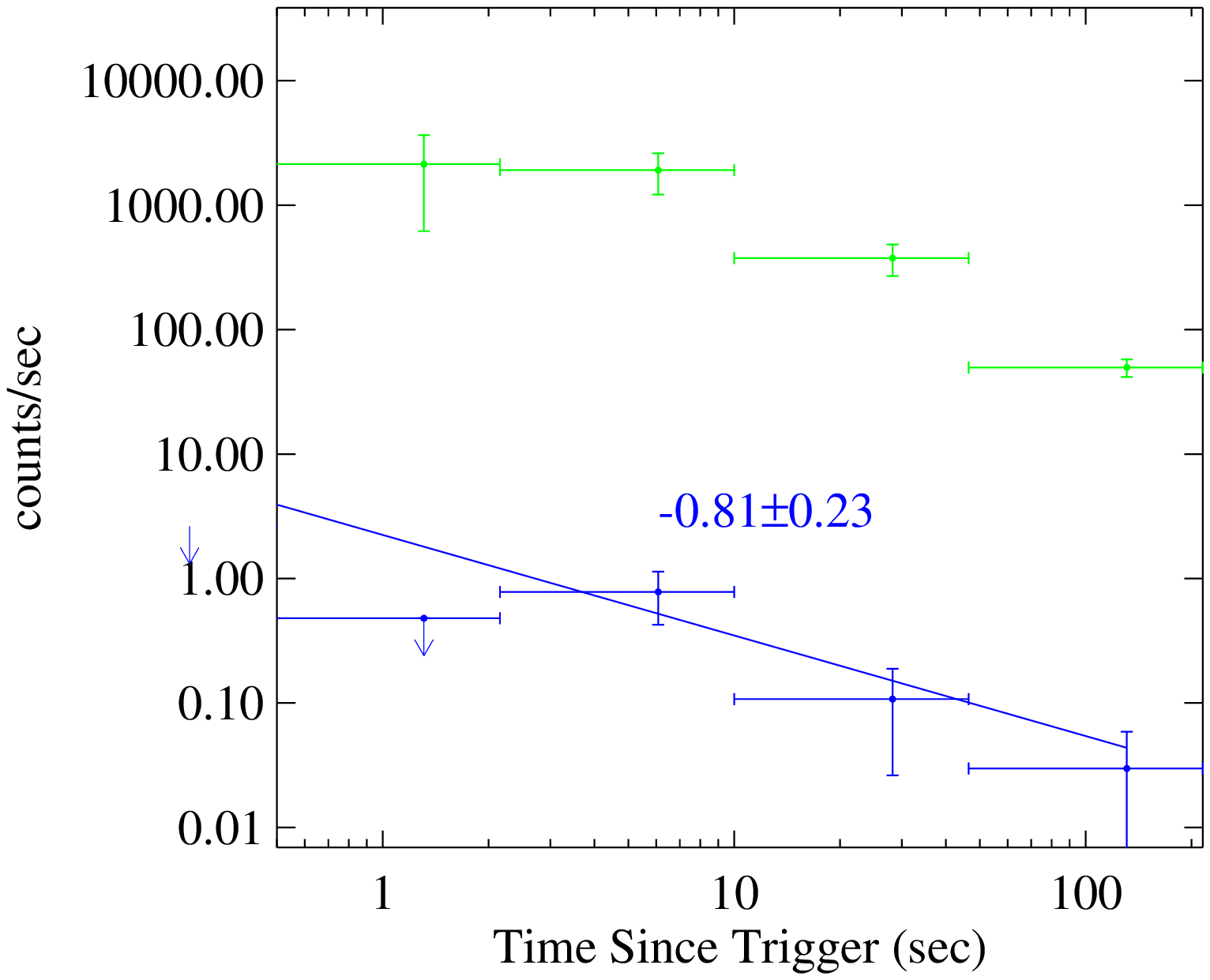}  \\
 \end{tabular}

\caption{Same as Figure 1, but for GRB 090217.}
\label{090217}
\end{figure}

\begin{figure}
\begin{tabular}{ll}
\multirow{3}{*}{ \includegraphics[angle=0,scale=0.78]{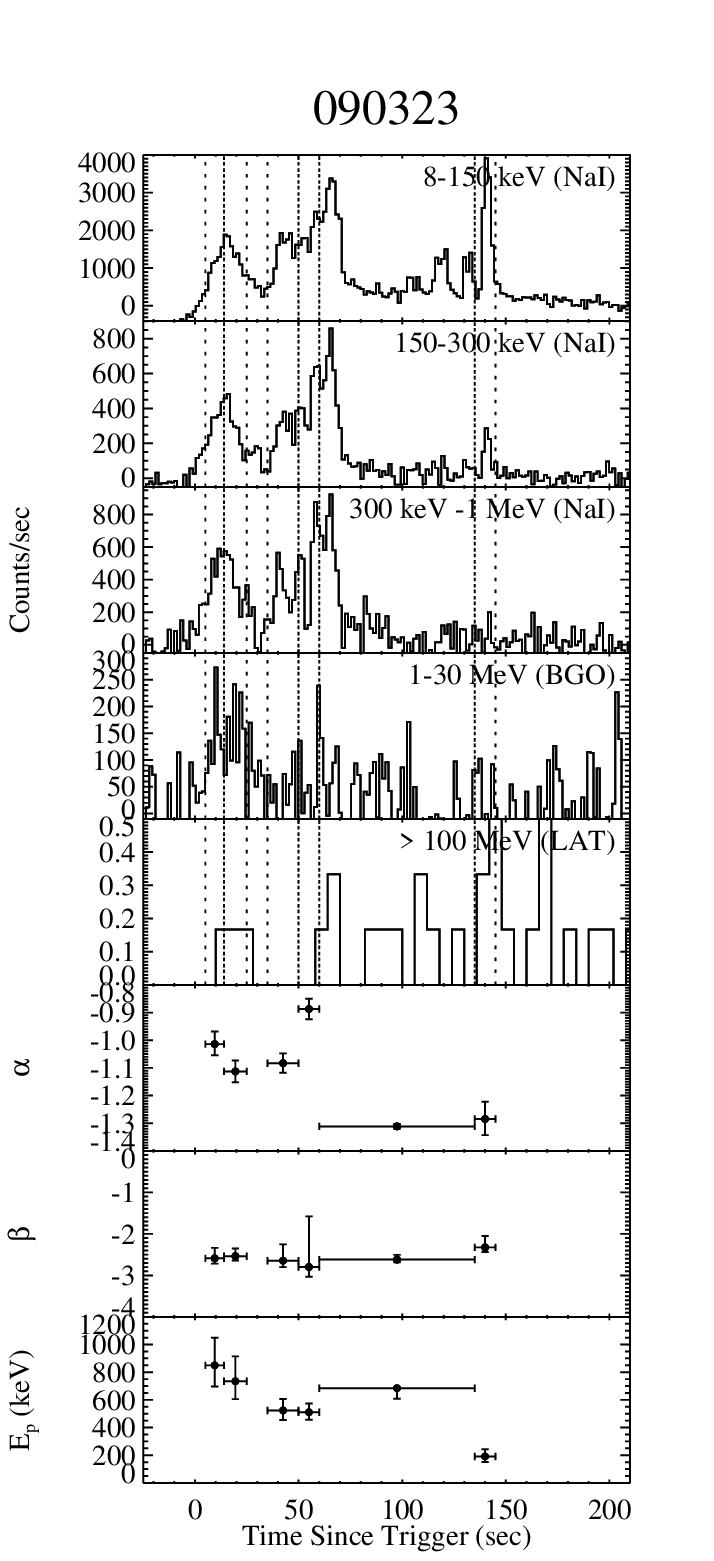}}
 & \includegraphics[angle=270,scale=0.2]{f6b.ps} \\
 & \includegraphics[angle=0,scale=.33]{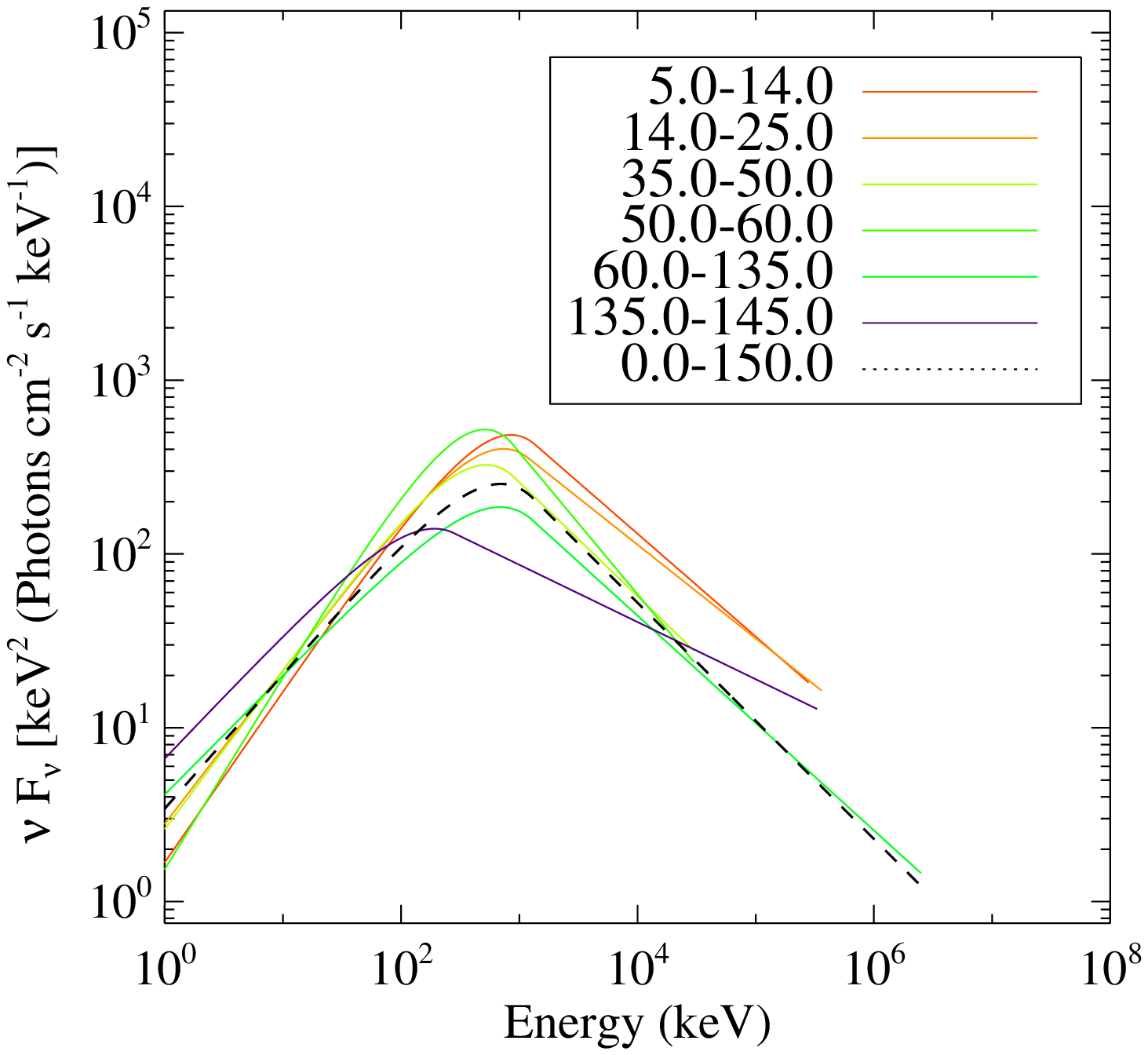} \\
  &  \includegraphics[angle=0,scale=.33]{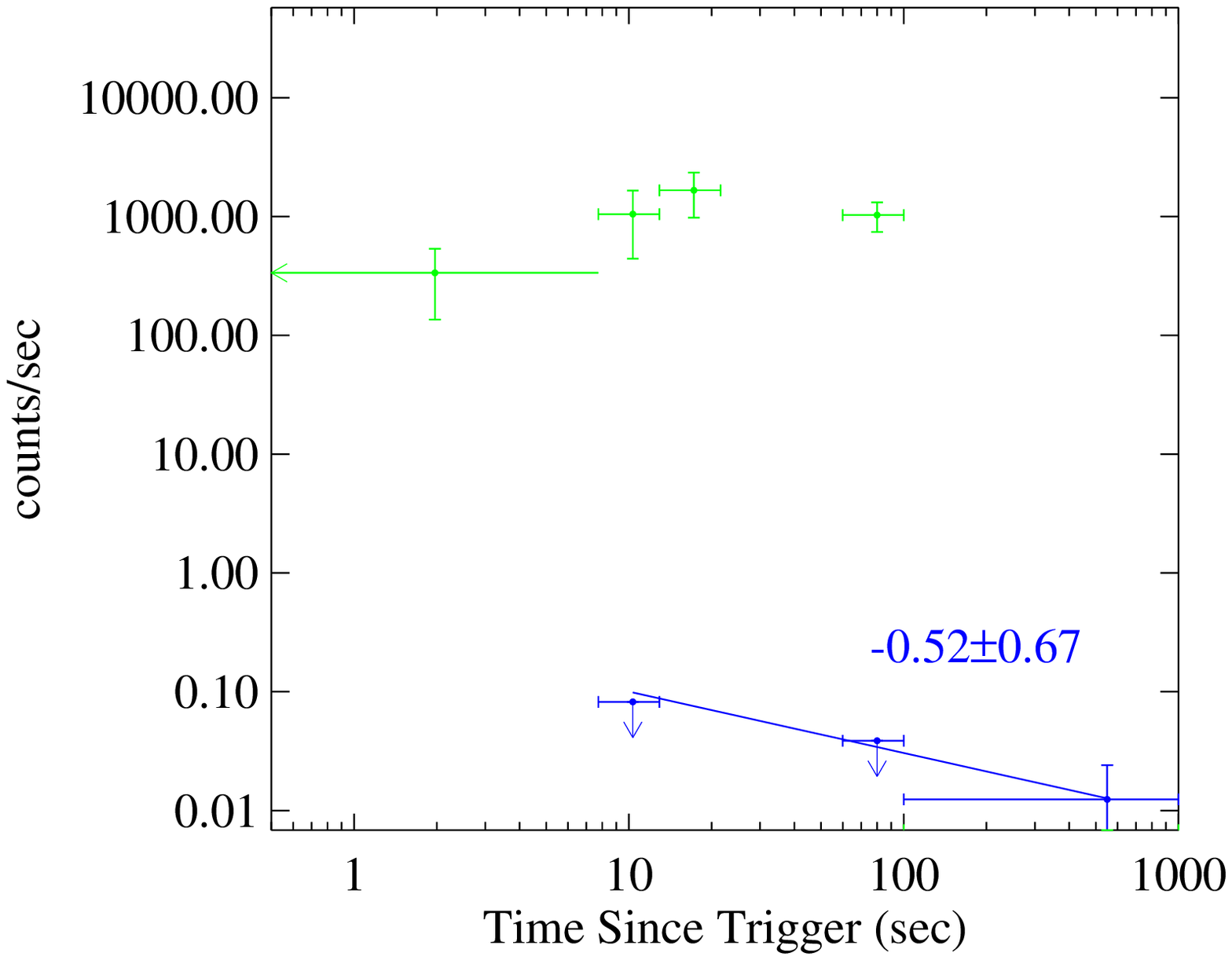}  \\
 \end{tabular}
\caption{Same as Figure 1, but for GRB 090323.}
\label{090323}
\end{figure}

\begin{figure}
\begin{tabular}{ll}
\multirow{3}{*}{ \includegraphics[angle=0,scale=0.78]{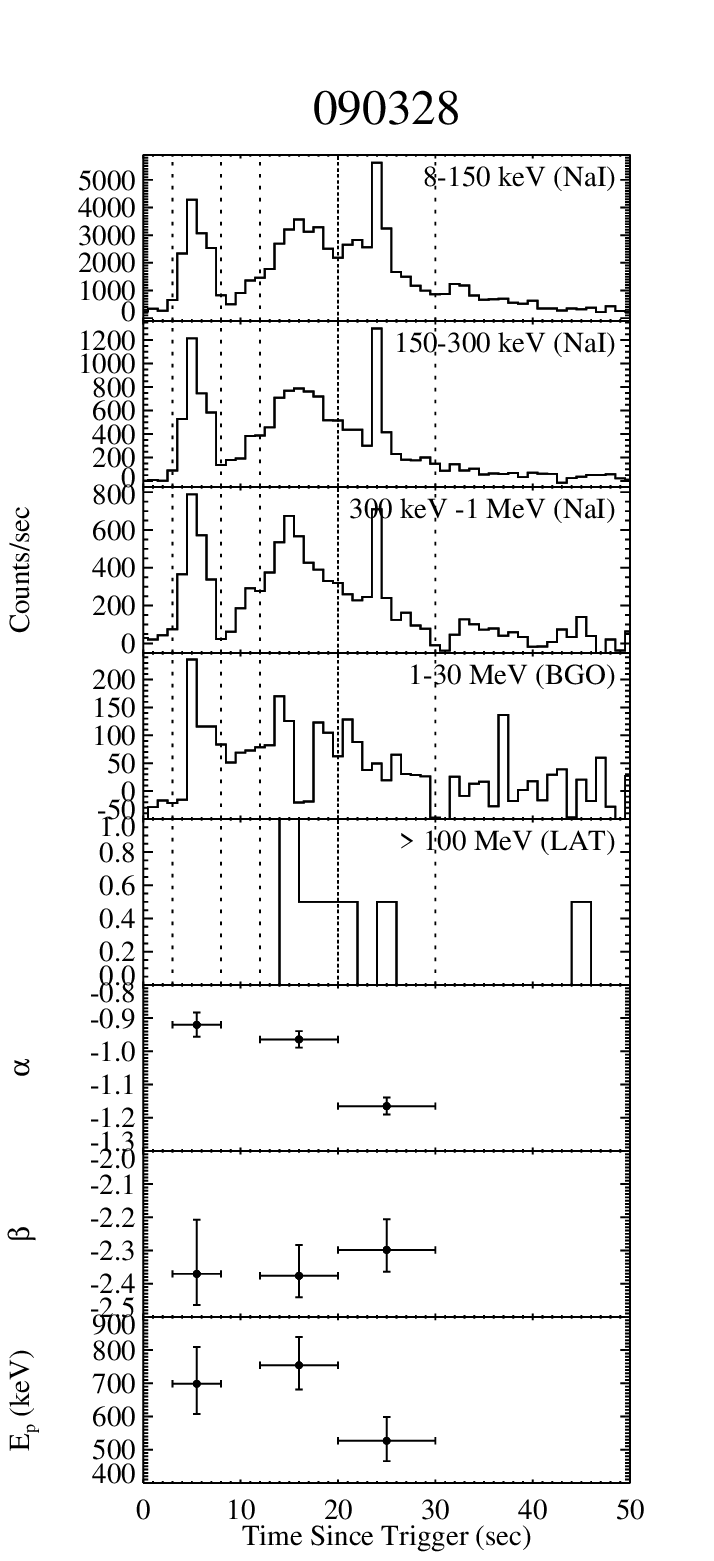}}
 & \includegraphics[angle=270,scale=0.2]{f7b.ps} \\
 & \includegraphics[angle=0,scale=.33]{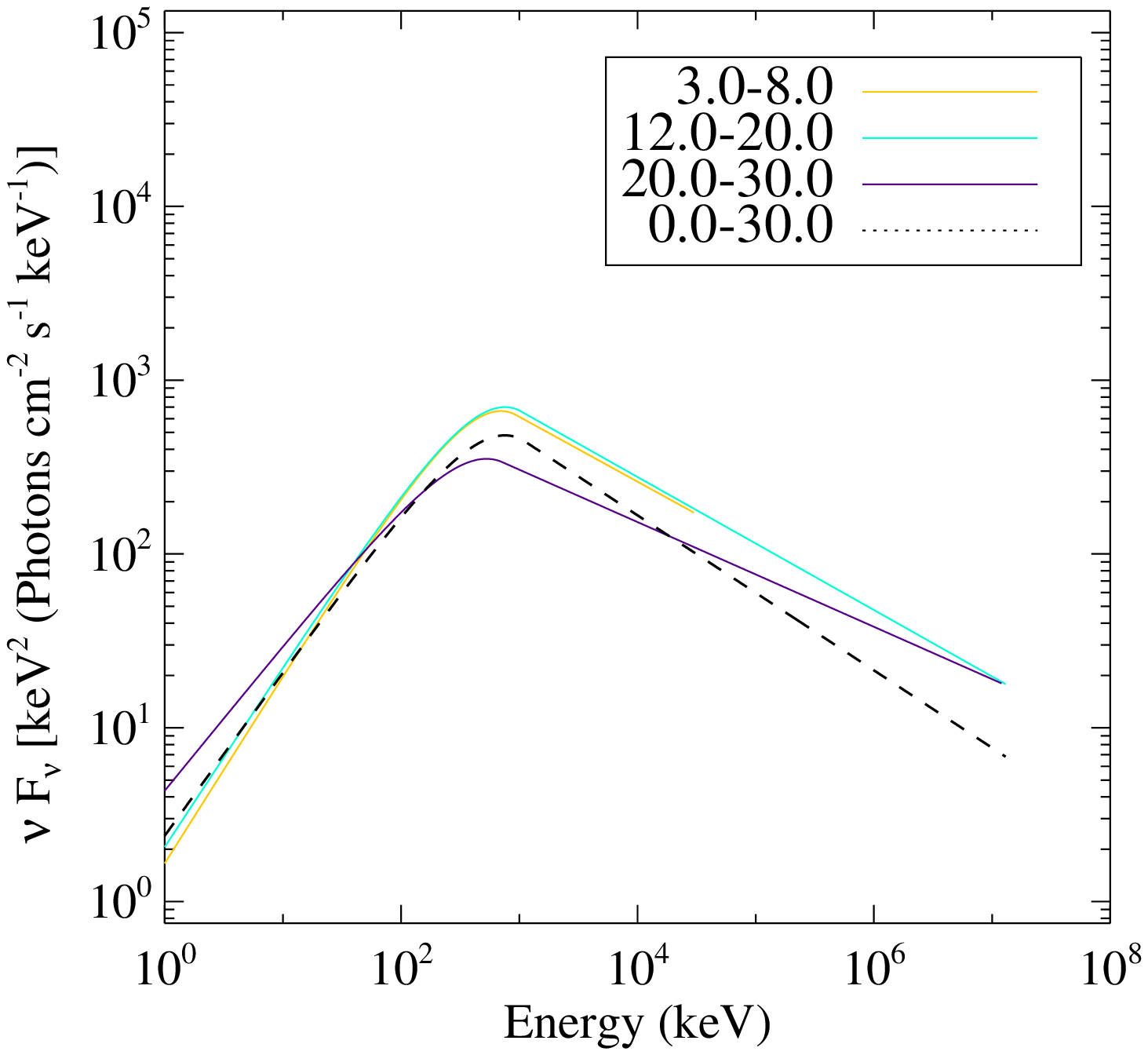} \\
  &  \includegraphics[angle=0,scale=.33]{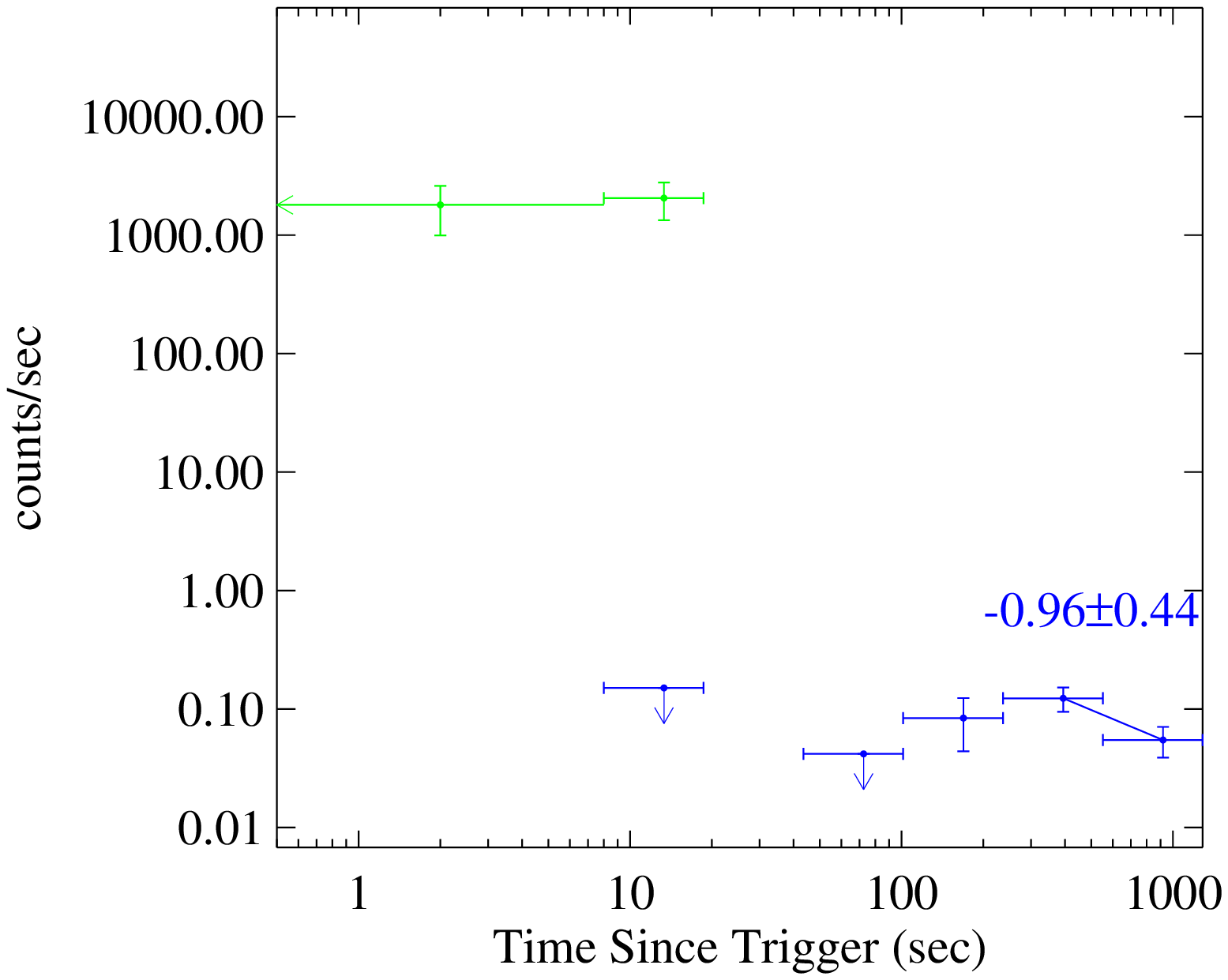}  \\
 \end{tabular}
\caption{Same as Figure 1, but for GRB 090328.}
\label{090328}
\end{figure}

\begin{figure}
\begin{tabular}{ll}
\multirow{3}{*}{ \includegraphics[angle=0,scale=0.78]{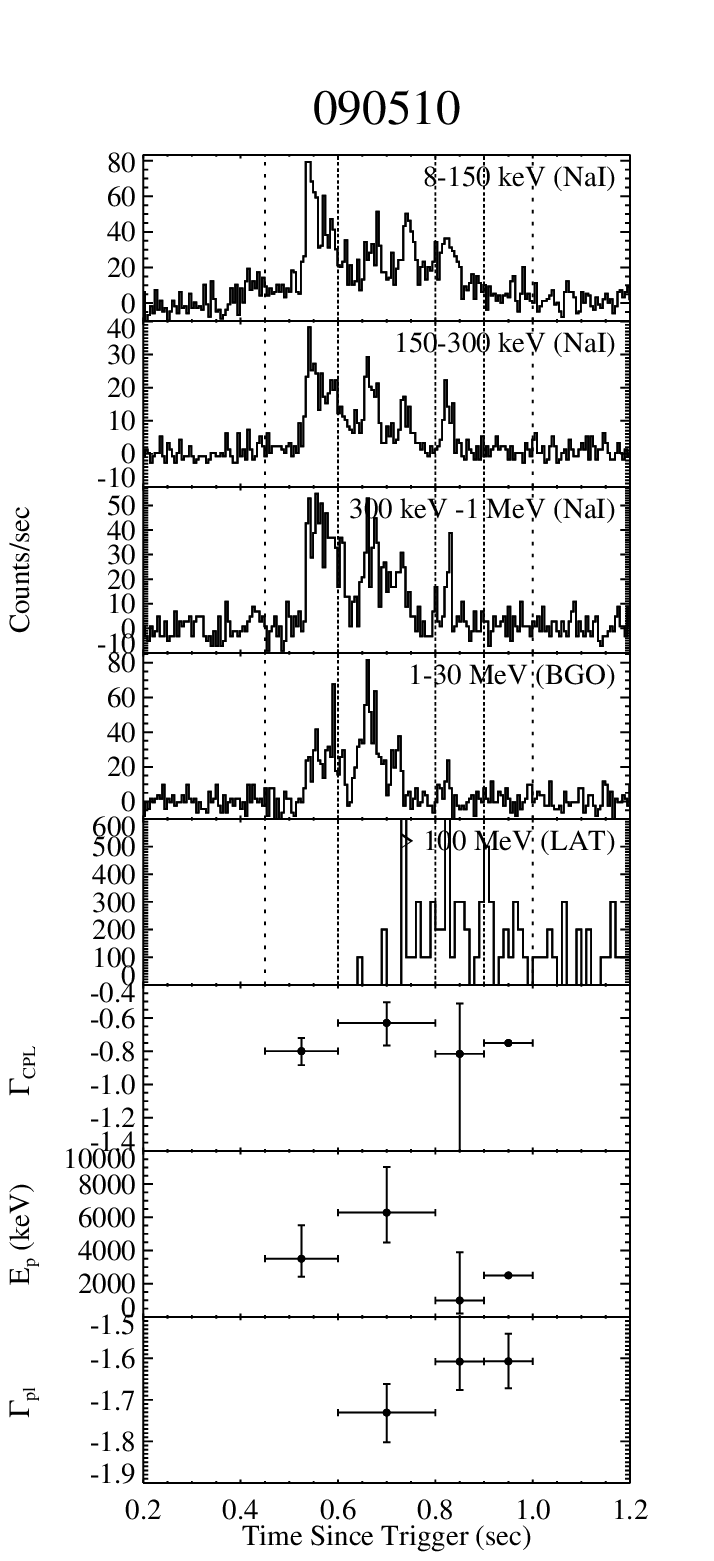}}
 & \includegraphics[angle=270,scale=0.2]{f8b.ps} \\
 & \includegraphics[angle=0,scale=.33]{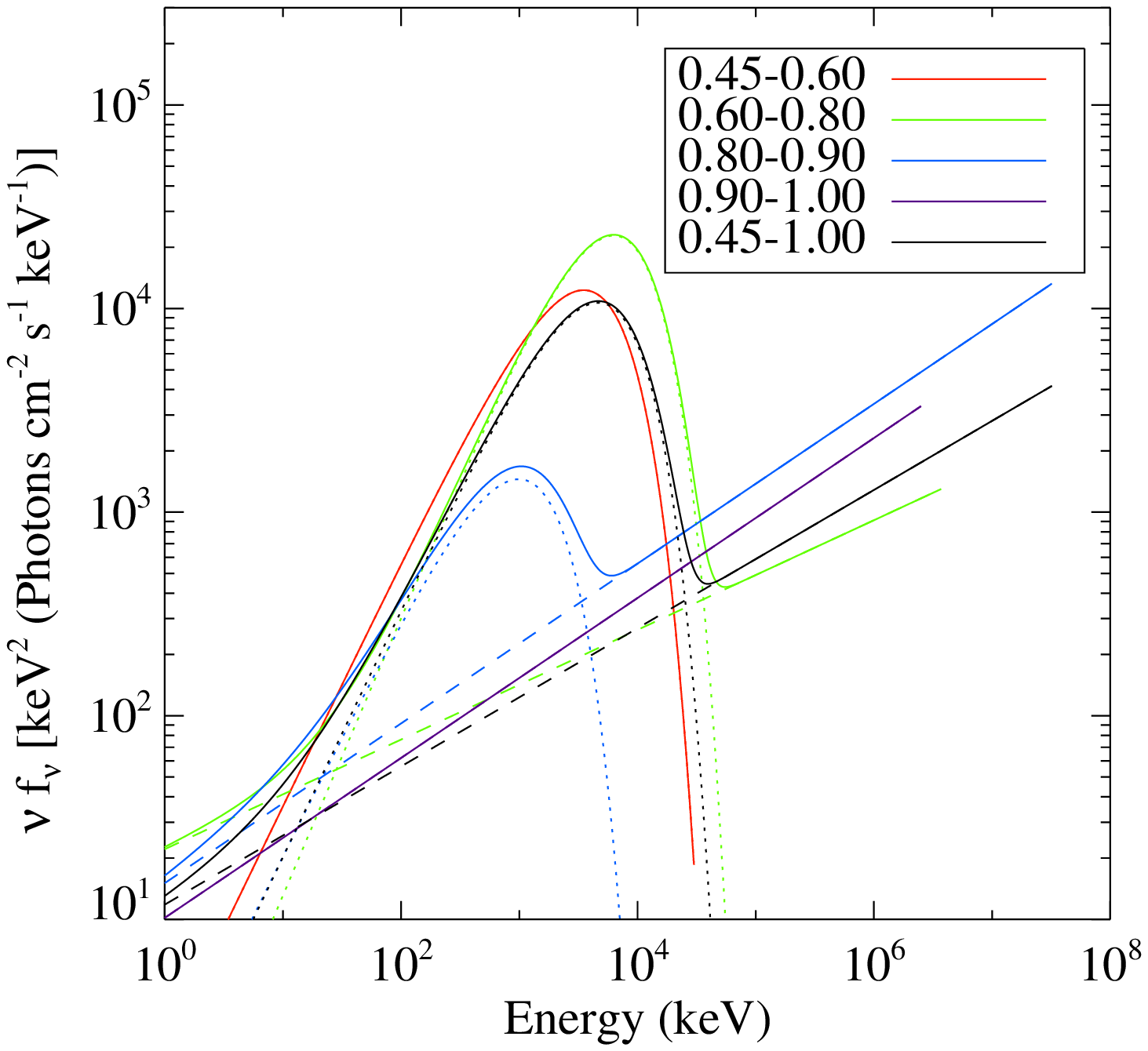} \\
  &  \includegraphics[angle=0,scale=.33]{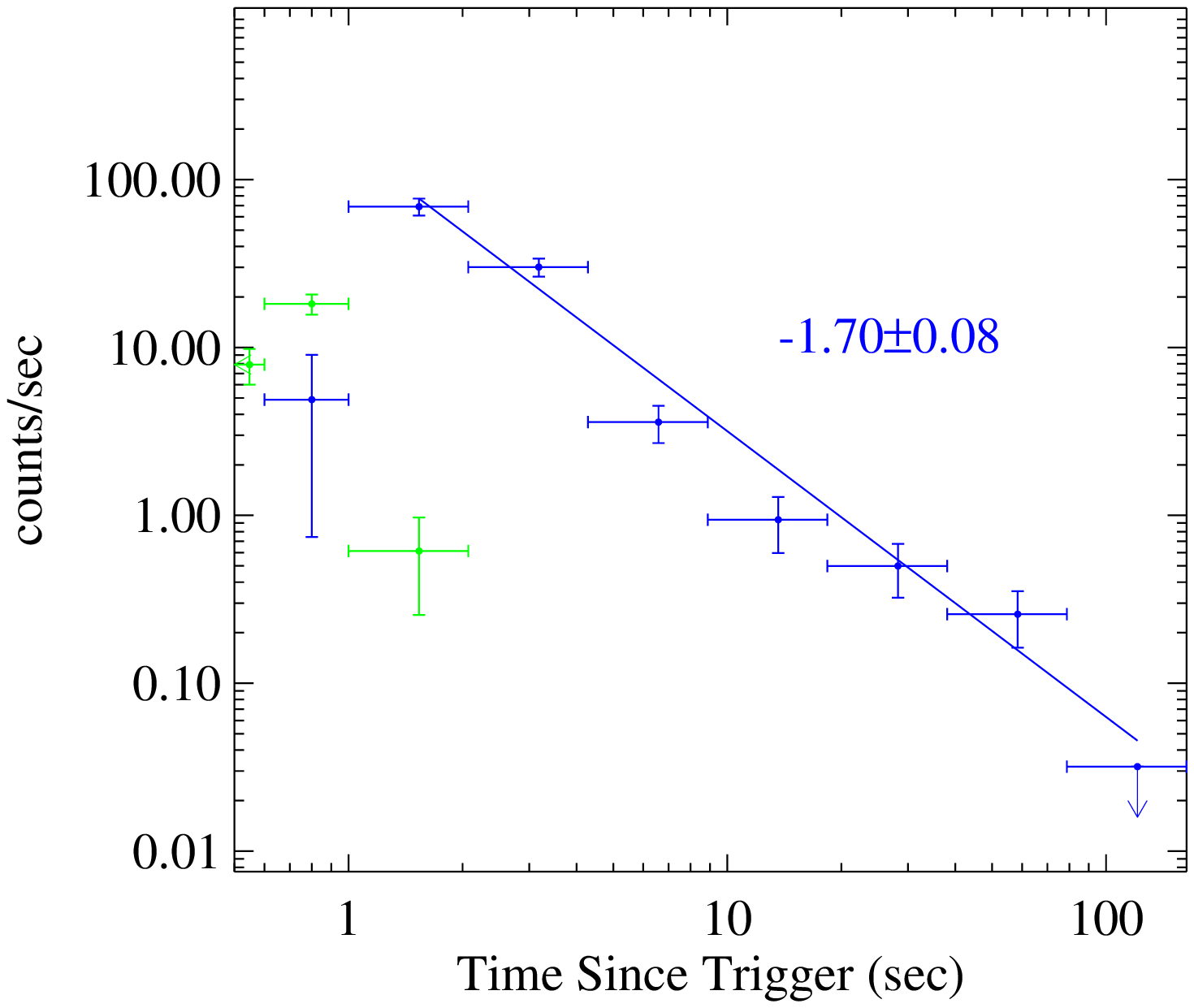}  \\
 \end{tabular}

\caption{Same as Figure 1, but for GRB 090510. The applied model is
cut-off power-law plus power-law (CPL + PL).}
\label{090510}
\end{figure}

\begin{figure}
\begin{tabular}{ll}
\multirow{3}{*}{ \includegraphics[angle=0,scale=0.78]{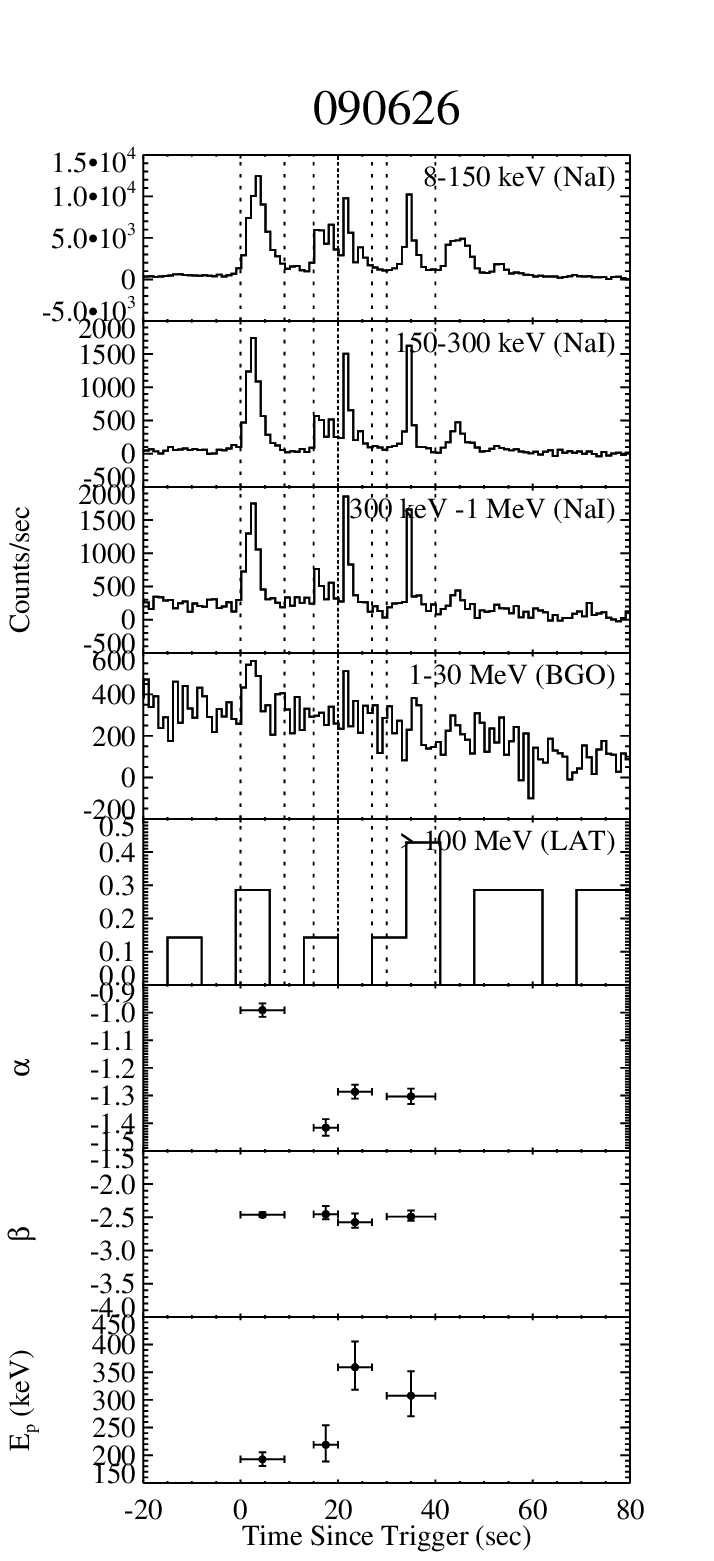}}
 & \includegraphics[angle=270,scale=0.2]{f9b.ps} \\
 & \includegraphics[angle=0,scale=.33]{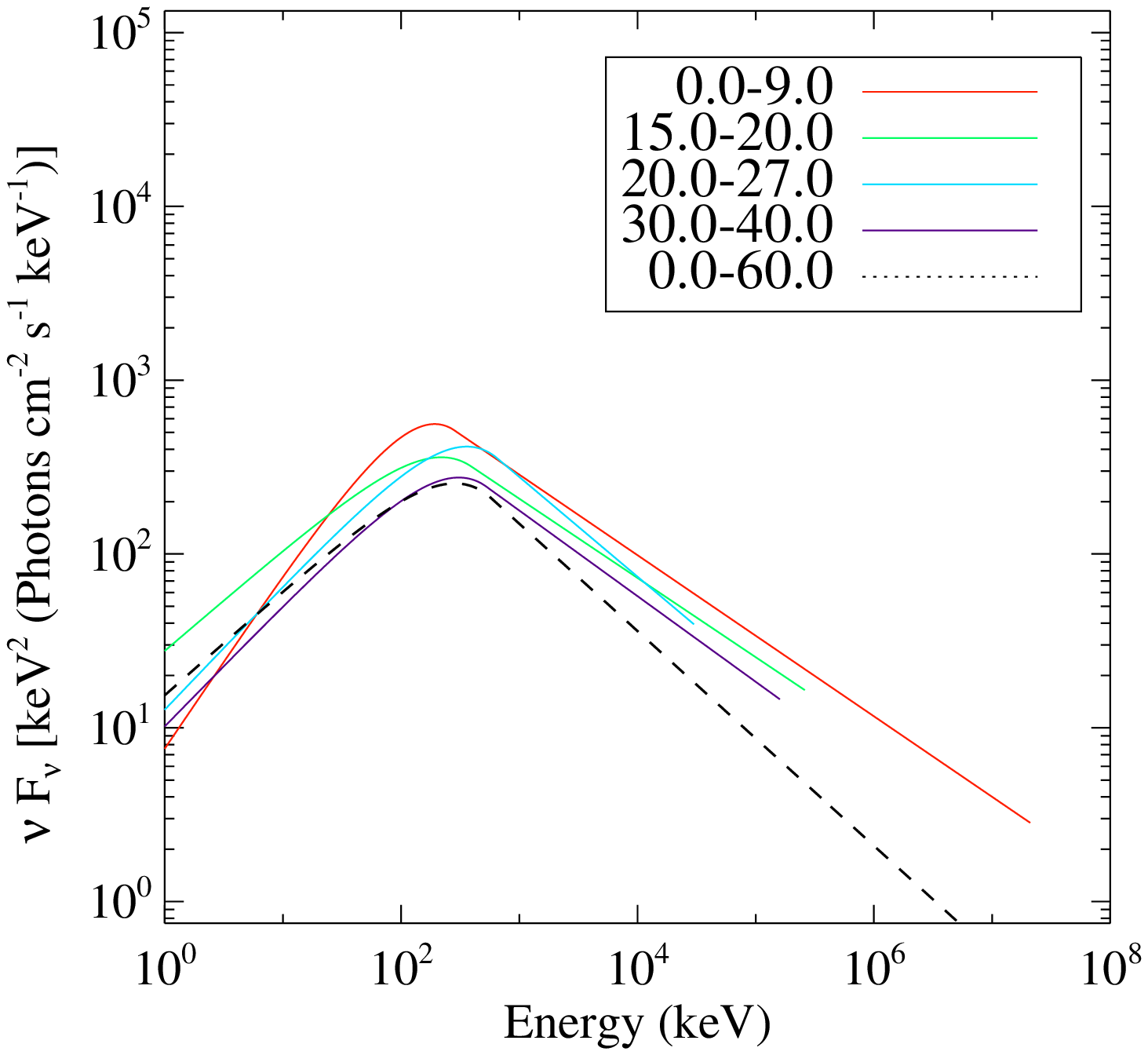} \\
  &  \includegraphics[angle=0,scale=.33]{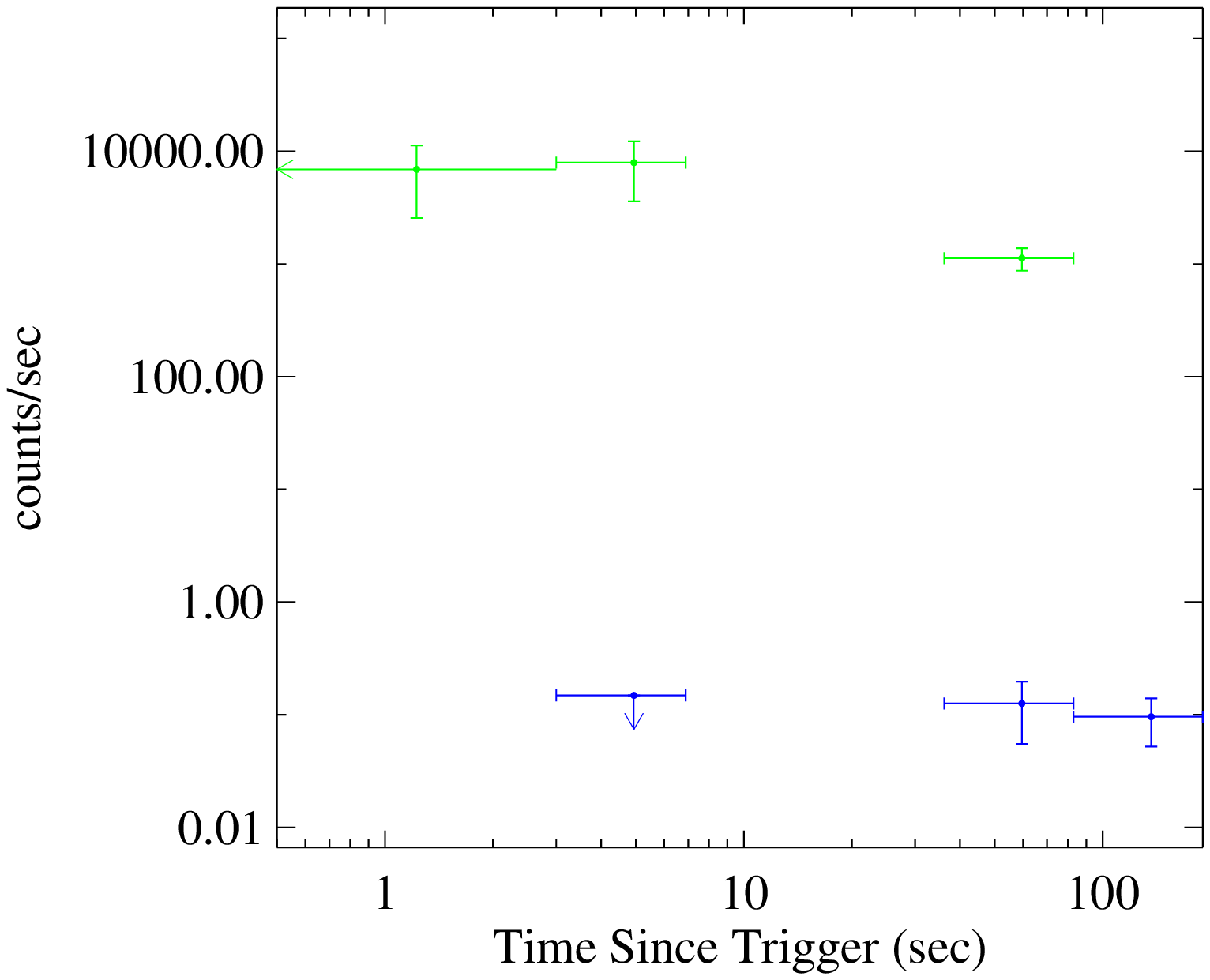}  \\
 \end{tabular}

\caption{Same as Figure 1, but for GRB 090626.}
\label{090626}
\end{figure}

\begin{figure}
\begin{tabular}{ll}
\multirow{3}{*}{ \includegraphics[angle=0,scale=0.78]{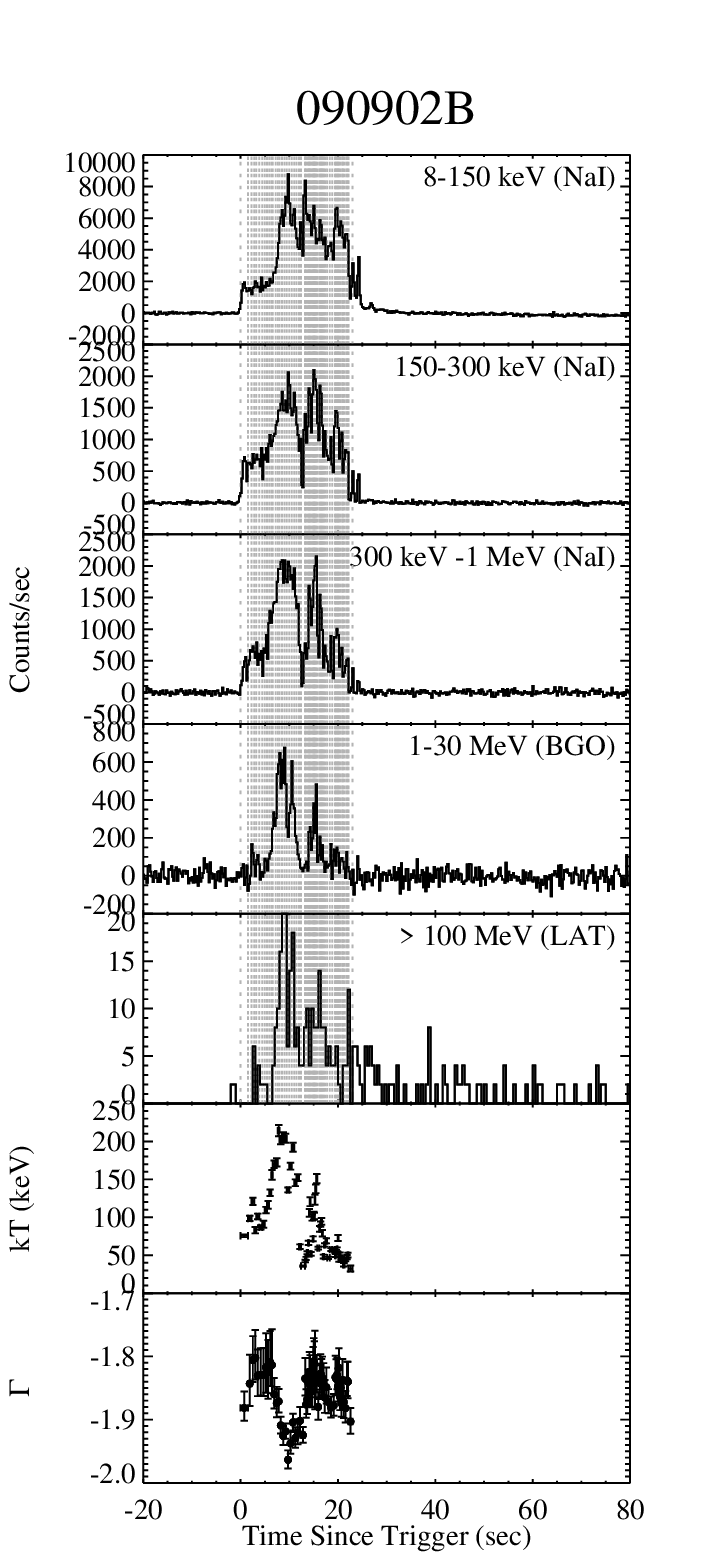}}
 & \includegraphics[angle=270,scale=0.2]{f10b.ps} \\
 & \includegraphics[angle=0,width=2.05in,height=2in]{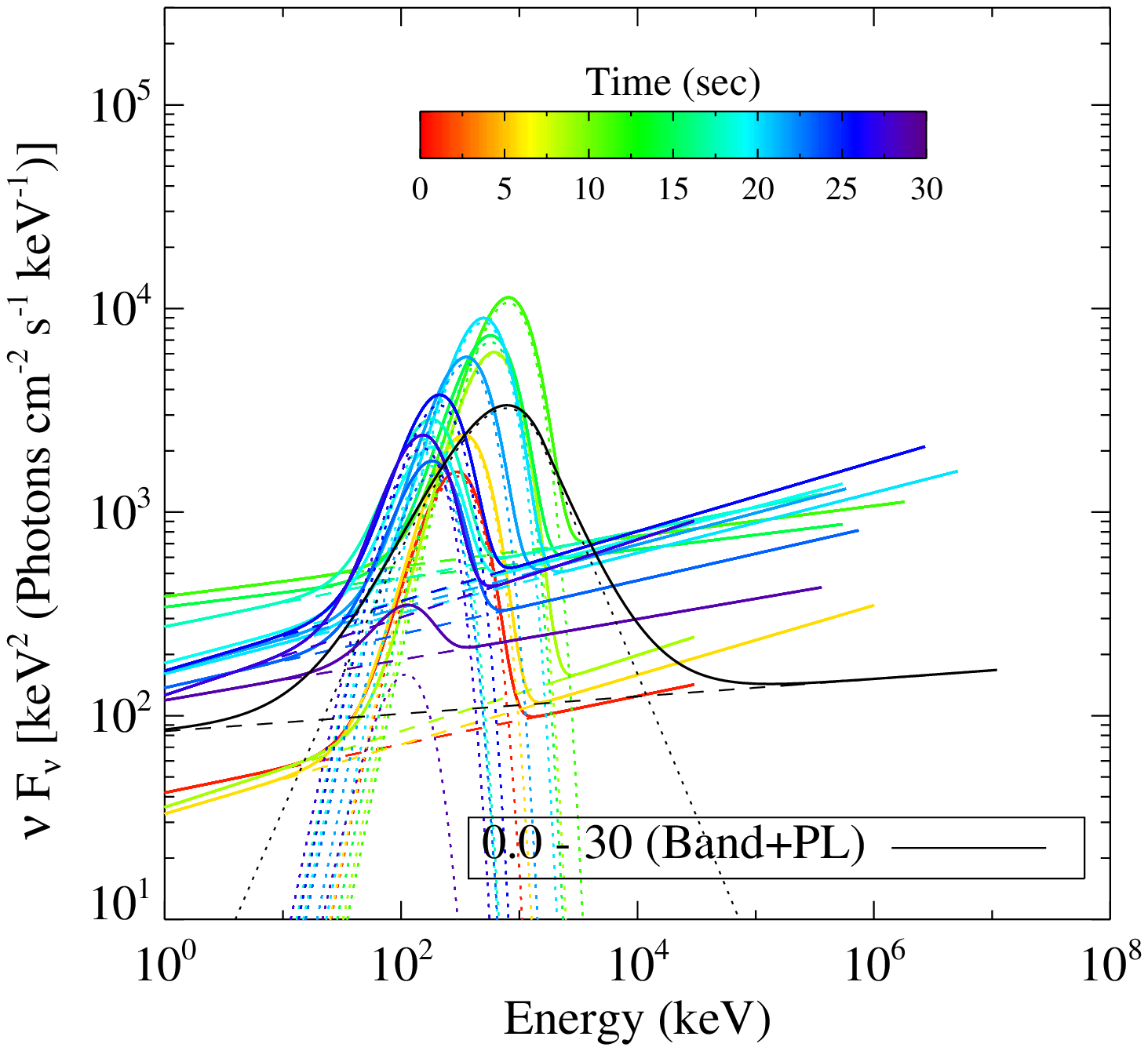} \\
  &  \includegraphics[angle=0,scale=.33]{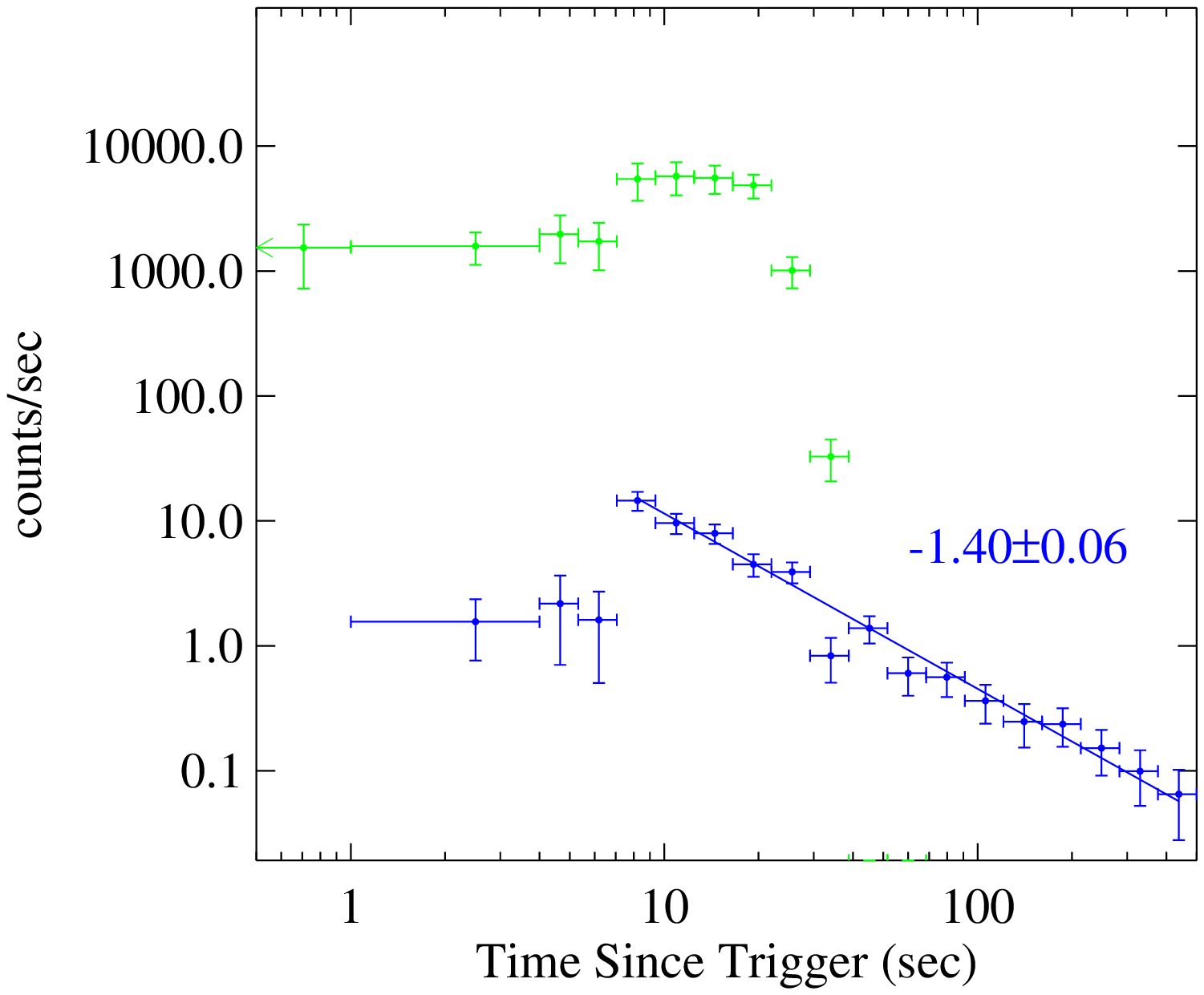}  \\
 \end{tabular}

\caption{Same as Fig. 1, but for GRB 090902B. The applied model is
blackbody plus power law (BB + PL). }
\label{090902B}
\end{figure}

\begin{figure}
\begin{tabular}{ll}
\multirow{3}{*}{ \includegraphics[angle=0,scale=0.78]{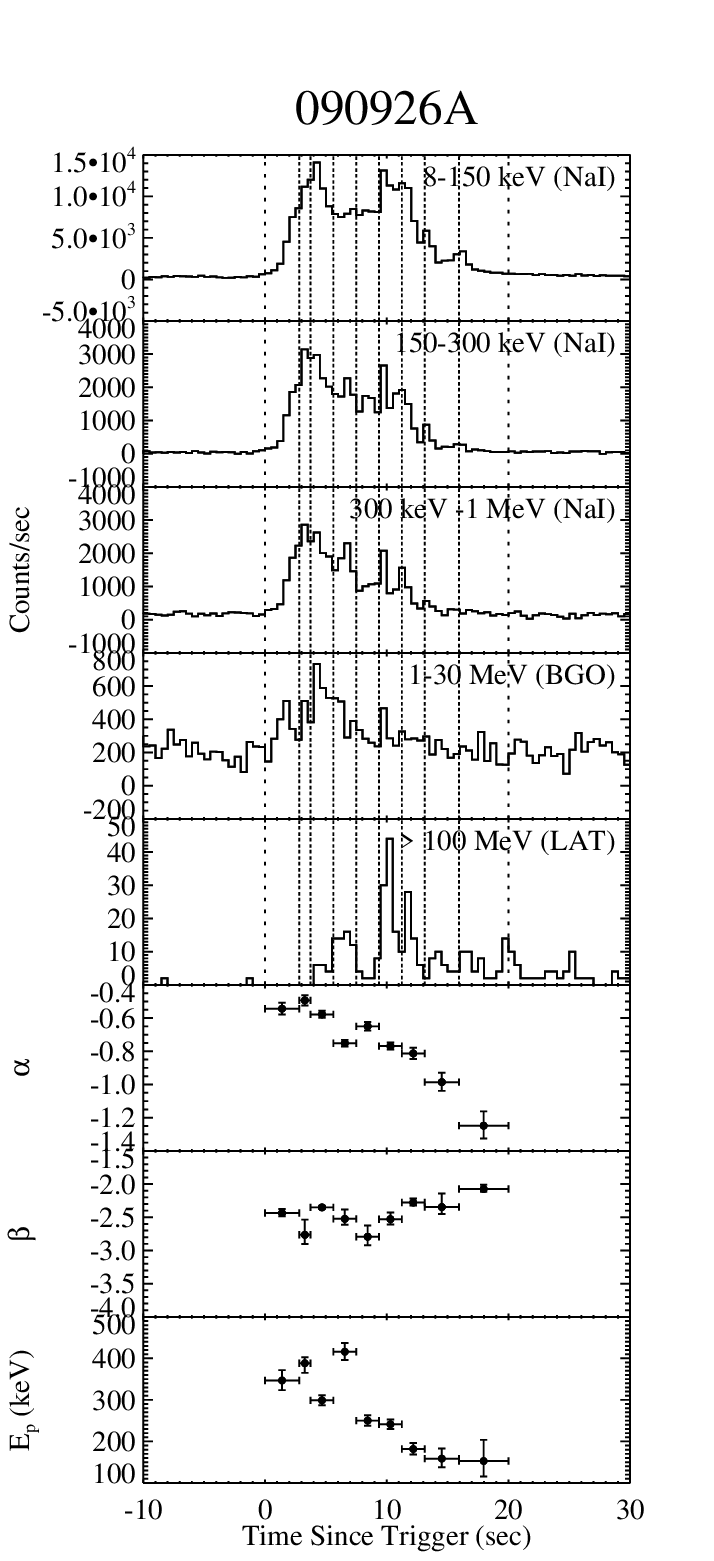}}
 & \includegraphics[angle=270,scale=0.2]{f11b.ps} \\
 & \includegraphics[angle=0,scale=.33]{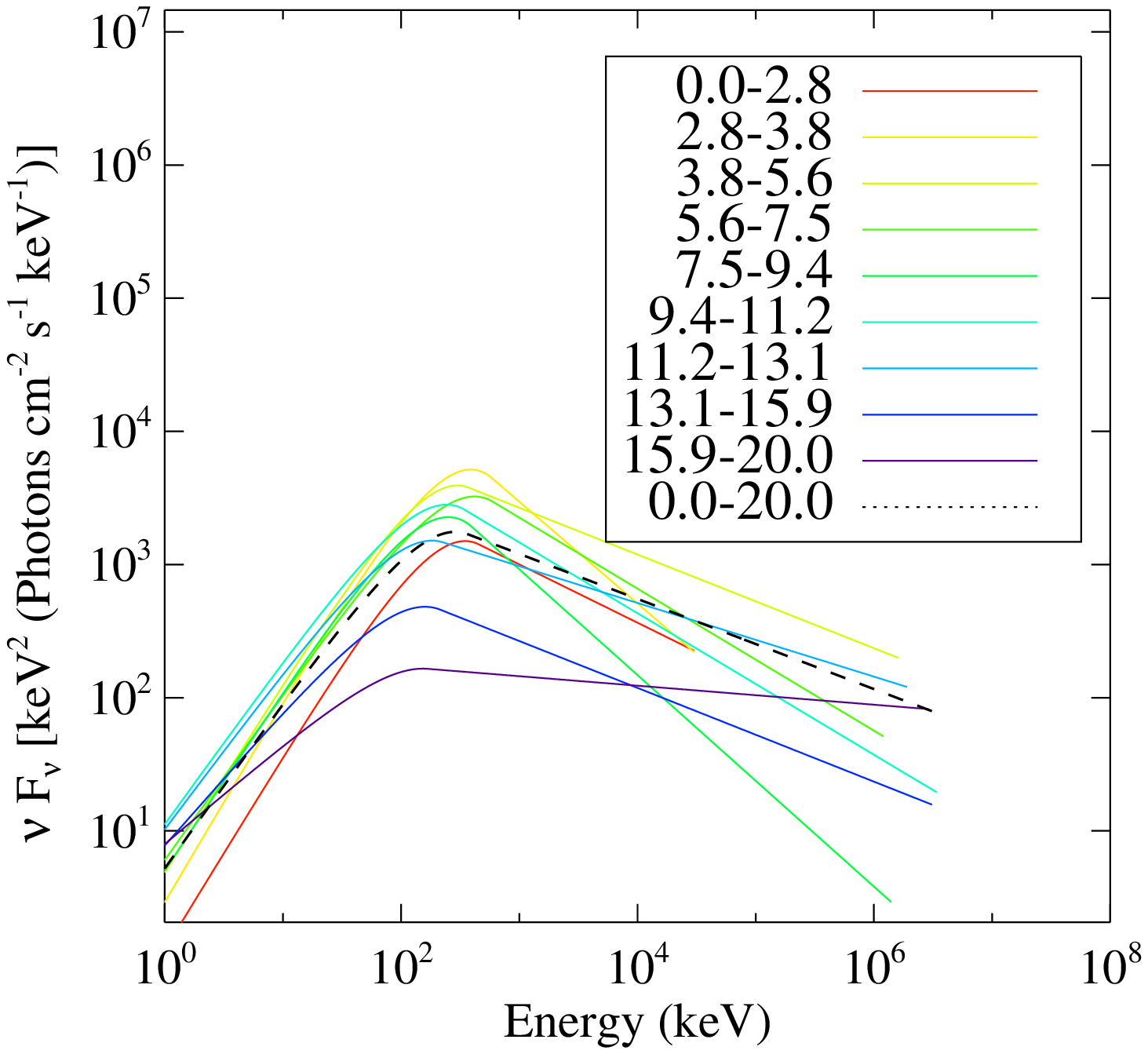} \\
  &  \includegraphics[angle=0,scale=.33]{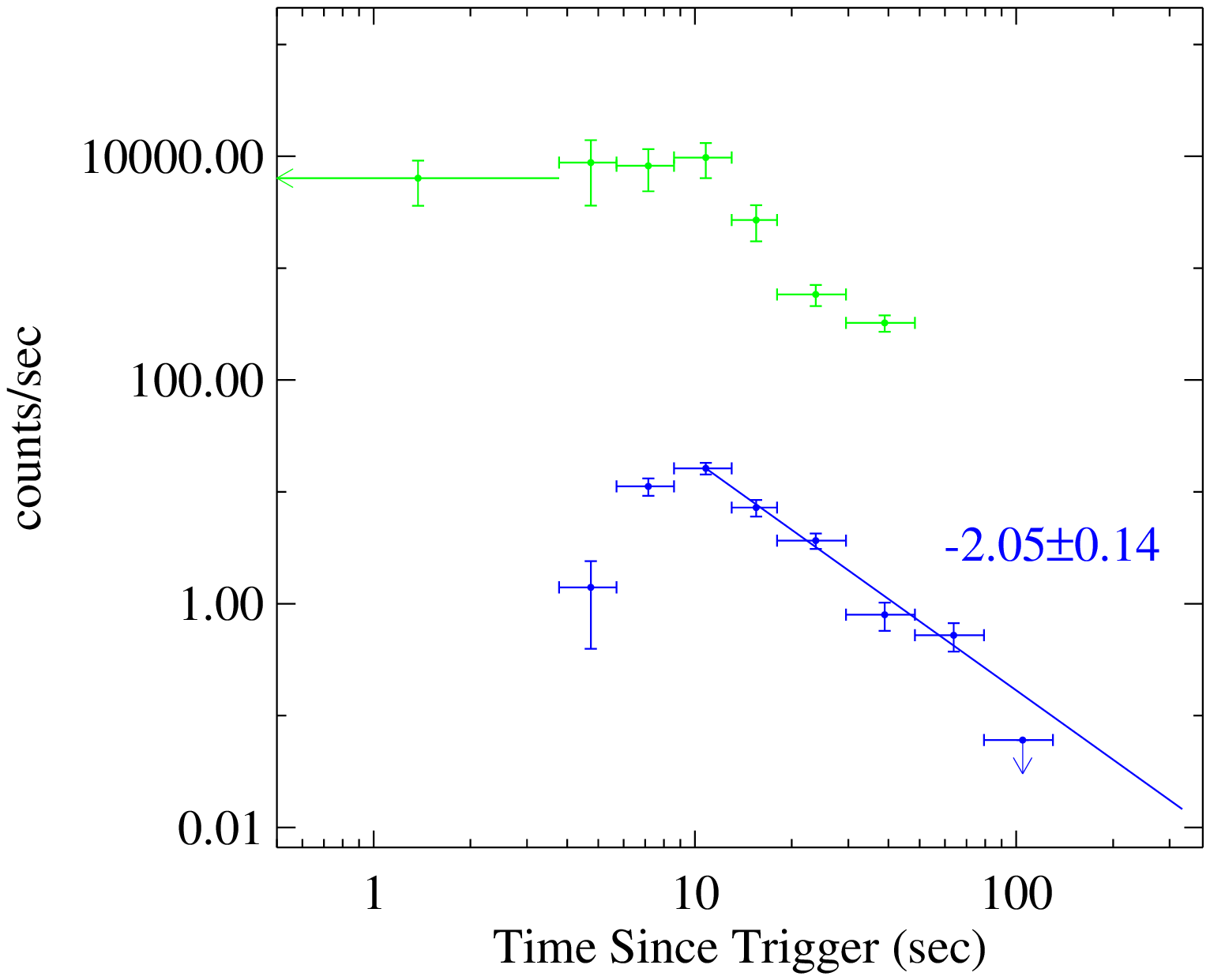}  \\
 \end{tabular}

\caption{Same as Figure 1, but for GRB 090926A.}
\label{090926A}
\end{figure}

\begin{figure}
\begin{tabular}{ll}
\multirow{3}{*}{ \includegraphics[angle=0,scale=0.78]{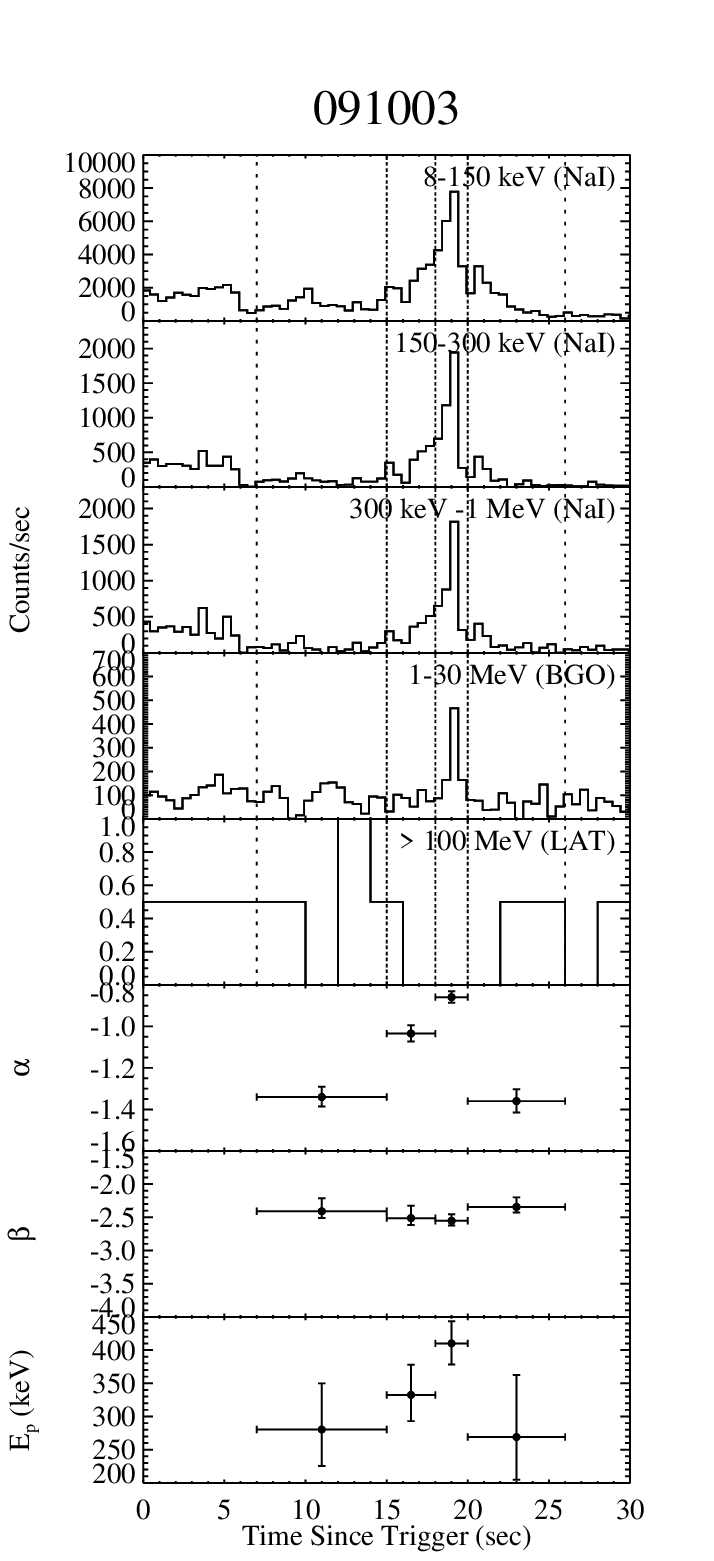}}
 & \includegraphics[angle=270,scale=0.2]{f12b.ps} \\
 & \includegraphics[angle=0,scale=.33]{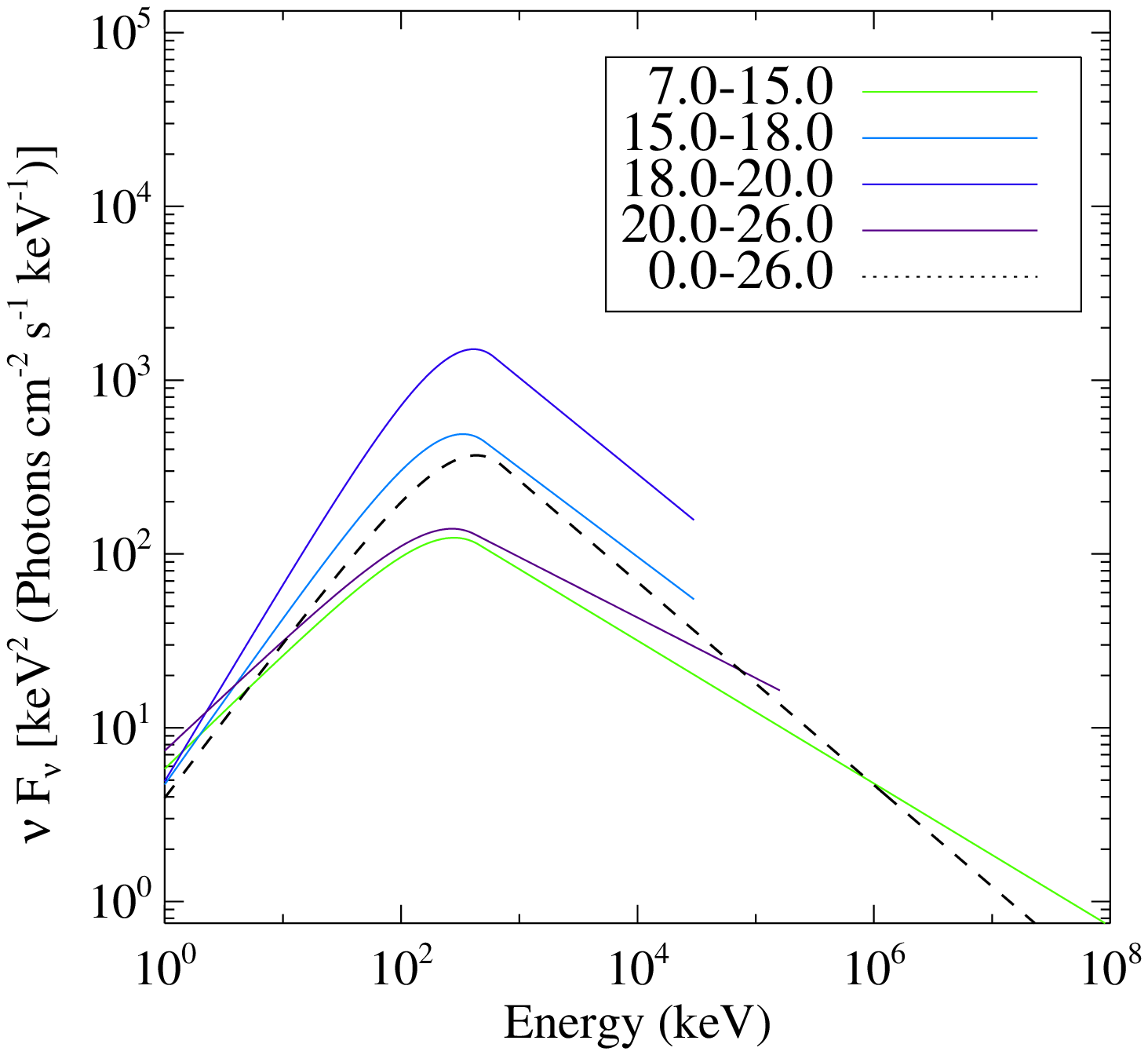} \\
  &  \includegraphics[angle=0,scale=.33]{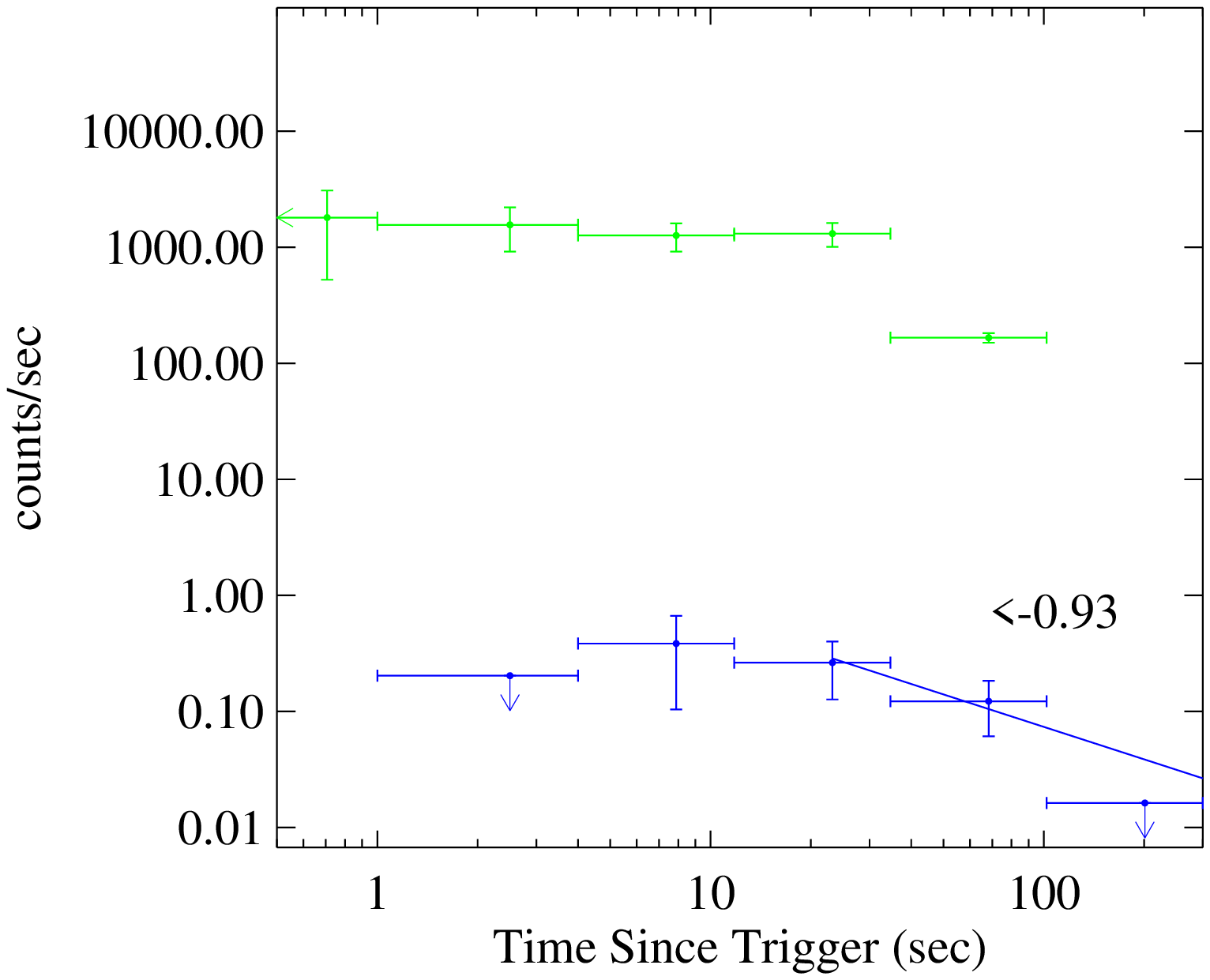}  \\
 \end{tabular}

\caption{Same as Figure 1, but for GRB 091003.}
\label{091003}
\end{figure}

\begin{figure}
\begin{tabular}{ll}
\multirow{3}{*}{ \includegraphics[angle=0,scale=0.78]{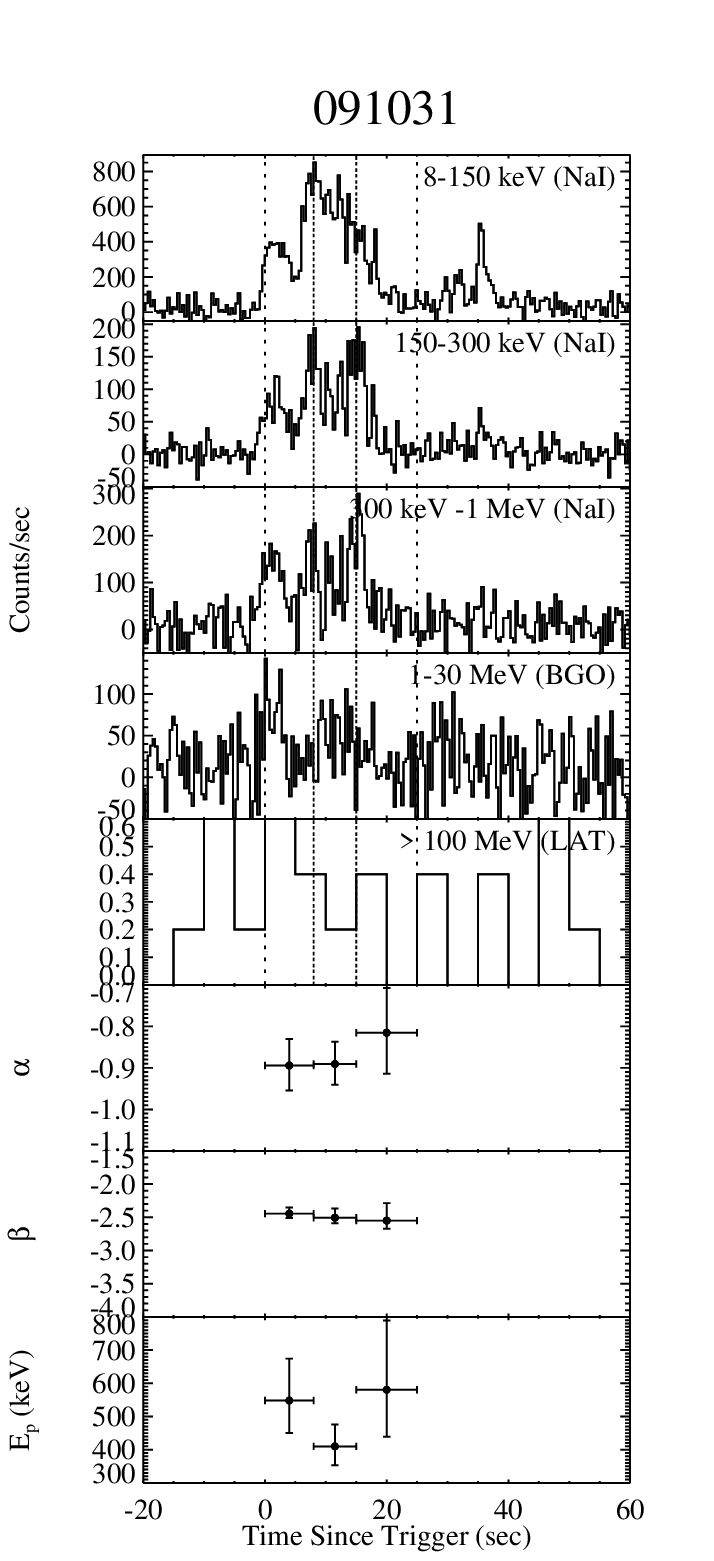}}
 & \includegraphics[angle=270,scale=0.2]{f13b.ps} \\
 & \includegraphics[angle=0,scale=.33]{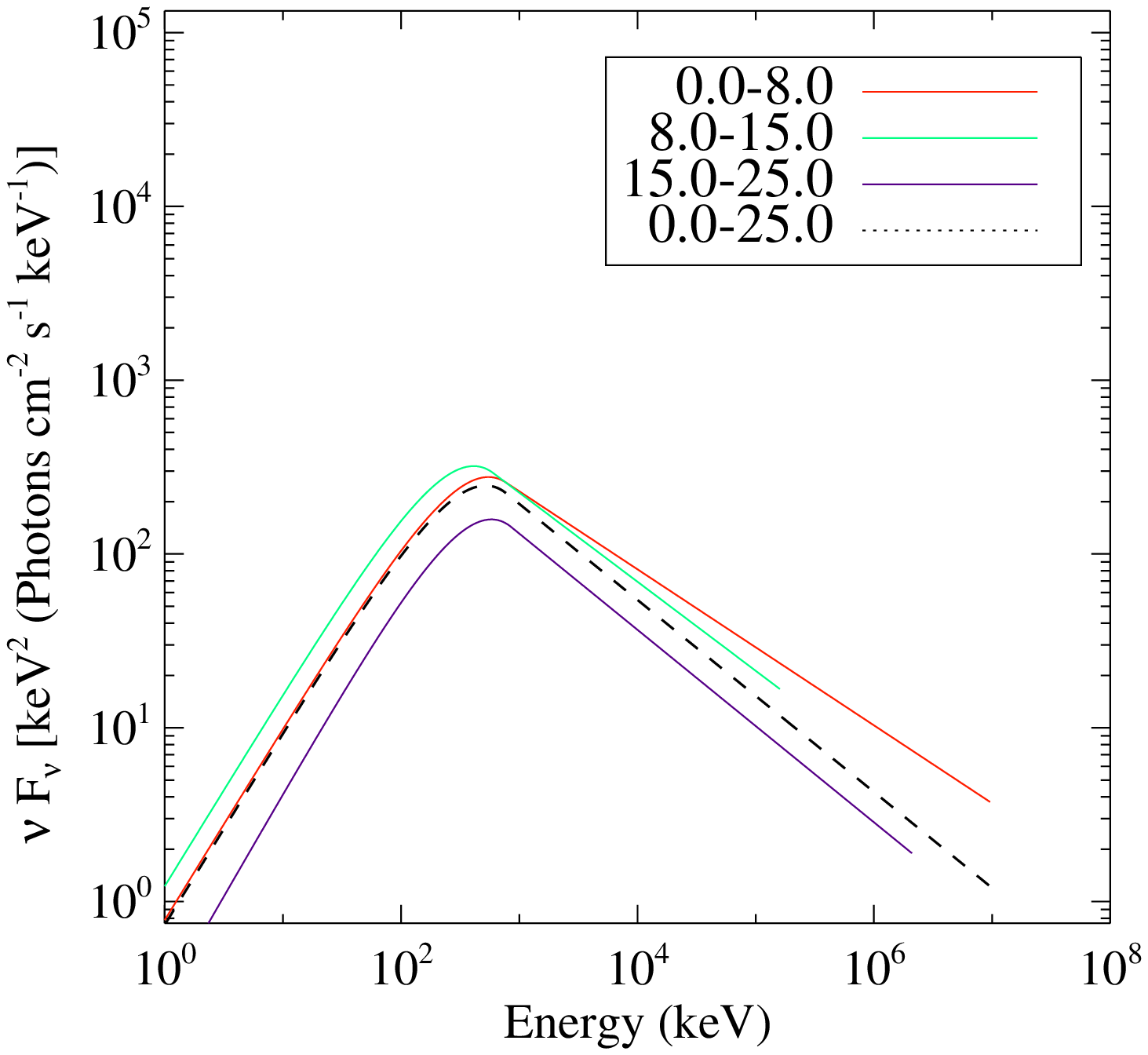} \\
  &  \includegraphics[angle=0,scale=.33]{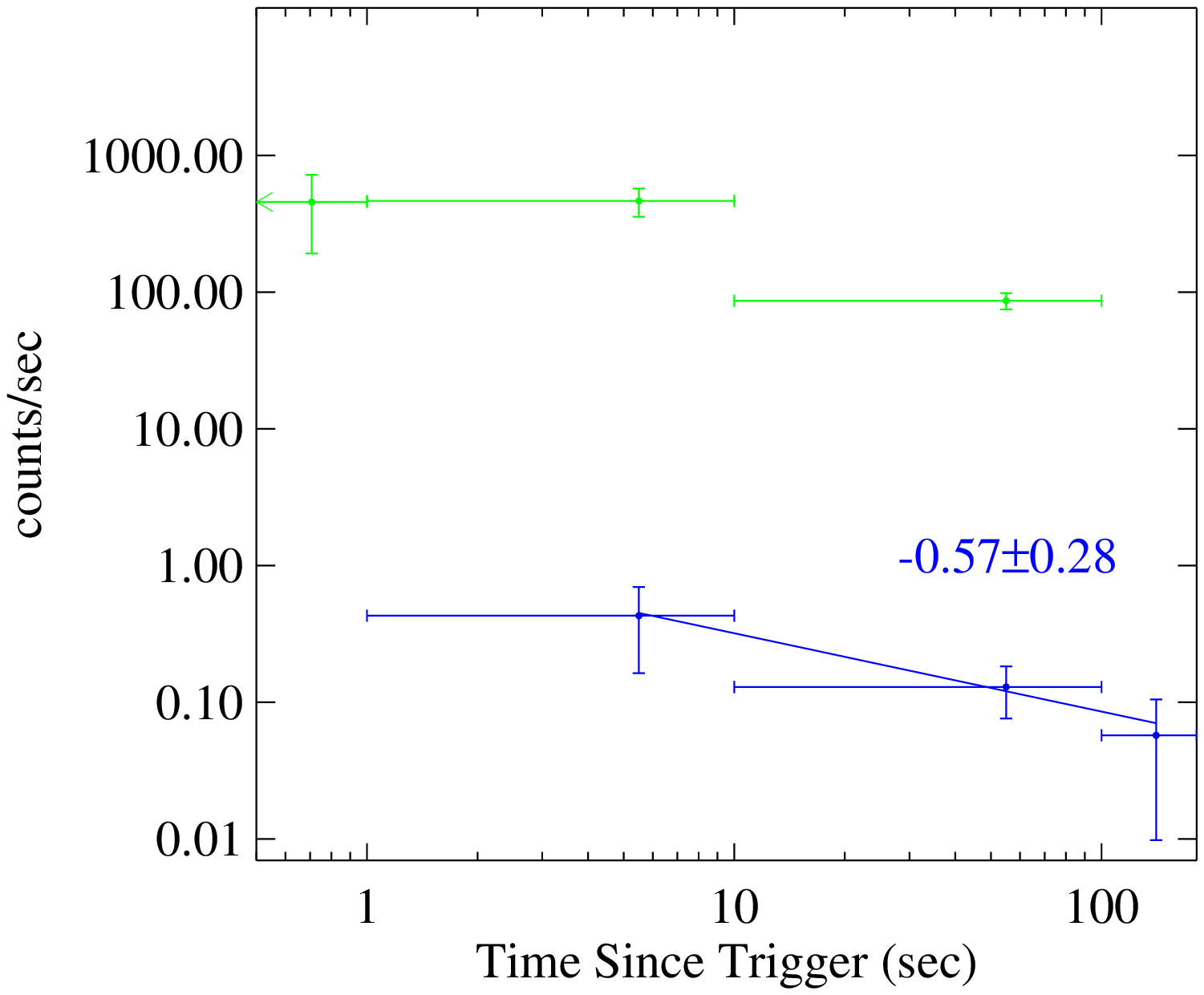}  \\
 \end{tabular}

\caption{Same as Figure 1, but for GRB 091031.}
\label{091031}
\end{figure}

\begin{figure}
\begin{tabular}{ll}
\multirow{3}{*}{ \includegraphics[angle=0,scale=0.78]{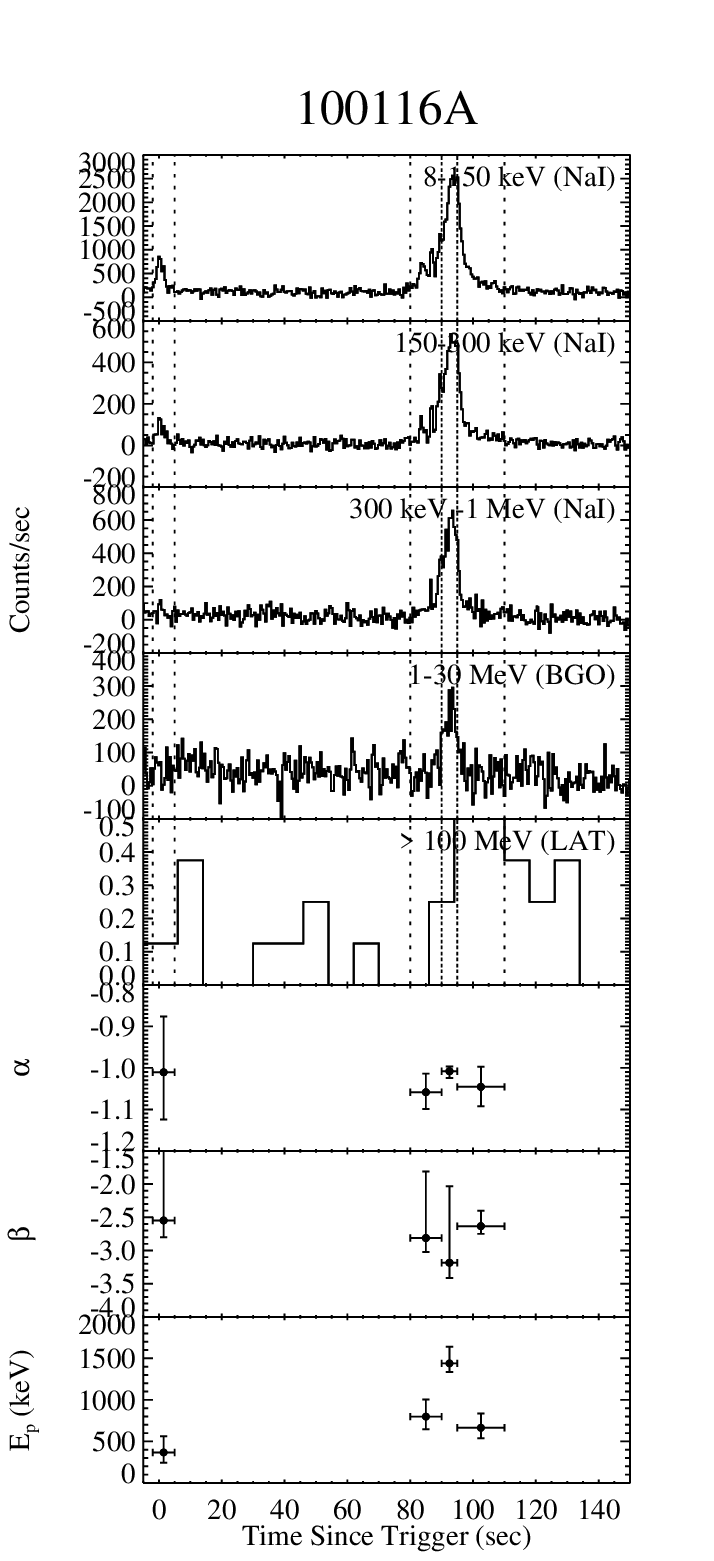}}
 & \includegraphics[angle=270,scale=0.2]{f14b.ps} \\
 & \includegraphics[angle=0,scale=.33]{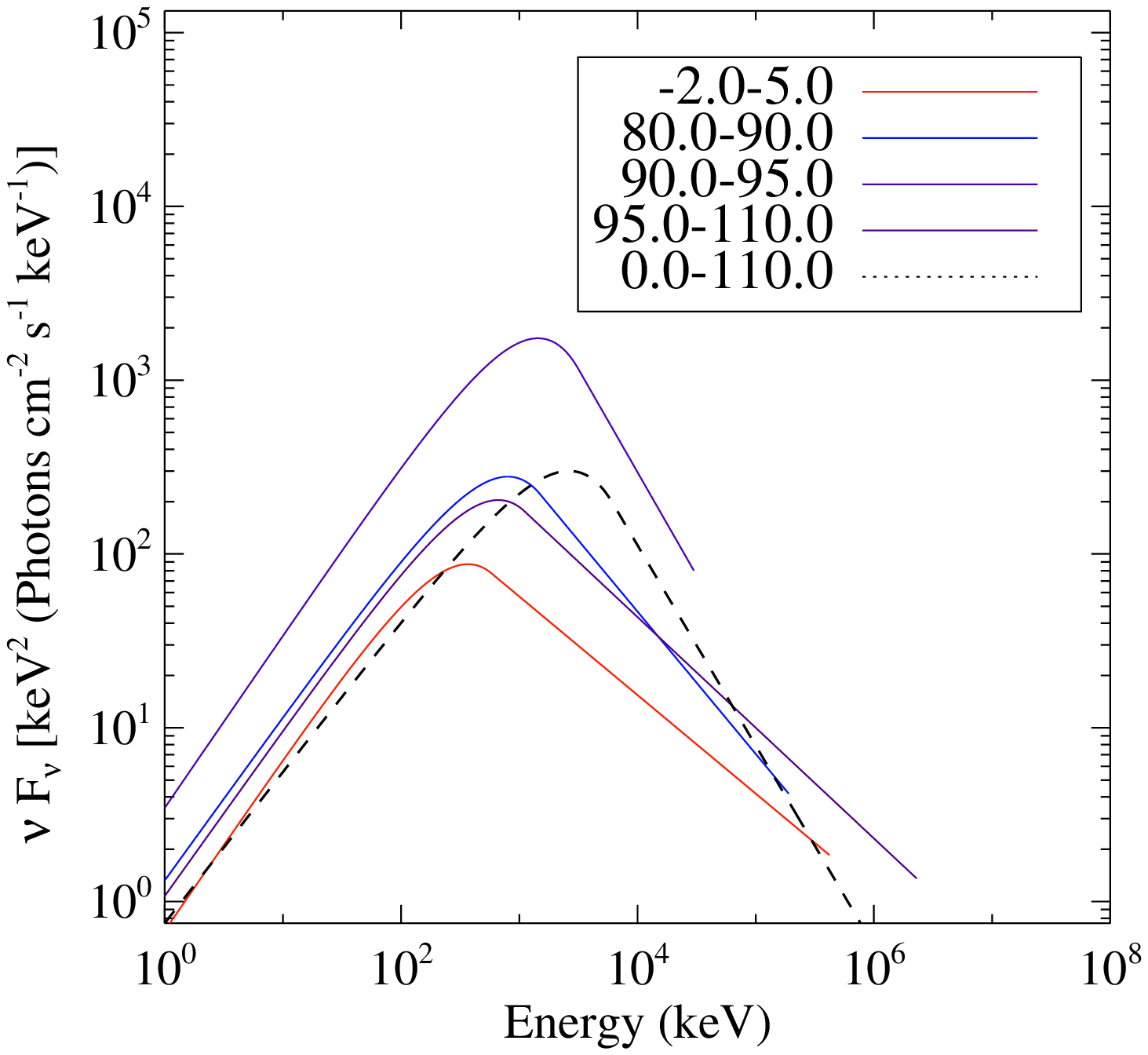} \\
  &  \includegraphics[angle=0,scale=.33]{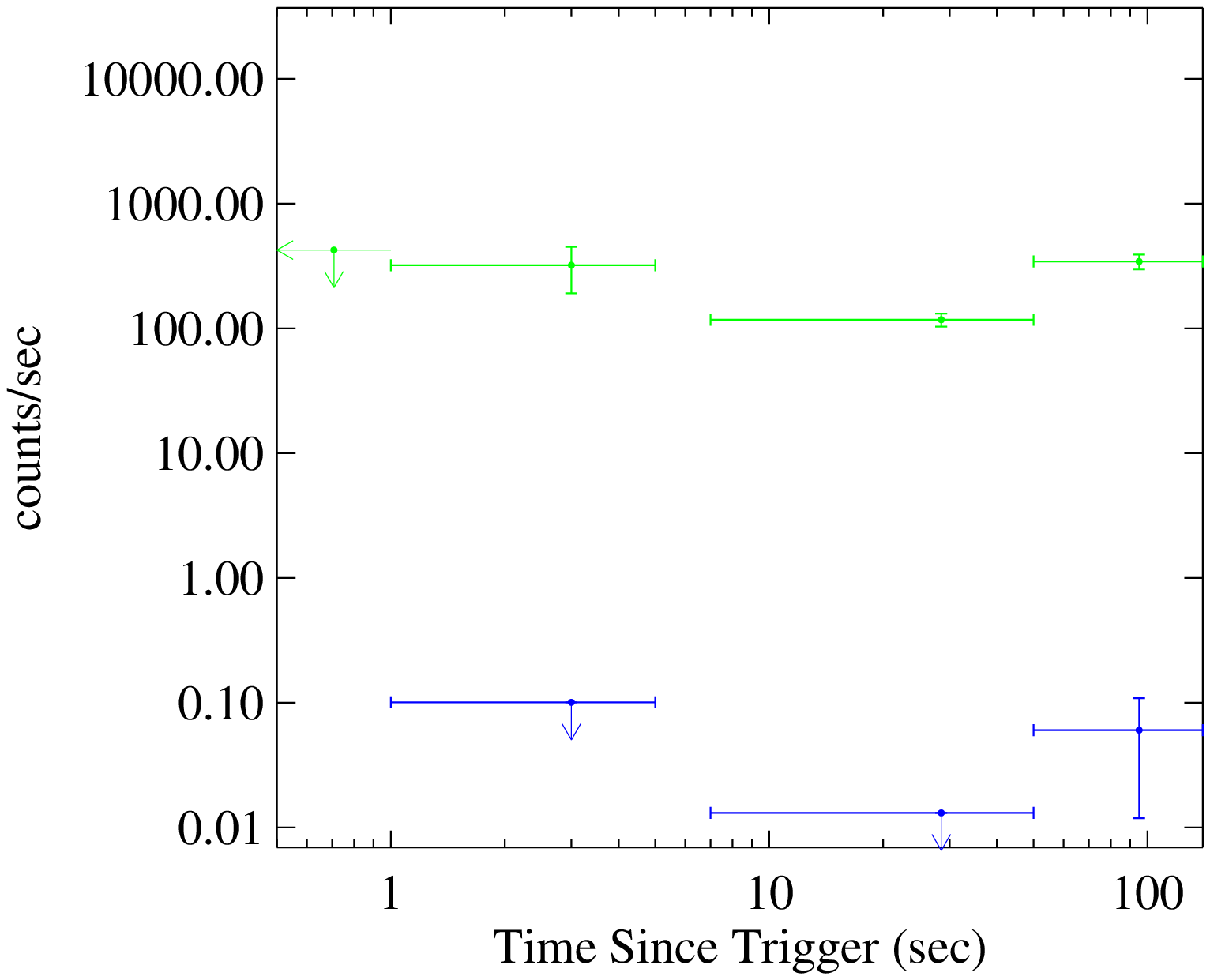}  \\
 \end{tabular}

\caption{Same as Figure 1, but for GRB 100116A.}
\label{100116A}
\end{figure}

\begin{figure}
\begin{tabular}{ll}
\multirow{3}{*}{ \includegraphics[angle=0,scale=0.78]{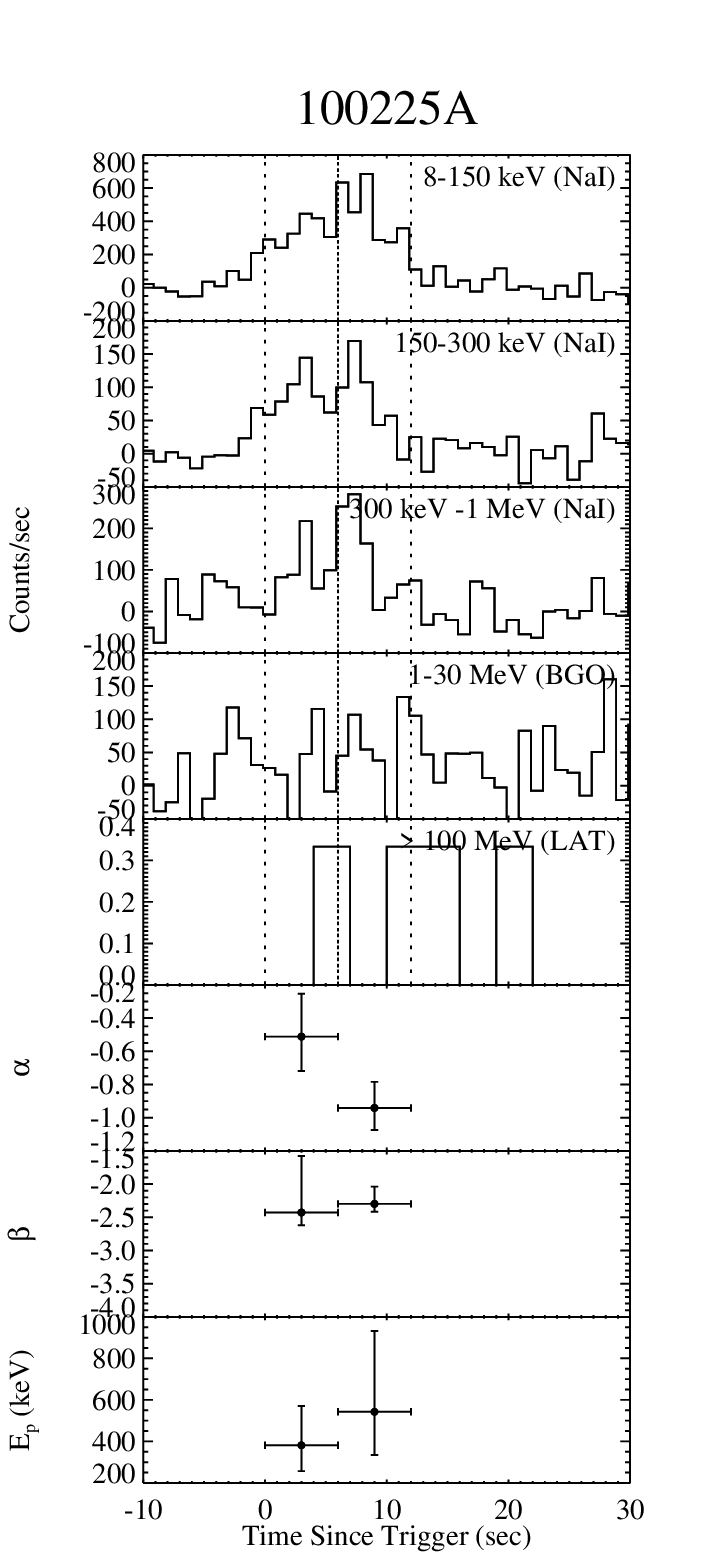}}
 & \includegraphics[angle=270,scale=0.2]{f15b.ps} \\
 & \includegraphics[angle=0,scale=.33]{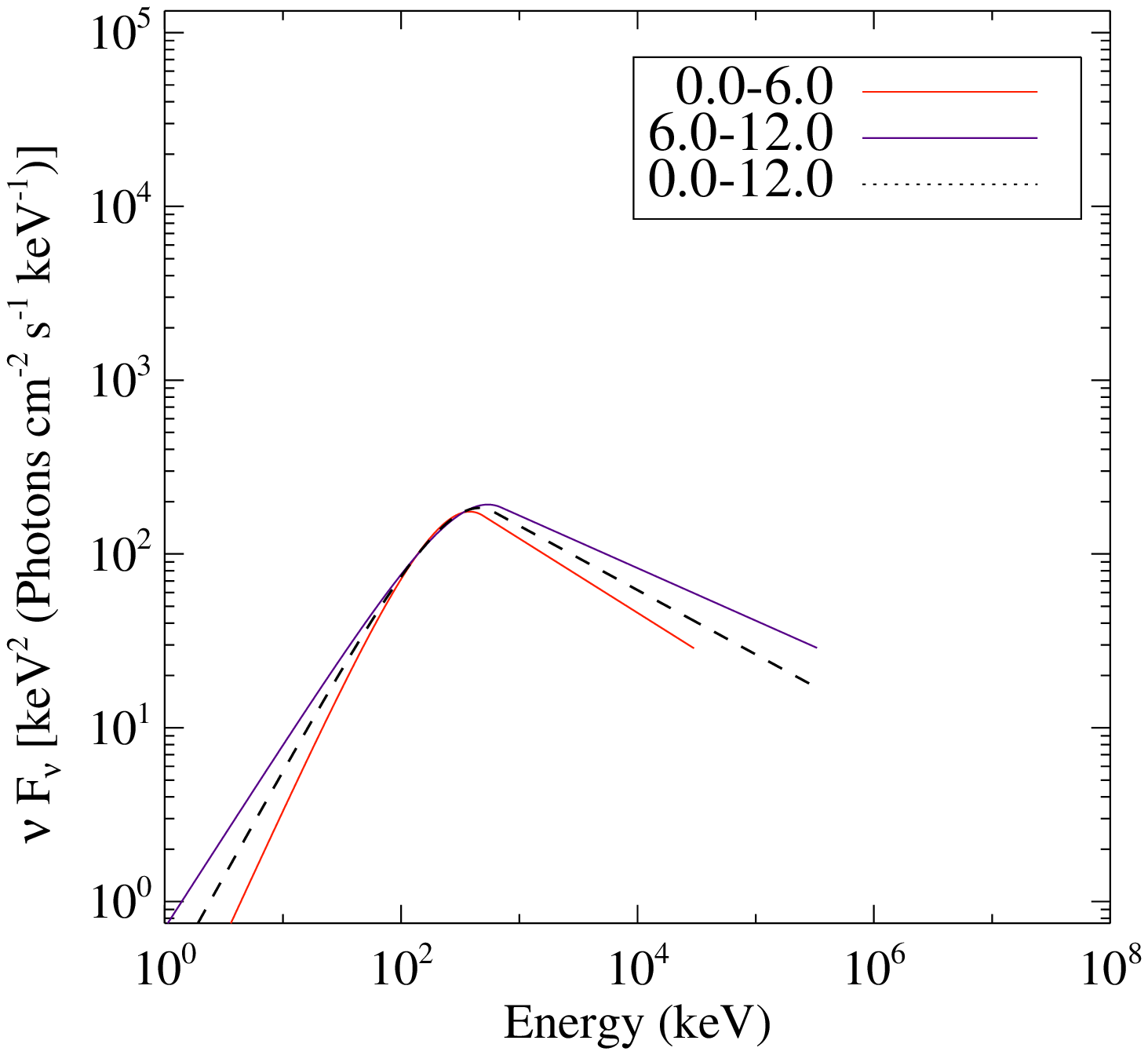} \\
  &  \includegraphics[angle=0,scale=.33]{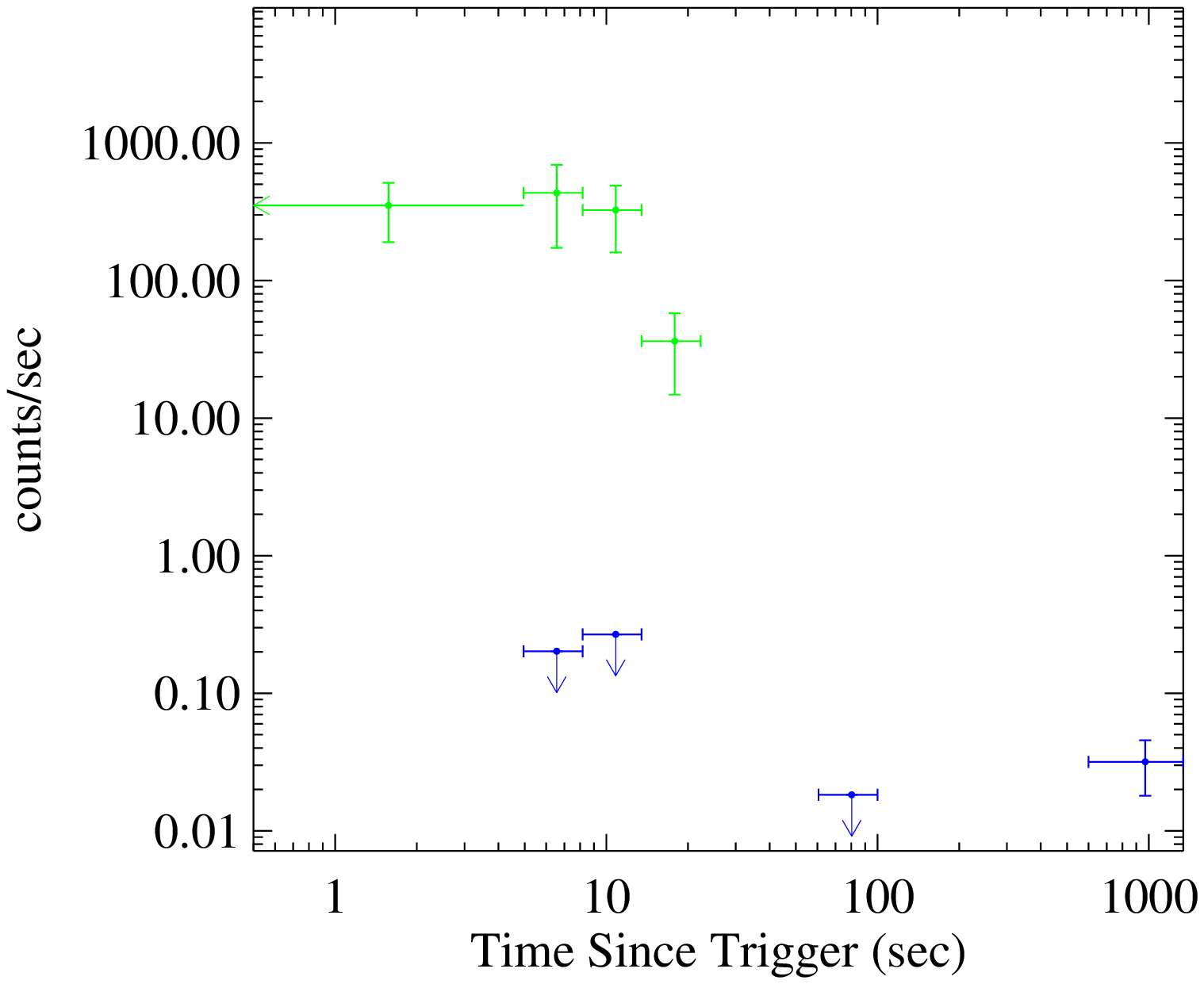}  \\
 \end{tabular}

\caption{Same as Figure 1, but for GRB 100225A.}
\label{100225A}
\end{figure}

\begin{figure}
\begin{tabular}{ll}
\multirow{3}{*}{ \includegraphics[angle=0,scale=0.78]{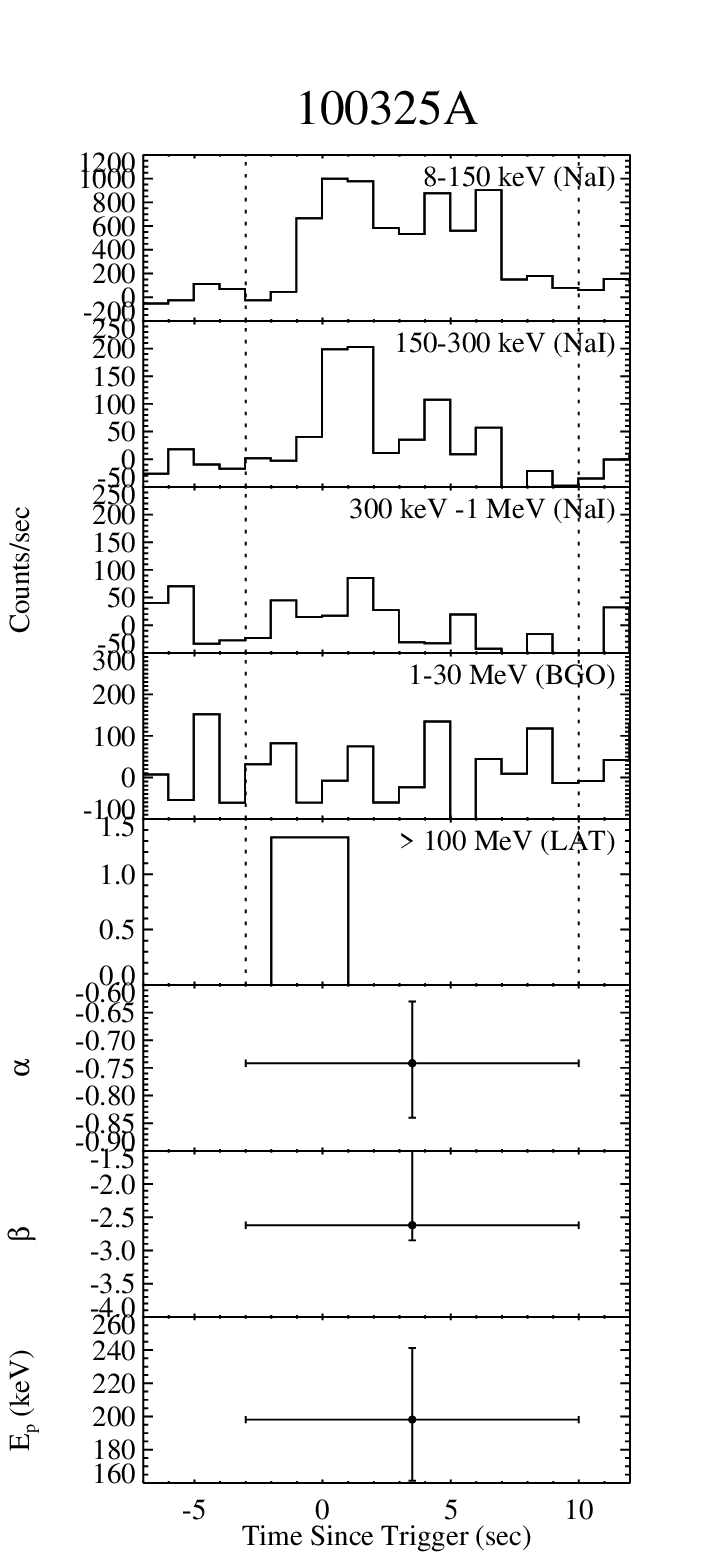}}
 & \includegraphics[angle=270,scale=0.2]{f16b.ps} \\
 & \includegraphics[angle=0,scale=.33]{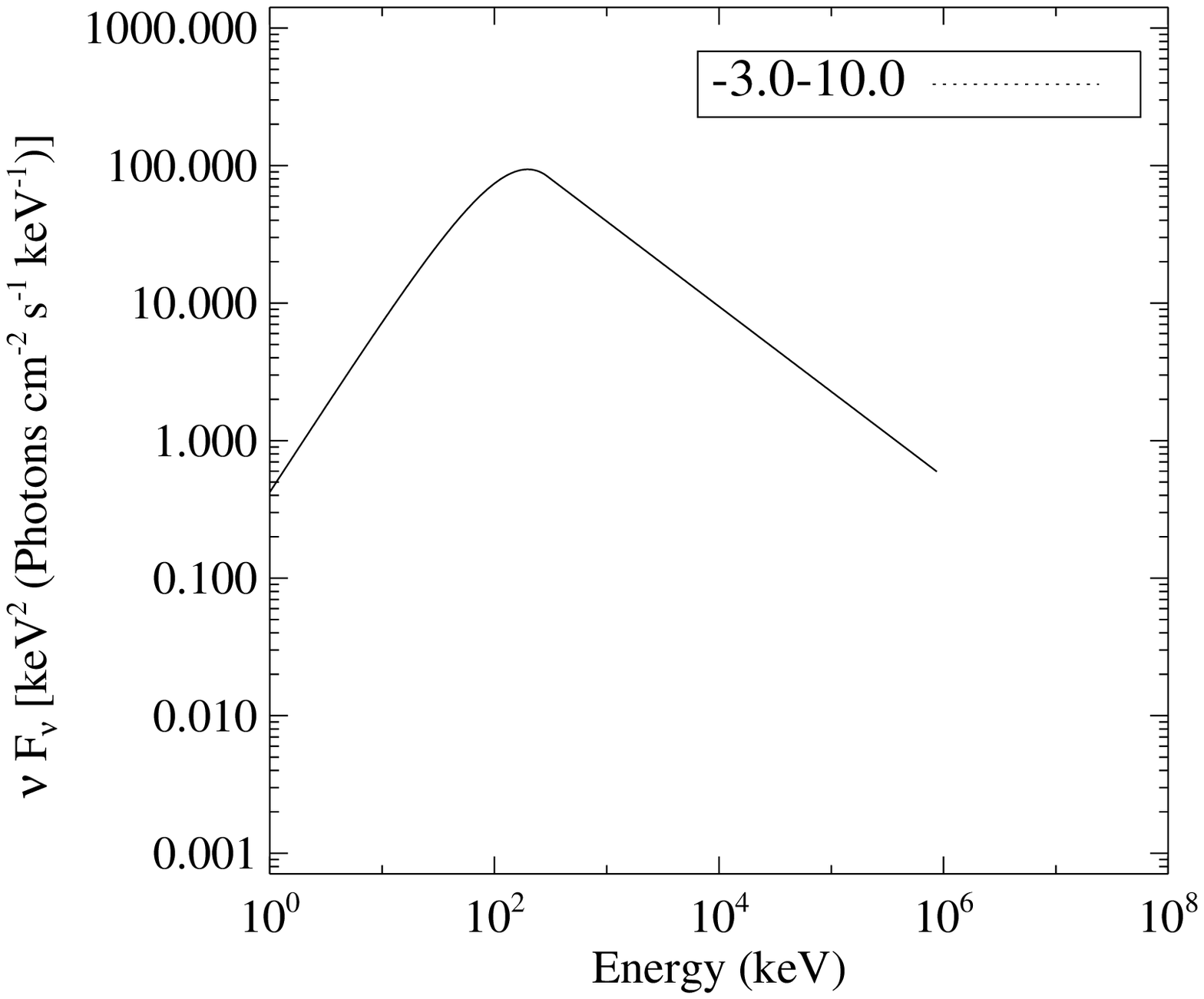} \\
  &  \includegraphics[angle=0,scale=.33]{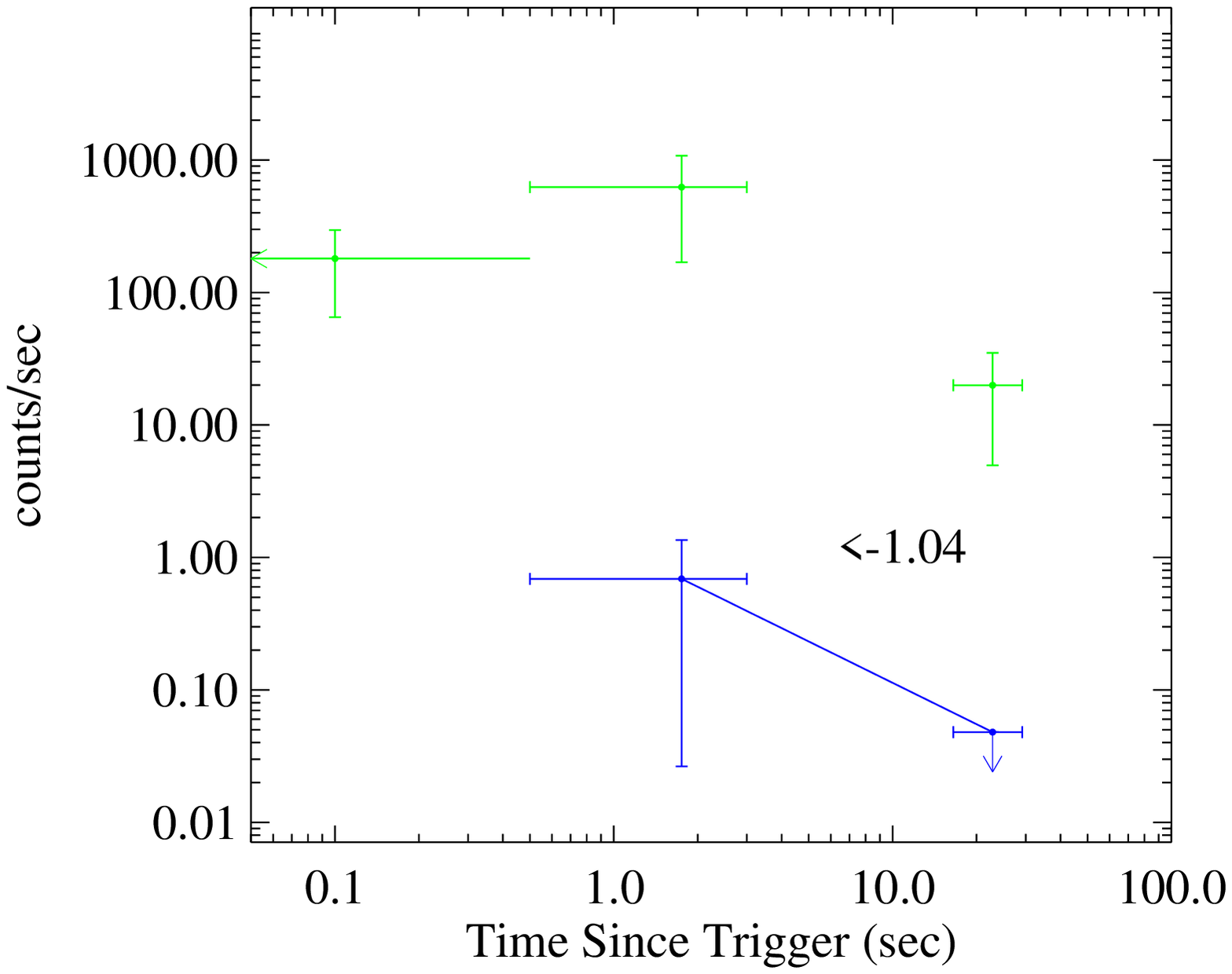}  \\
 \end{tabular}

\caption{Same as Figure 1, but for GRB 100325A.}
\label{100325A}
\end{figure}

\begin{figure}
\begin{tabular}{ll}
\multirow{3}{*}{ \includegraphics[angle=0,scale=0.78]{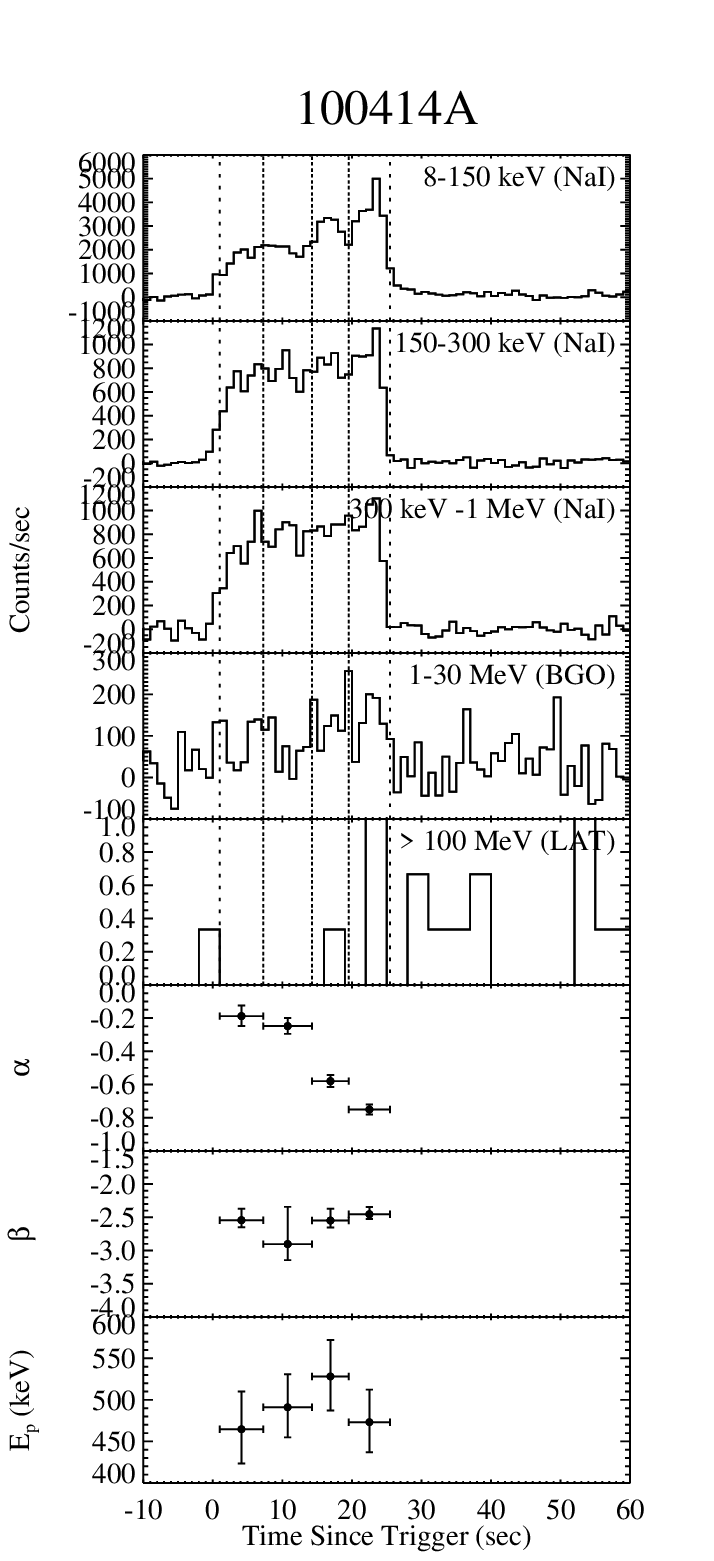}}
 & \includegraphics[angle=270,scale=0.2]{f17b.ps} \\
 & \includegraphics[angle=0,scale=.33]{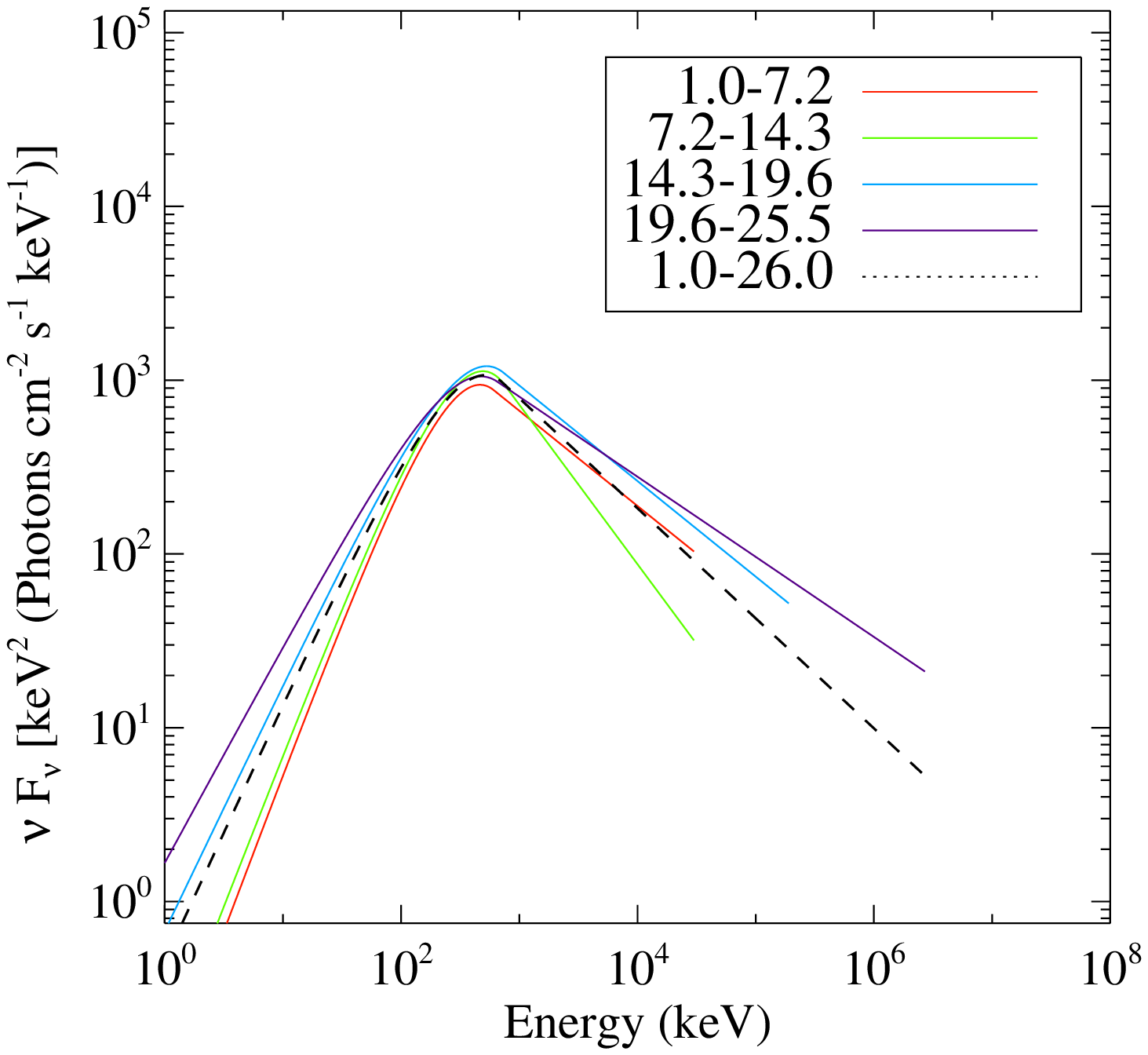} \\
  &  \includegraphics[angle=0,scale=.33]{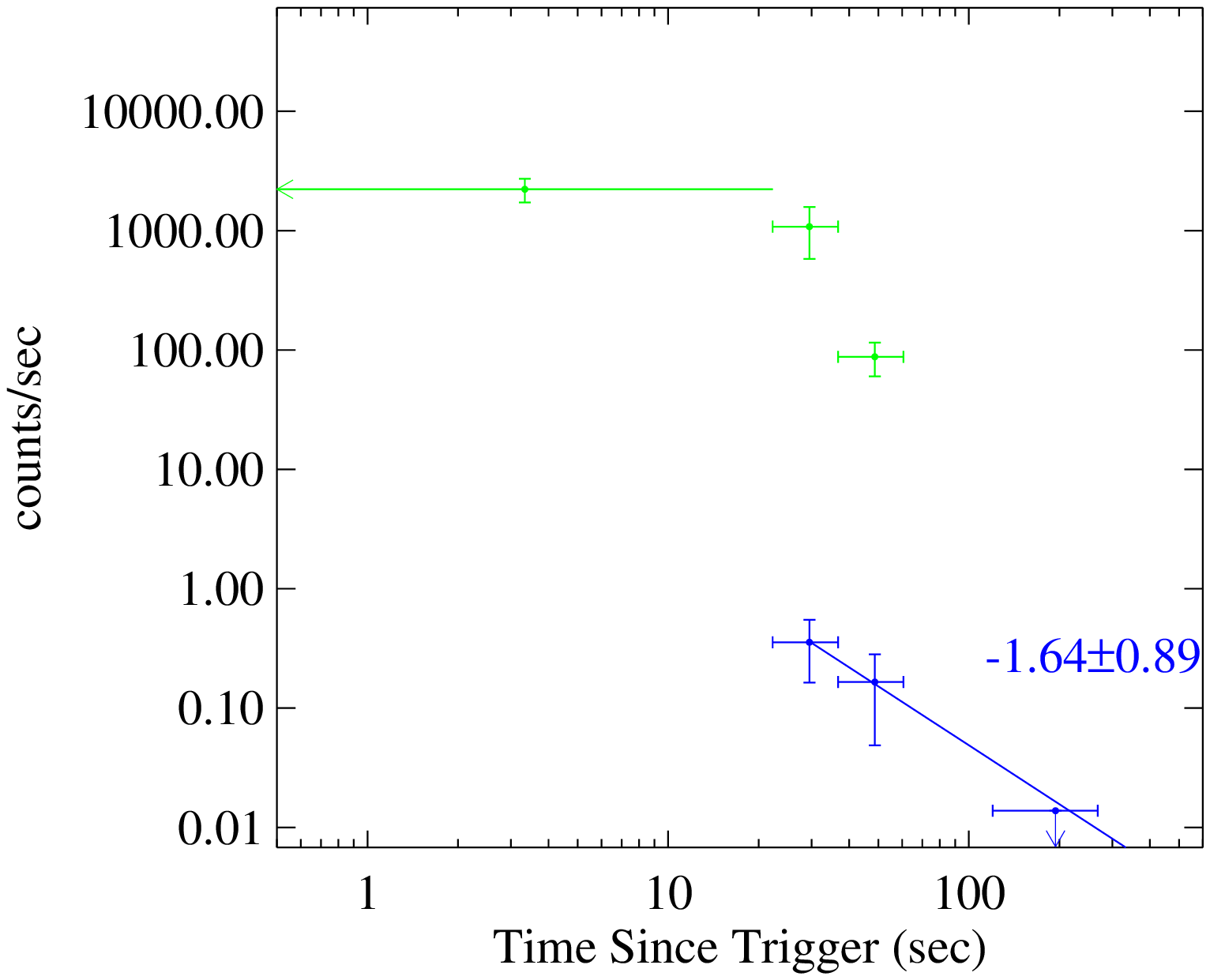}  \\
 \end{tabular}

\caption{Same as Figure 1, but for GRB 100414A.}
\label{100414A}
\end{figure}

\begin{figure}
\begin{tabular}{lll}
 \includegraphics[angle=0,scale=.33]{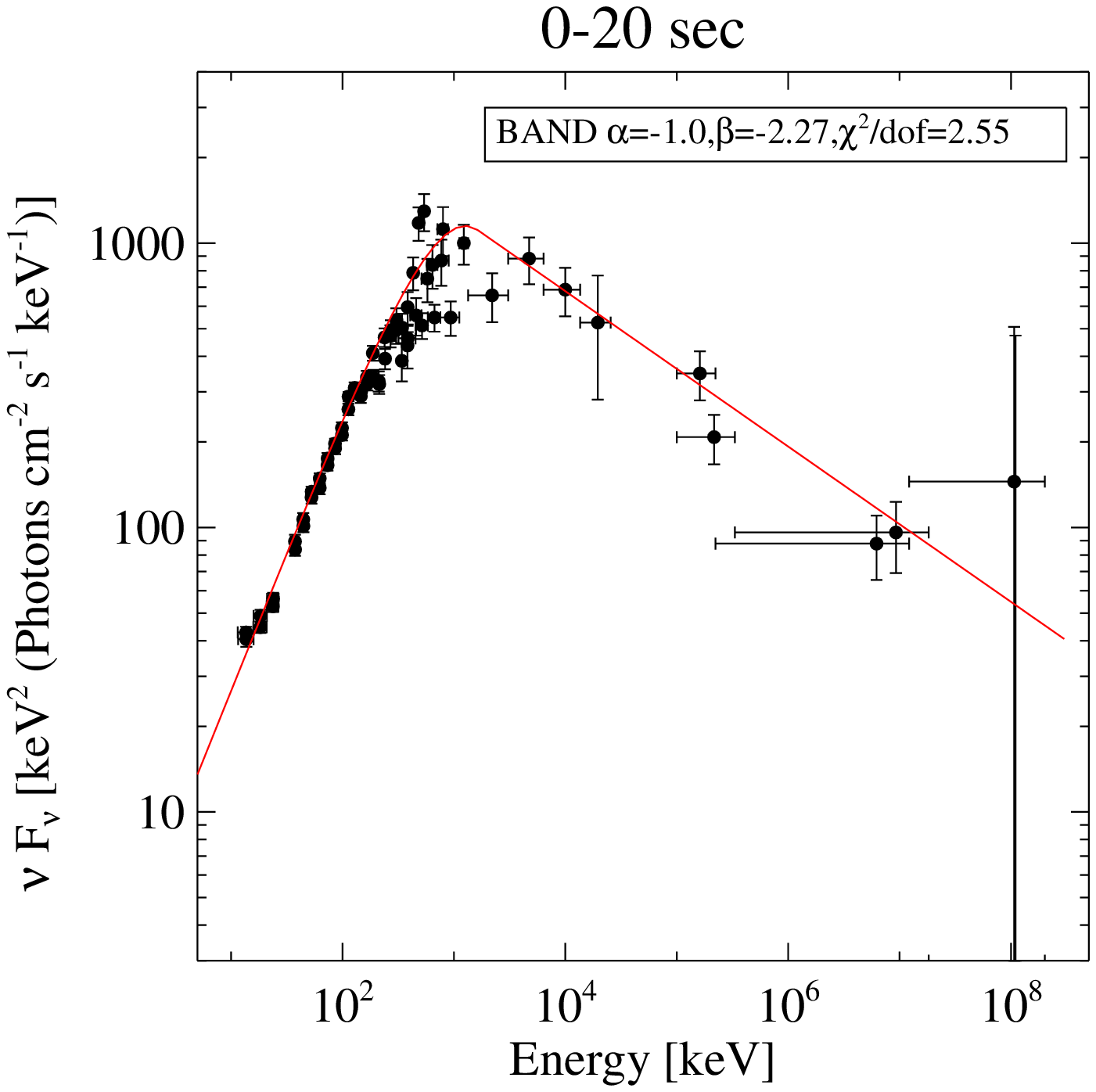}
 & \includegraphics[angle=0,scale=.33]{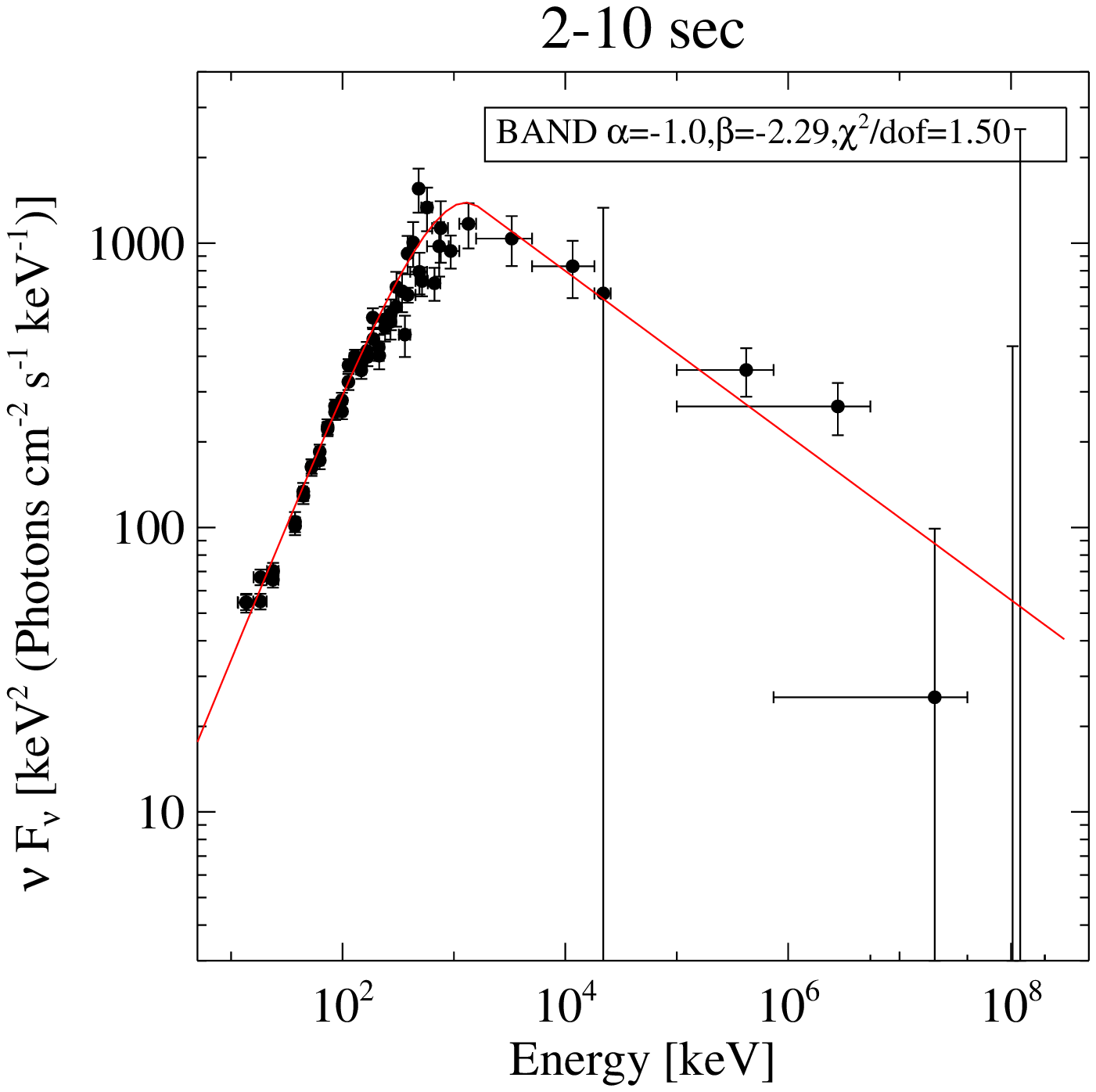}
 & \includegraphics[angle=0,scale=.33]{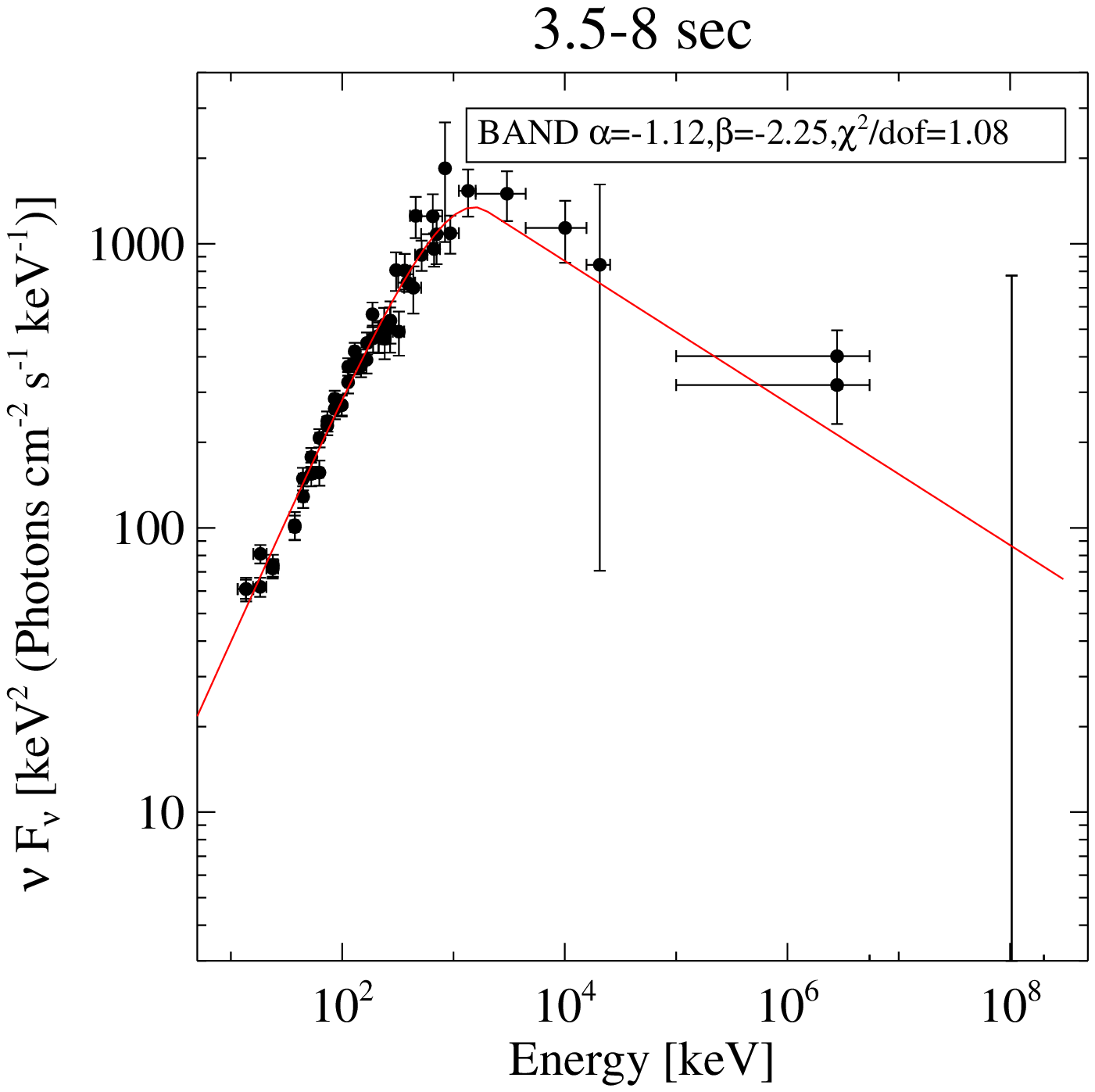} \\
 \includegraphics[angle=0,scale=.33]{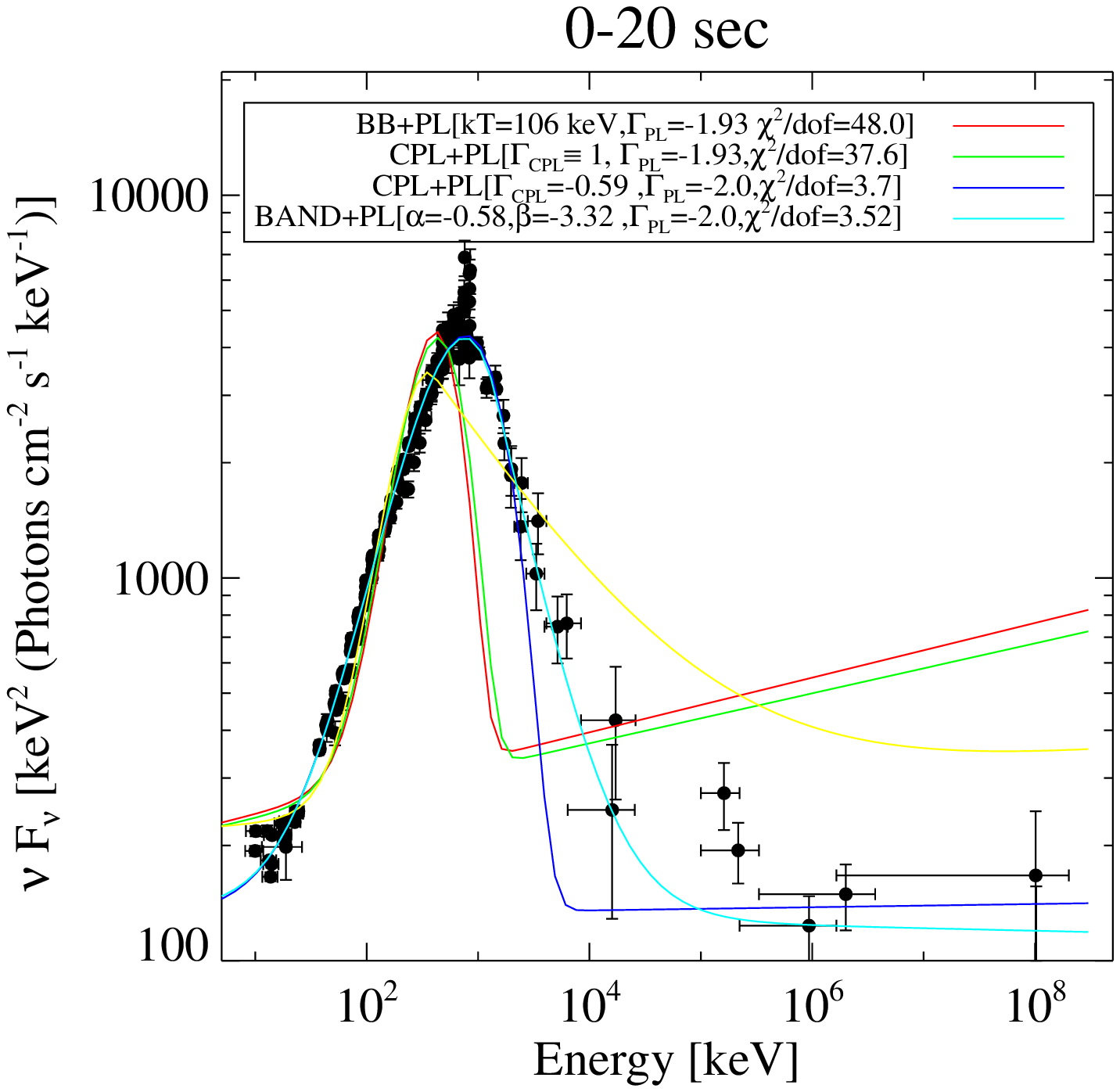}
 & \includegraphics[angle=0,scale=.33]{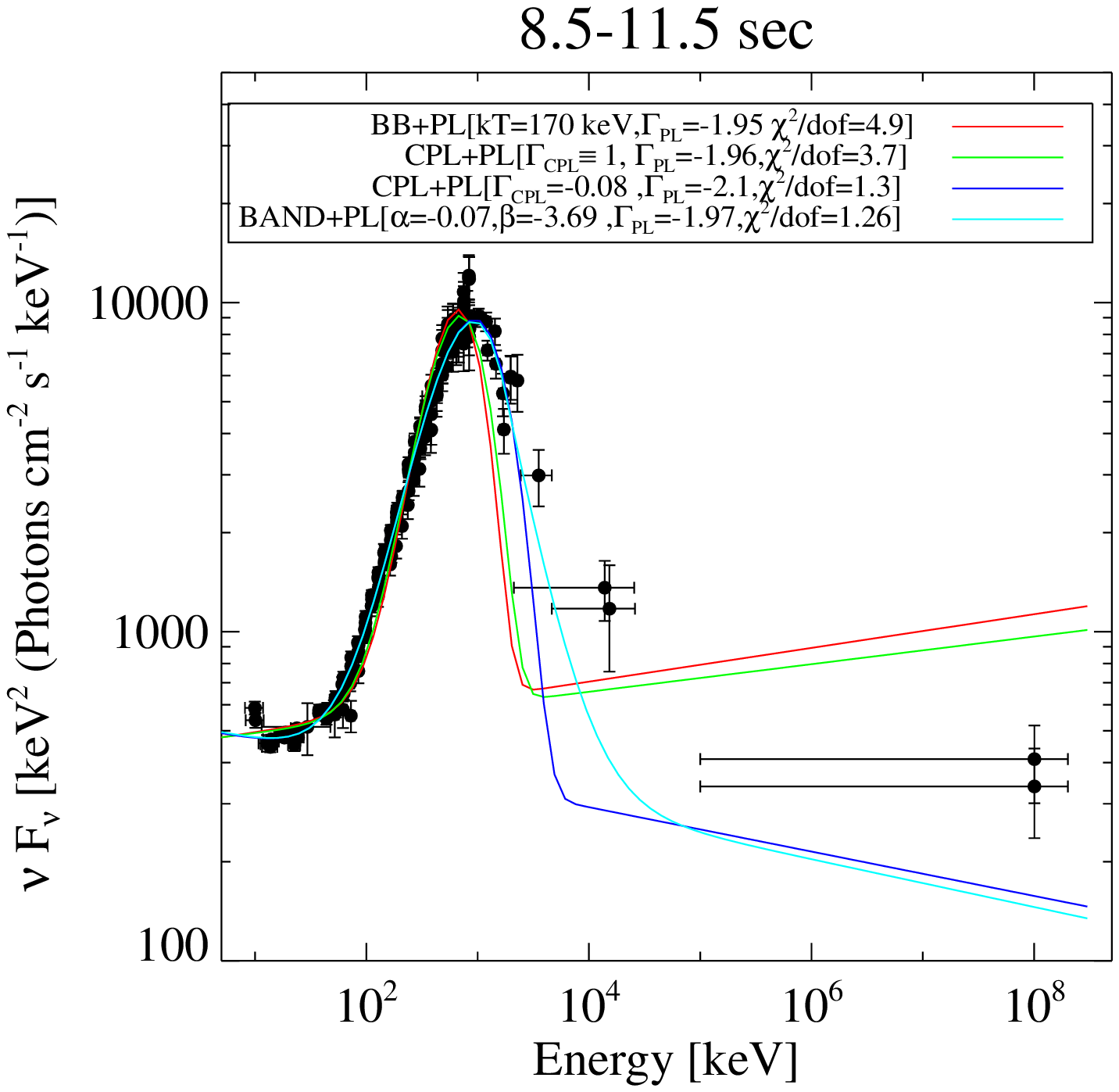}
 & \includegraphics[angle=0,scale=.33]{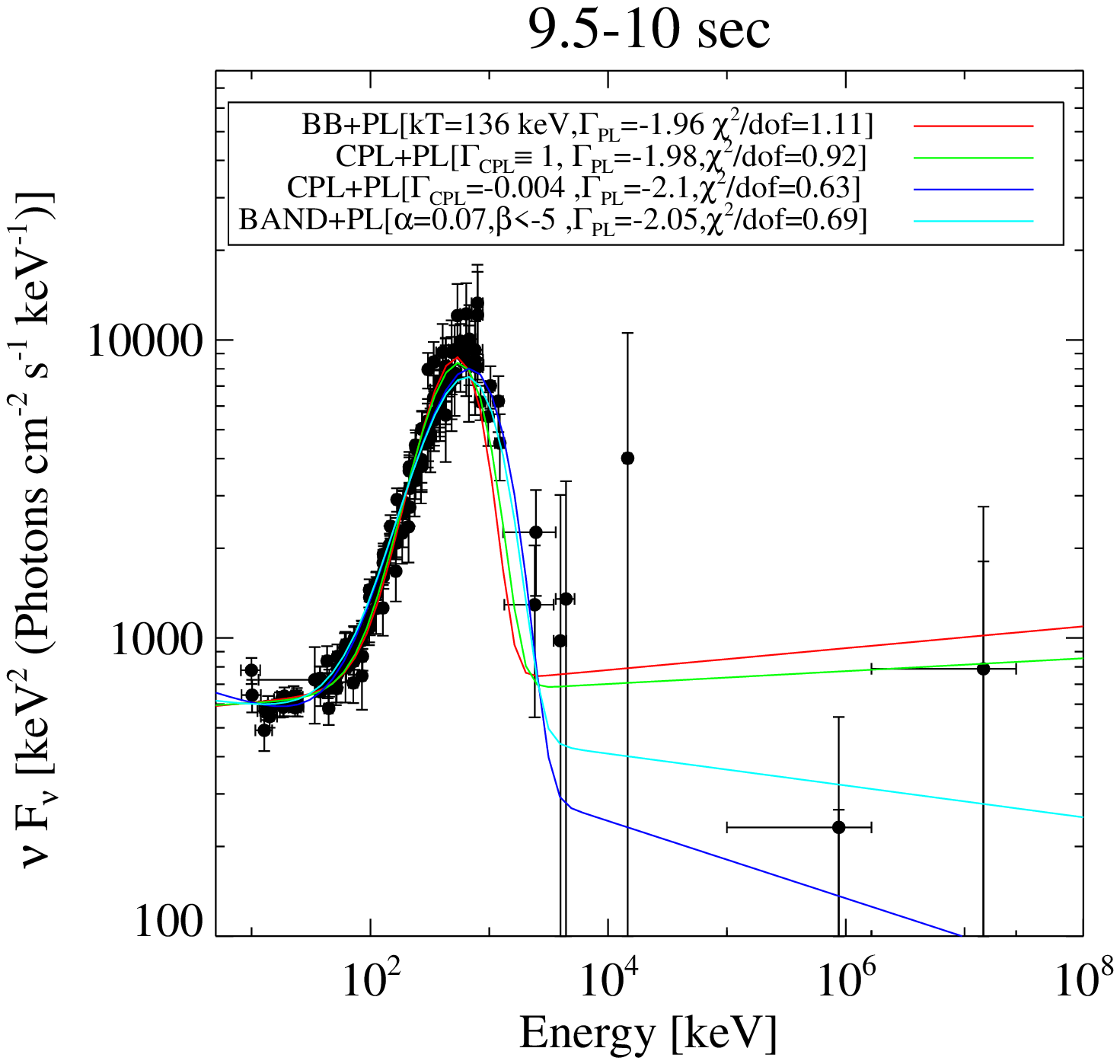}
\end{tabular}
\caption{A comparison between GRB 080916C and GRB 090902B.
{\em Upper panel}: The case of GRB 080916C. The Band parameters are
$(\alpha, \beta)= (-1.0, -2.27) (-1.0, -2.29), (-1.12, -2.25)$
for 0-20 s, 2-10 s, and 3.5-8 s, respectively. Little spectral parameter
variation is seen with reducing time bins.
{\em Lower panel}: The case of GRB 090902B. (1) For 0-20 s, the
Band+PL model ($\alpha=-0.58$, $\beta=-3.32$, $\Gamma_{\rm PL} =-2.0$
with $\chi^2/{\rm dof}=3.52$) and the CPL+PL model ($\Gamma_{\rm CPL}=-0.59$,
$\Gamma_{\rm PL}=-2.0$ with $\chi^2{\rm dof}=3.7$)
give marginally acceptable fits to the data.
The CPL+PL model with $\Gamma_{\rm CPL}=1$
(Rayleigh-Jeans) and the BB+PL model give unacceptable fits.
(2) For 8.5-11.5 s, the Band+PL model ($\alpha=-0.07$, $\beta=-3.69$,
$\Gamma_{\rm PL} =-1.97$ with $\chi^2/{\rm dof}=1.26$) and the CPL+PL model
($\Gamma_{\rm CPL}=-0.08$, $\Gamma_{\rm PL}=-2.1$ with $\chi^2{\rm dof}=1.3$)
give acceptable fits to the data. The CPL+PL model with $\Gamma_{\rm CPL}=1$
($\chi^2/{\rm dof}=3.7$) and the BB+PL model
($\chi^2/{\rm dof}=4.9$) give marginally acceptable fits.
(3) 9.5-10 s, the Band+PL model ($\alpha=0.07$, $\beta<-5$,
$\Gamma_{\rm PL} =-2.05$ with $\chi^2/{\rm dof}=0.69$) can only give
an upper limit on $\beta$. The CPL+PL model
($\Gamma_{\rm CPL}=-0.0004$, $\Gamma_{\rm PL}=-2.1$ with $\chi^2{\rm dof}=0.63$)
give marginally acceptable fit to the data. On the other hand,
the CPL+PL model with $\Gamma_{\rm CPL}=1$
($\chi^2/{\rm dof}=0.92$) and the BB+PL model
($\chi^2/{\rm dof}=1.11$) give acceptable fits.
Clear narrowing trend is seen when the time bins get smaller.}
\label{080916C-090902B}
\end{figure}

\begin{figure}
\begin{tabular}{lll}
 \includegraphics[angle=0,scale=.33]{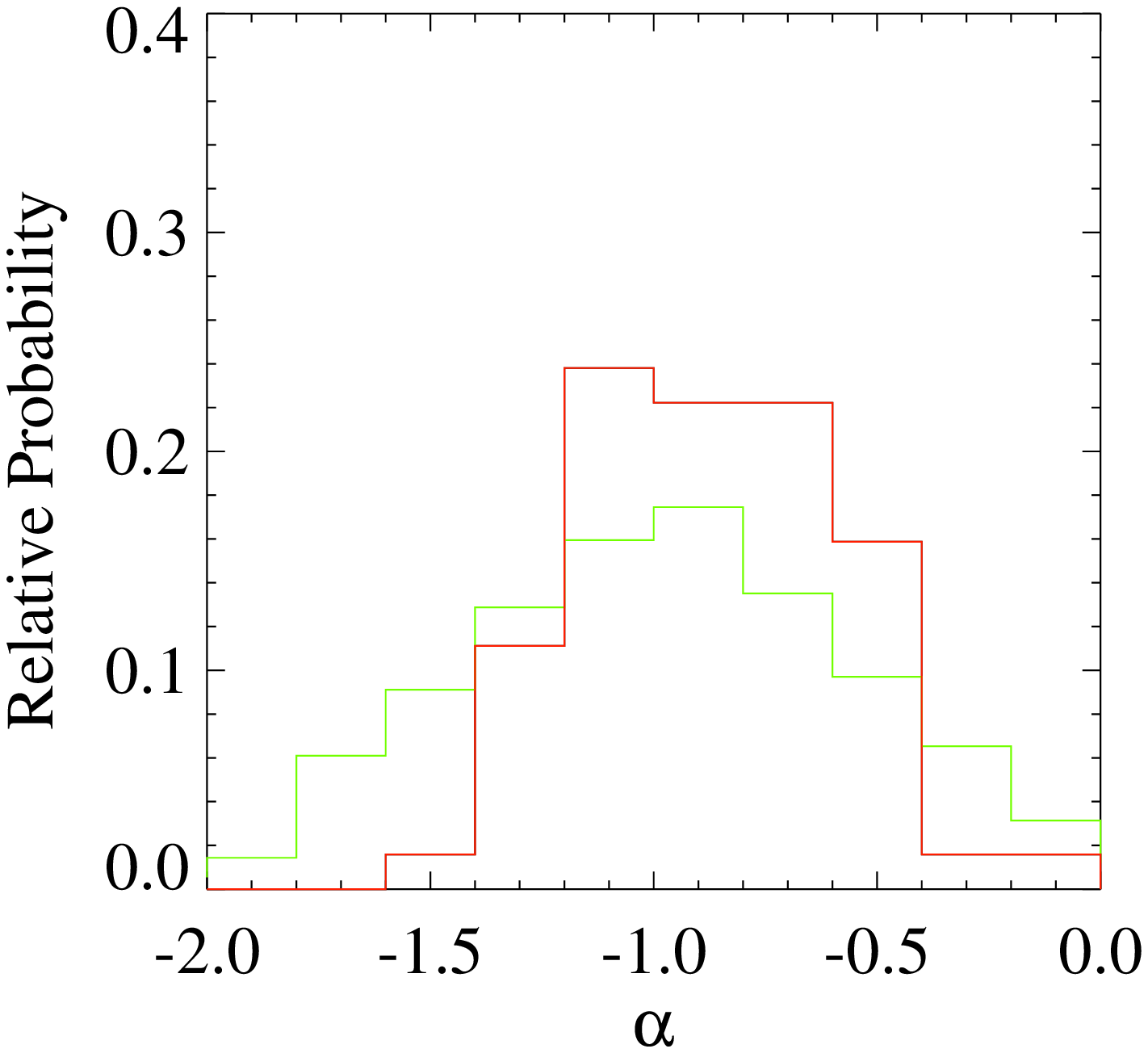}
 & \includegraphics[angle=0,scale=.33]{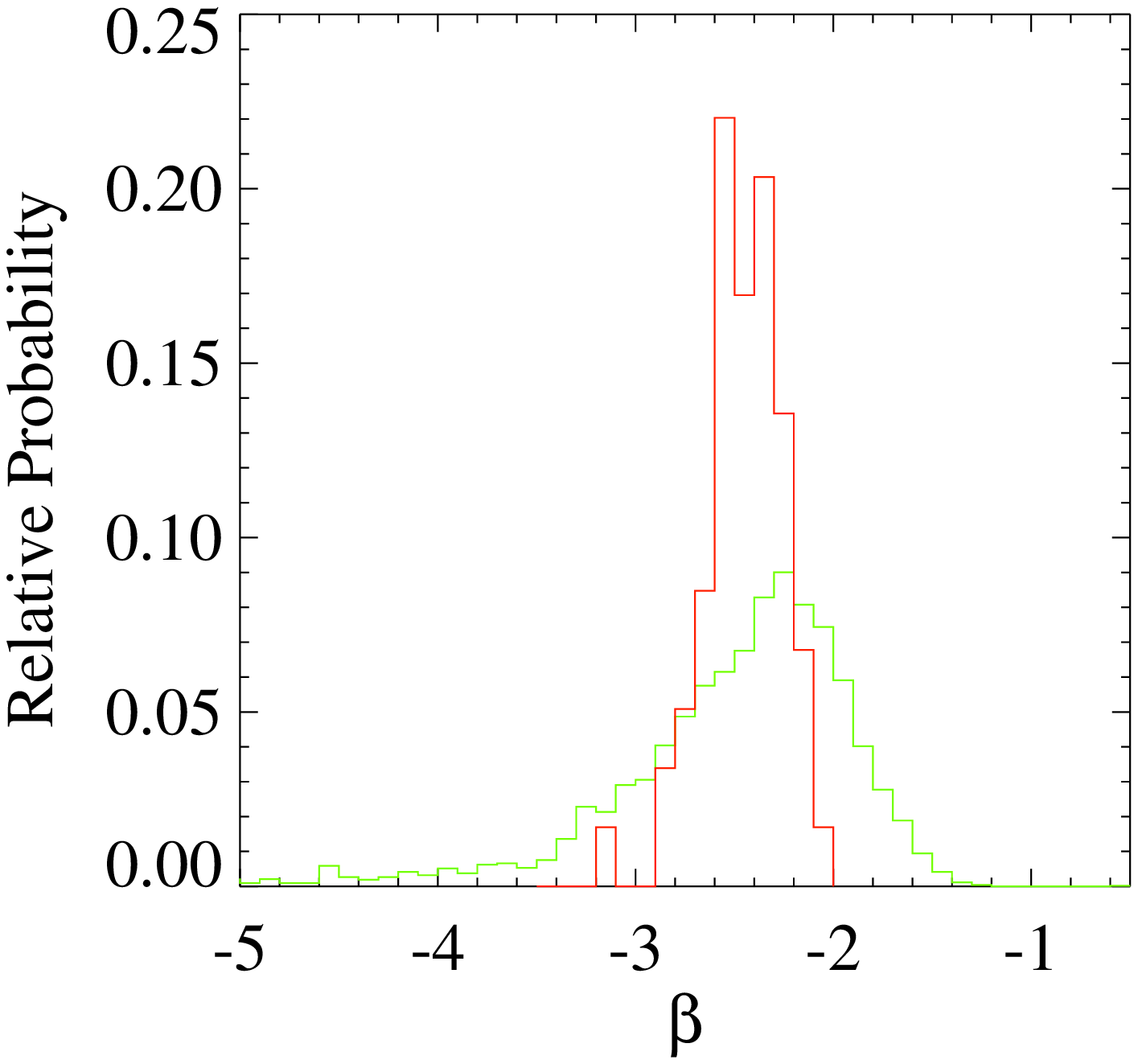}
 & \includegraphics[angle=0,scale=.33]{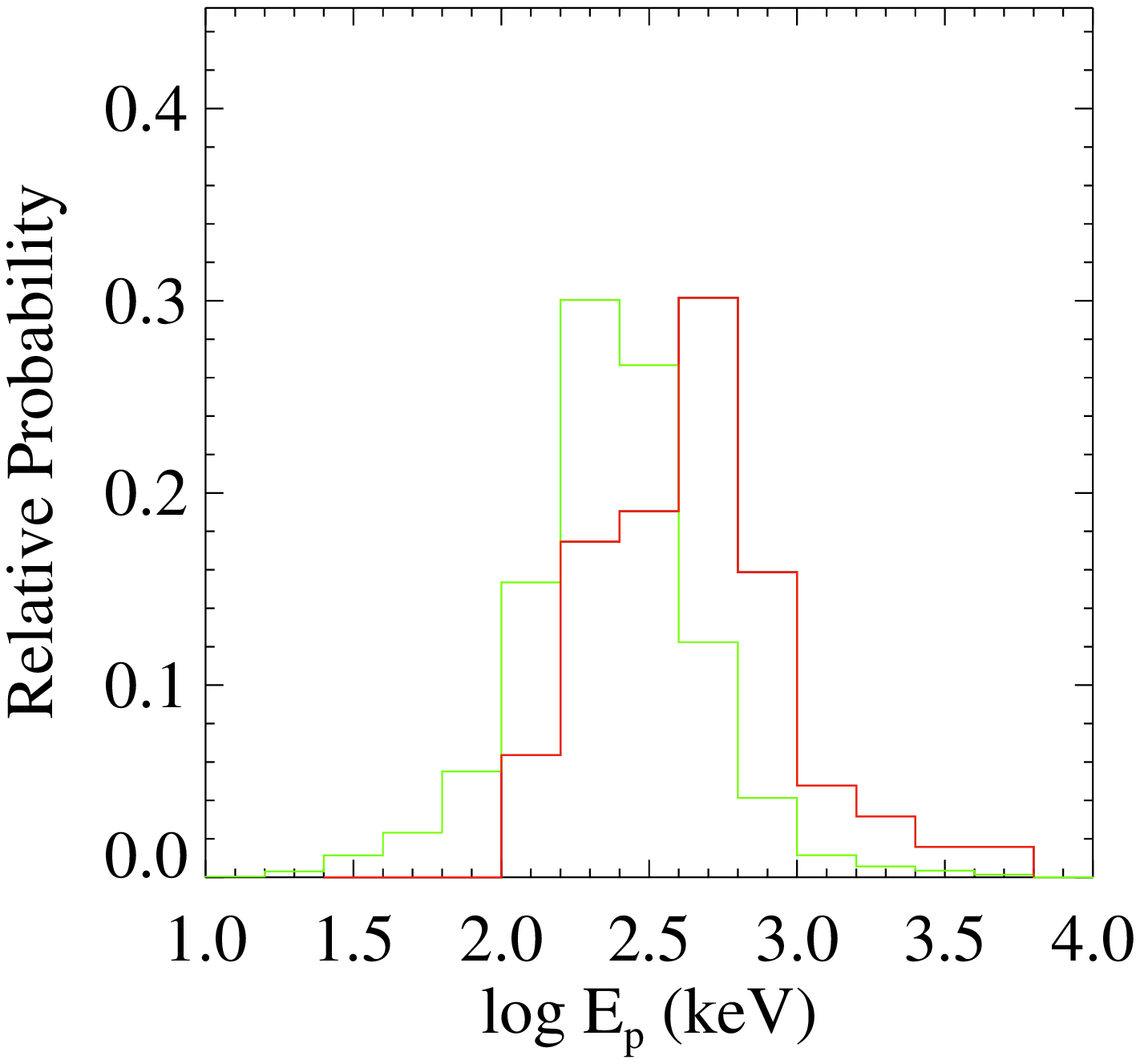}
\end{tabular}
\caption{Distributions of the Band-function parameters $\alpha$, $\beta$, and $E_p$
in our sample (red) in comparison with the BATSE bright sources sample (green).
The BATSE sample is adopted from Preece et al. (2000).}
\label{distributions}
\end{figure}

\begin{figure}
  \includegraphics[angle=0,scale=.56]{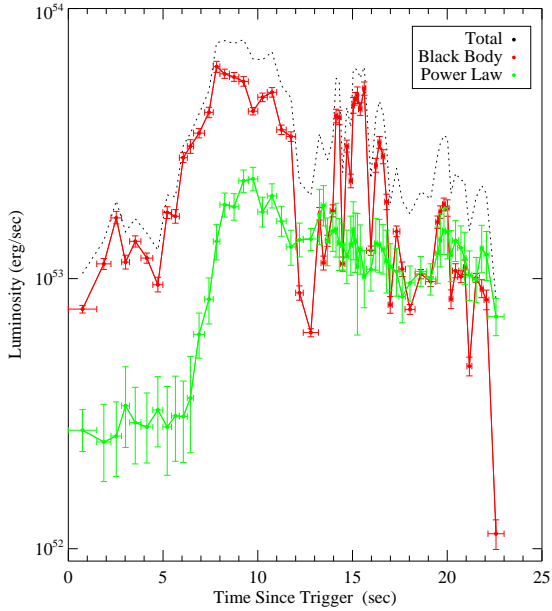}

\caption{A comparison between the lightcurves of the blackbody component
(red) and the power-law component (green) in GRB 090902B. The total
lightcurve (the sum of the two components, {\em
dotted} line) is also shown for comparison.}
\label{090902B-lightcurves}
\end{figure}

\begin{figure}
  \includegraphics[angle=0,scale=0.6]{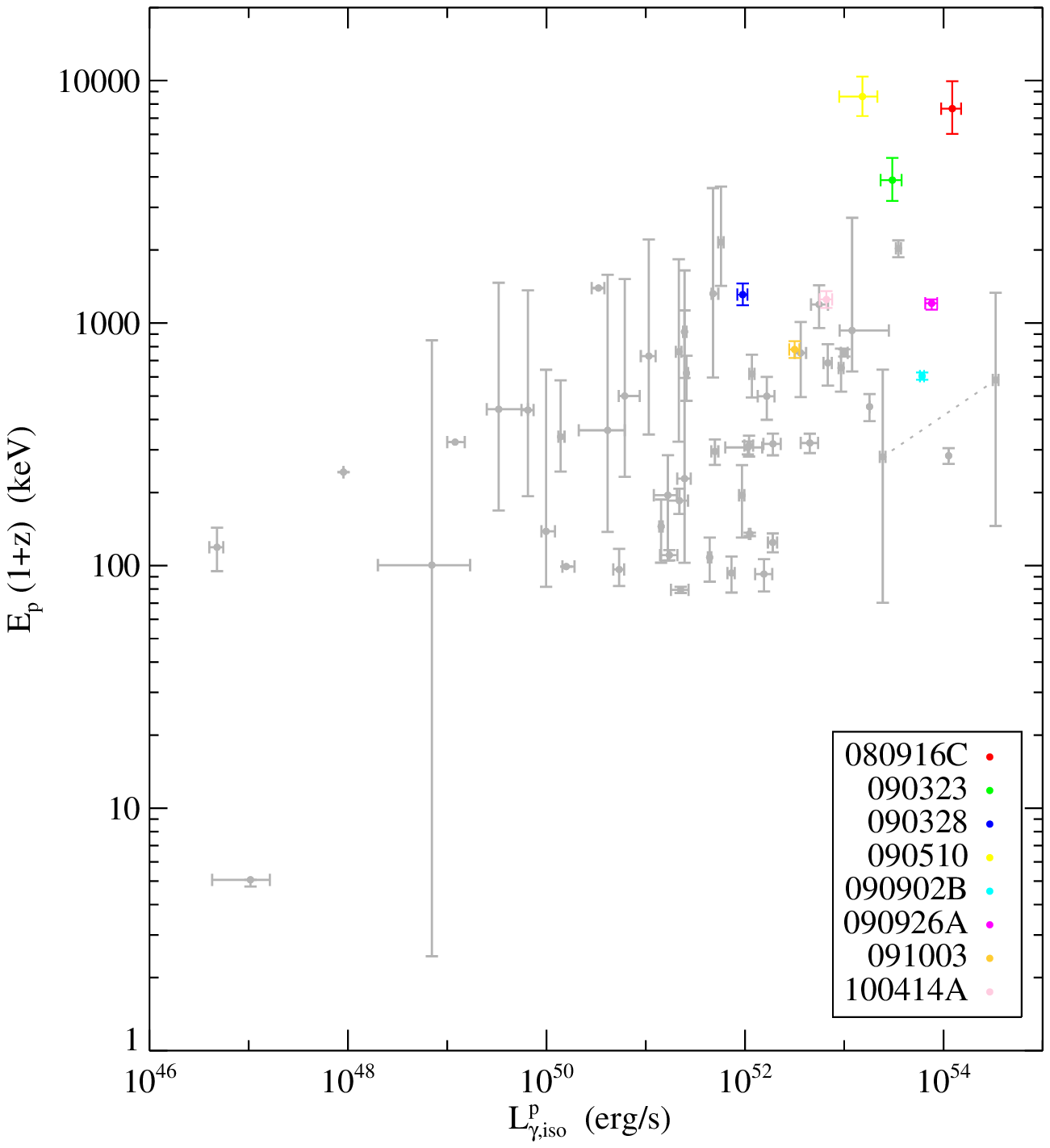}
  \includegraphics[angle=0,scale=0.6]{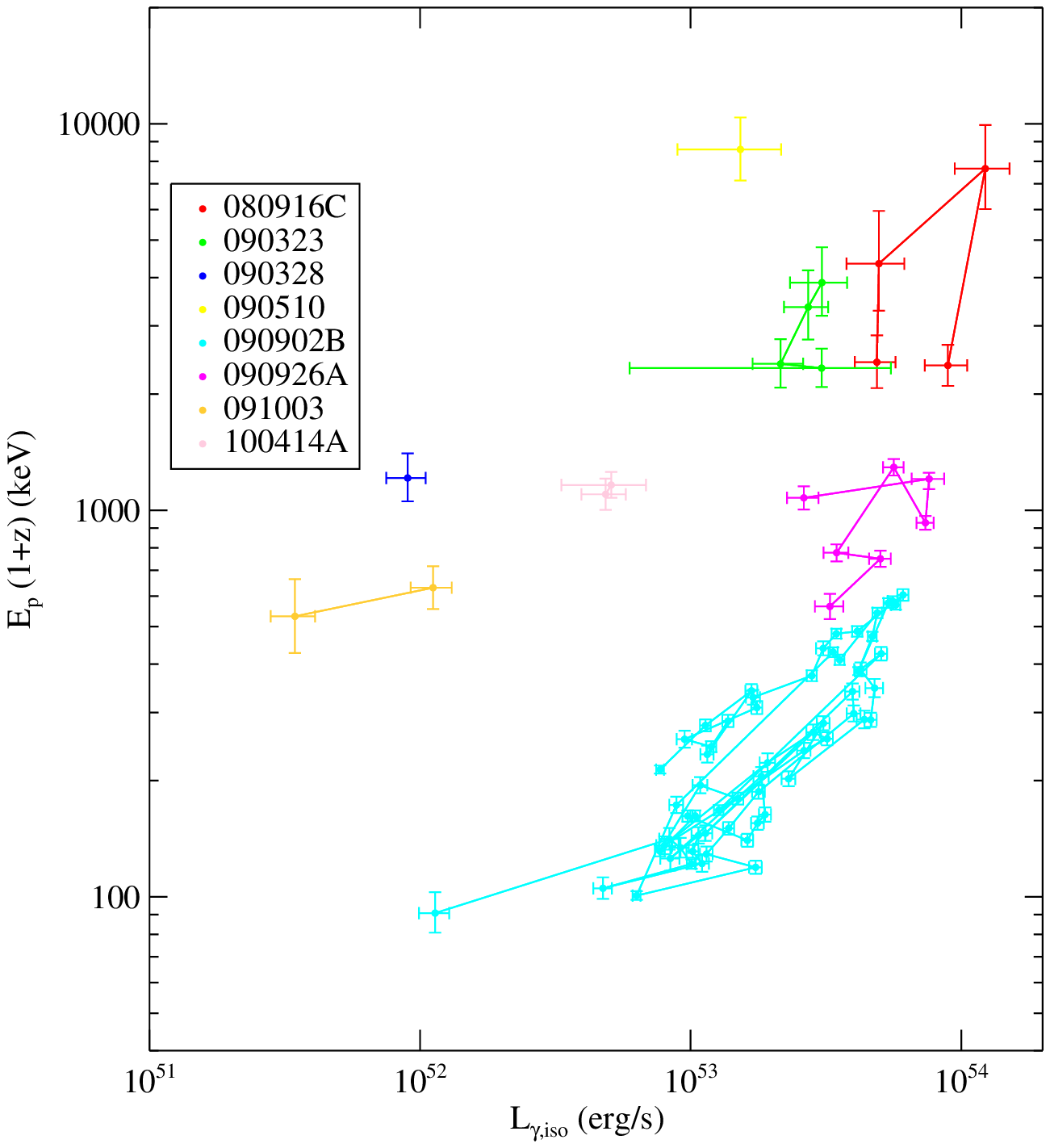}
\caption{The global $L_{\gamma,\rm iso}^{p}$ vs. $E_p(1+z)$ correlation (panel a)
and internal $L_{\gamma,\rm iso}$ vs. $E_p(1+z)$ correlation (panel b) for the
8 {\em Fermi}/LAT GRBs with known redshifts. The grey dots in (a) are previous bursts
taken from Zhang et al. (2009).}\label{Lp_Ep}
\label{L-Ep}
\end{figure}

\begin{figure}
  \includegraphics[angle=0,scale=0.51]{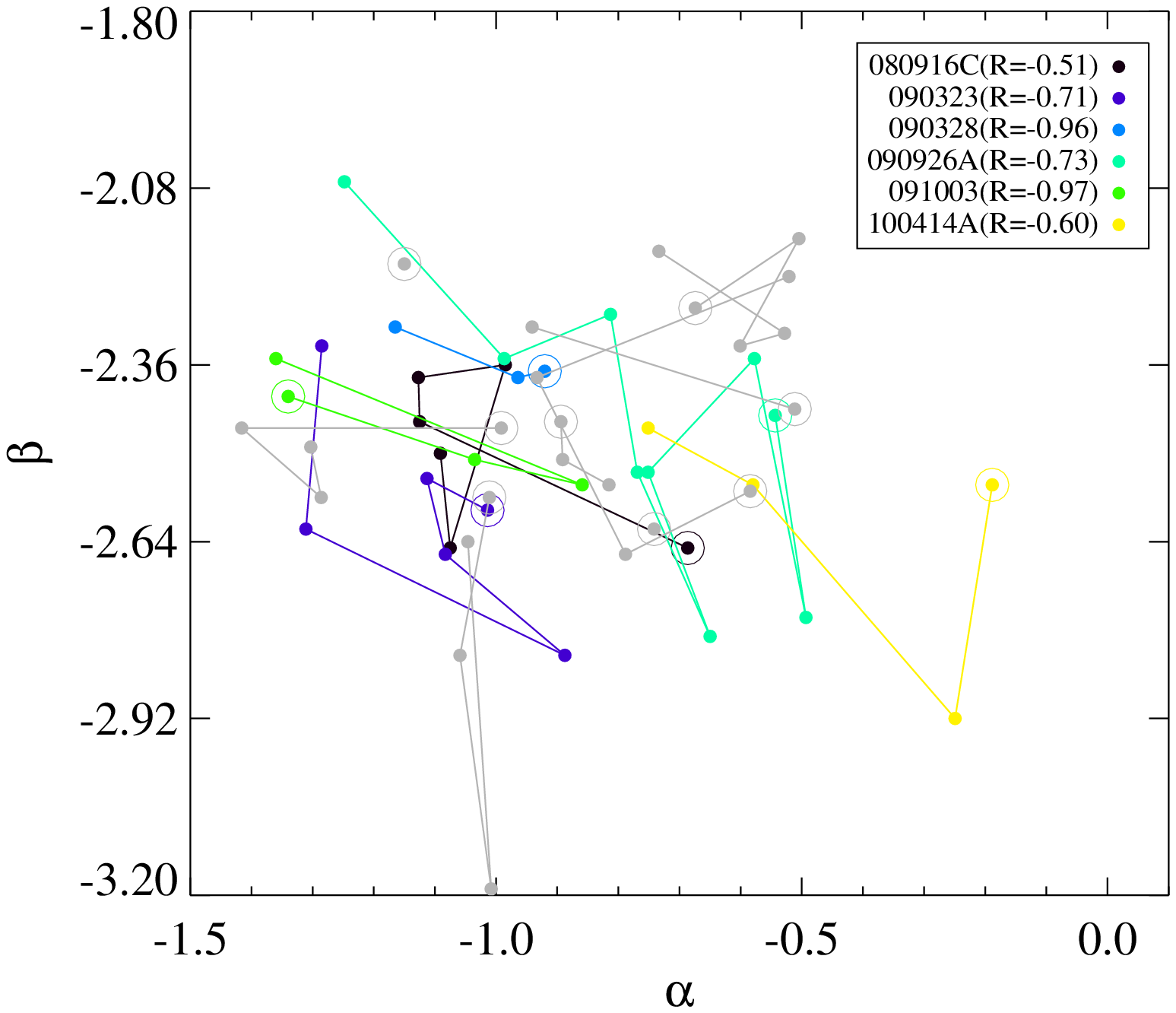}\ \  
  \includegraphics[angle=0,scale=0.51]{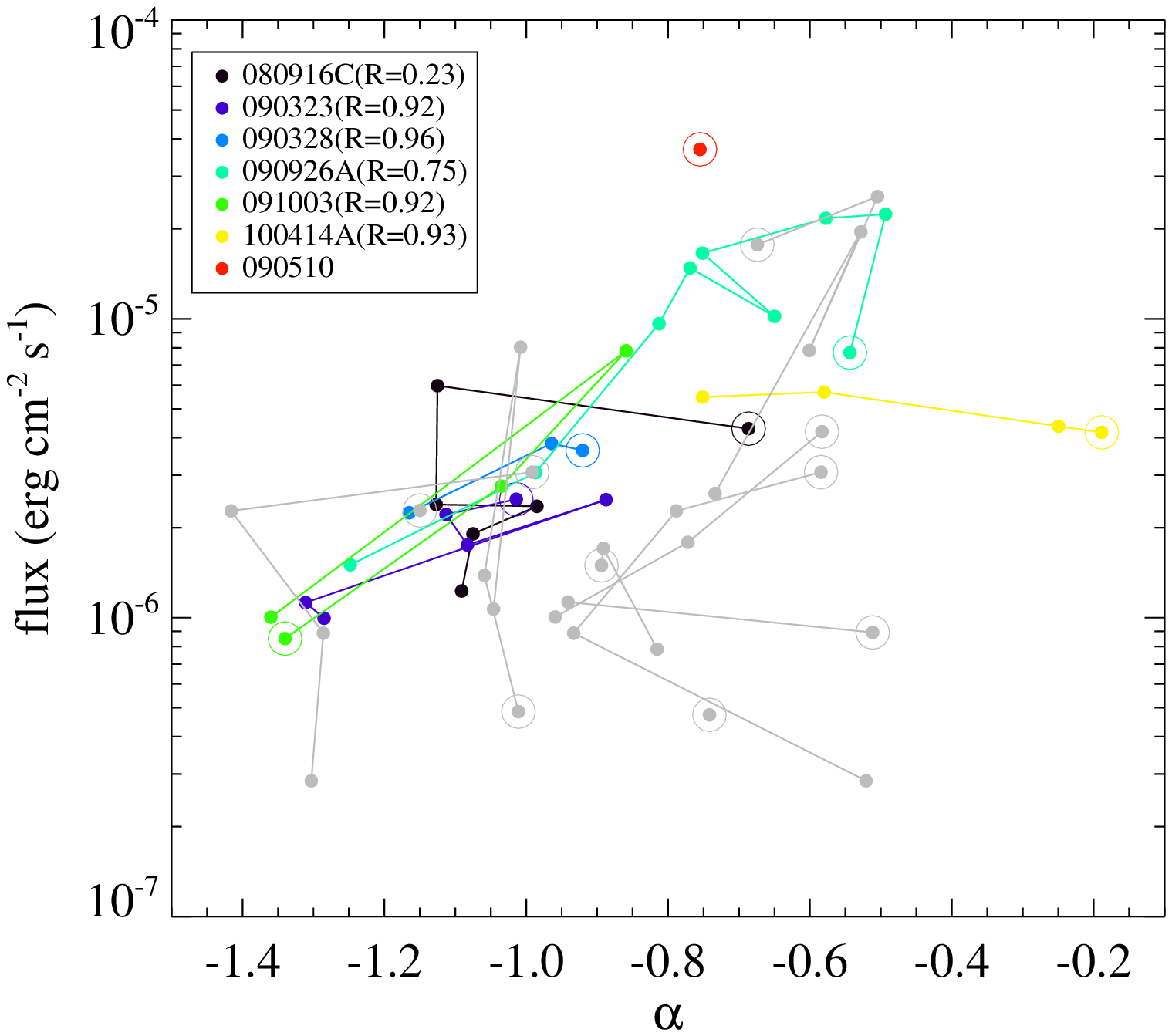}\\
  \includegraphics[angle=0,scale=0.51]{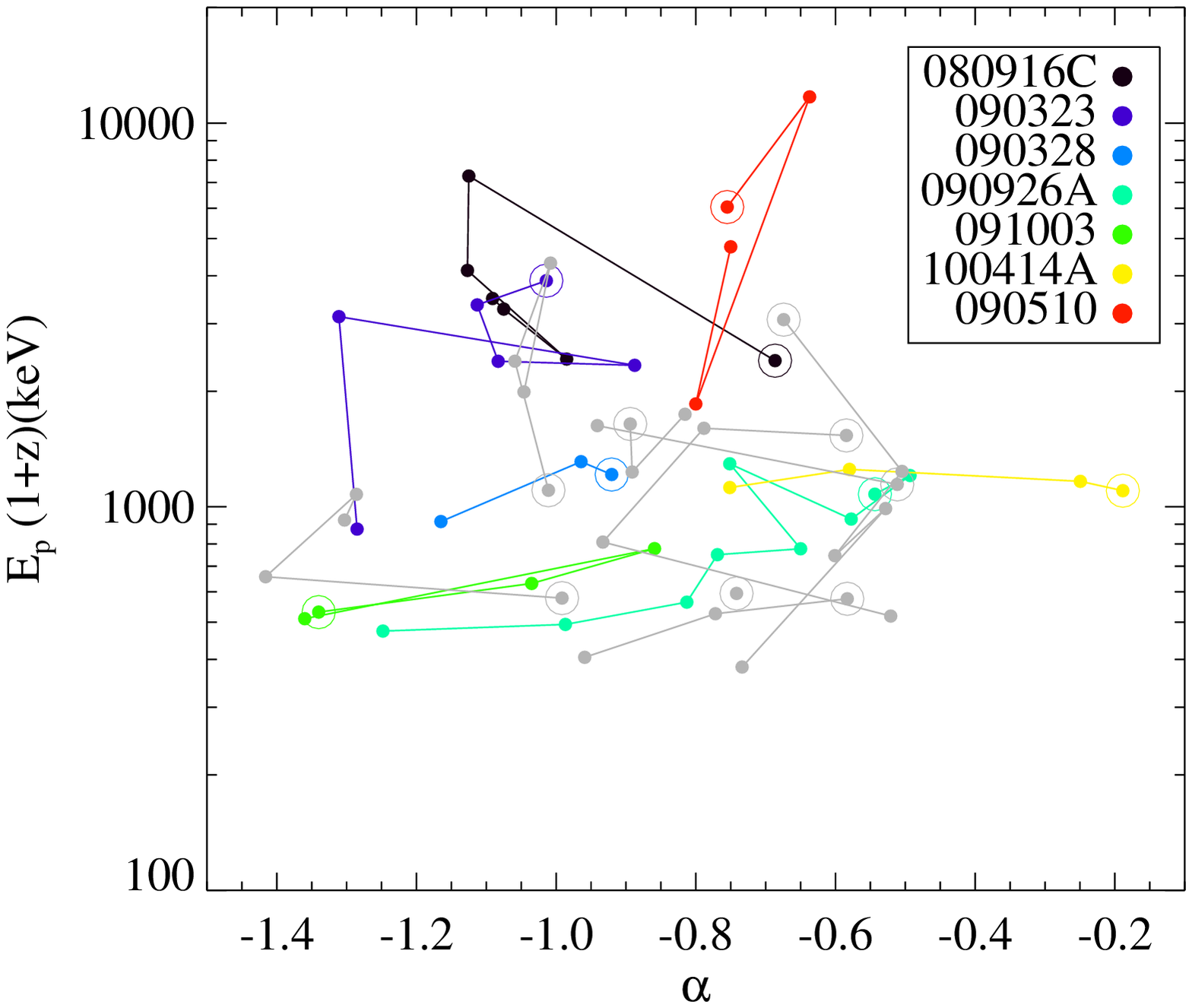}\ \ \ 
  \includegraphics[angle=0,scale=0.51]{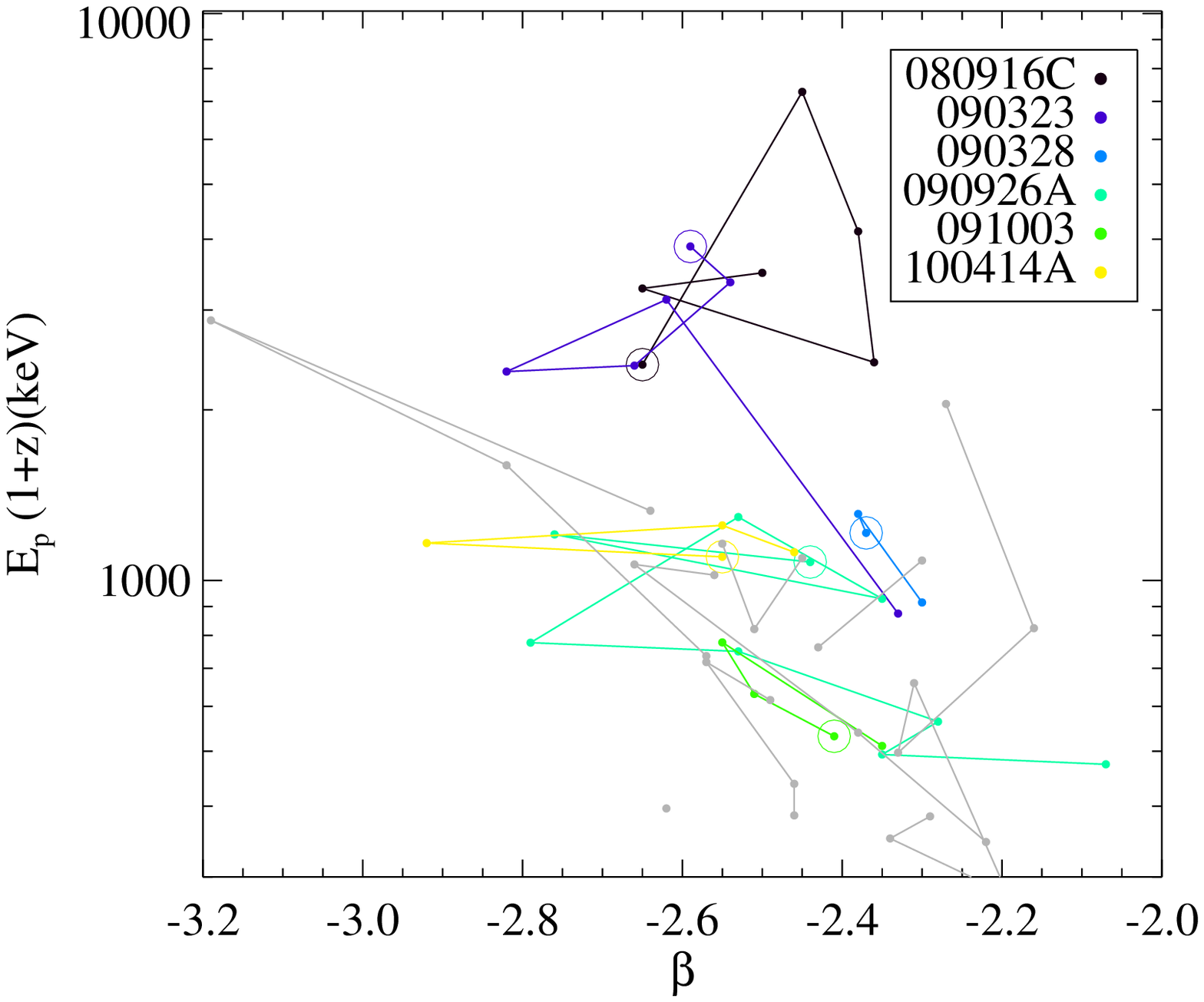}\\
\caption{The two dimension plots of various pairs of spectral 
parameters. (a) $\alpha-\beta$, with linear Pearson correlation coeffcients 
for individual bursts marked in the inset; (b) $\alpha-$flux, with linear Pearson 
correlation coeffcients for individual bursts marked in the inset; (c) $E_p-\alpha$;
(d) $E_p-\beta$. For those burst without redshift, $z=2.0$ is assumed (grey symbols
and lines).}
\label{correlations}
\end{figure}

In order to study the long-term lightcurve behavior, we extract 
the GBM-band and LAT-band lightcurves in logarithmic scale and 
present them in the bottom right panel of Figs.\ref{080825C}-\ref{100414A}.
We unevenly bin the LAT lightcurves with bin
sizes defined by the requirement that the signal-to-noise ratio
must be $>5$. For a close comparison, we correspondingly re-bin the 
GBM lightcurves using the same bin sizes.
Some GRBs (e.g. 080916C, 090510, 090902B, and 090926A) have
enough photons to make a well sampled LAT lightcurve.

In several GRBs, LAT emission lasts longer than GBM emission
and decays as a single power law (Ghisellini et al. 2010).
The decay indices of LAT emission are marked in the last panel 
of Figs.\ref{080825C}-\ref{100414A}, which can be also found in Table 3.
{ Due to low photon numbers, it is impossible to carry
out a time resolved spectral analysis. In any case, the LAT-band
photon indices of long-term LAT emission are estimated and also 
presented in Table 3.}
In Table 1 we mark those GRBs with detected LAT emission longer
than GBM emission and those without.
The most prominent ones with long lasting LAT afterglow
are GRBs 080916C, 090510, 090902B, and
090926A. Spectral analyses suggest that the LAT emission in
GRBs 090510 and 090902B is a different spectral component from the MeV
emission. The GBM lightcurves of these GRBs indeed follow a different
trend by turning off sharply as compared with the extended
PL decay in the LAT band. GRB 090926A, on the
other hand, shows a similar decay trend in both GBM and LAT
bands. GRB 080916C is special. Although the
spectral analysis shows a single Band function component, the
GBM lightcurve turns over sharply around 70-80 seconds,
while the LAT emission keeps decaying with a single PL. 

One caveat of LAT long-term lightcurves is that they depend
on the level of background and time-bin selection. Due to the low
count rate at late times, the background uncertainty can enormously
change the flux level, and a different way of binning the data may change
the shape of the lightcurve considerably.
In our analysis, the background model is extracted from the
time interval prior to the GBM trigger in the same sky region that
contains the GRB. The bin-size is chosen to meet the $5\sigma$
statistics to reduce the uncertainty caused by arbitrary binning.

Our data analysis suggests a controversial picture regarding
the origin of this GeV afterglow. Spectroscopically, the LAT-band
emission is usually an extension of the GBM-band emission and forms 
a single Band-function component, suggesting a common physical origin
with the GBM-band emission. If one focuses on the prompt emission
lightcurves, the LAT-band activities seem to track the GBM-band
activities. Even for GRB 090902B which shows a clear second spectral 
component, the PL component variability tracks that of the BB 
component well (Fig.\ref{090902B-lightcurves}), suggesting a
physical connection between the two spectral components. These
facts tentatively suggest that at least during the prompt emission 
phase, the LAT-band emission is likely connected to the GBM-band
emission, and may be of an ``internal'' origin similar to the 
GBM-band emission.

It has been
suggested that the entire GeV emission originates from the
external shock (e.g. Kumar \& Barniol Duran 2009a, 2009b;
Ghisellini et al. 2010; Corsi et al. 2009). This idea is based on
the power law temporal decay law that follows the prompt
emission. Such a GeV afterglow scenario is not
straightforwardly expected for the following reasons. First,
before {\em Fermi}, afterglow modeling suggests that for typical
afterglow parameters, the GeV afterglow is initially dominated by
the synchrotron self-Compton component (M\'esz\'aros \& Rees 1994;
Dermer et al.  2000; Zhang \& M\'esz\'aros 2001; Wei \& Fan 2007; Gou
\& M\'esz\'aros 2007; Galli \& Piro 2007; Yu et al. 2007; Fan et
al. 2008), or by other IC processes invoking both forward and
reverse shock electrons (Wang et al. 2001).  For very
energetic GRBs such as GRB 080319B, one may expect a 
synchrotron-dominated afterglow all the way to an energy $\sim$ 
10 GeV (Zou et al. 2009; Fan et al. 2008). 
Second, the required
parameters for the external shock are abnormal to interpret the
data. For example, the magnetic field strength at the forward
shock needs to be much smaller than equipartition, consistent
with simply compressing the ISM magnetic field without shock
amplification (Kumar \& Barniol Duran 2010). This, in turn,
causes a problem in accelerating electrons to a high enough
energy to enable emission of GeV photons (Li 2010a; Piran \& Nakar
2010).  Moreover, the circumburst number density of these long
GRBs are required to be much lower than that of a typical
ISM (e.g., Kumar \& Barniol Duran 2010), which challenges the
collapsar model. 
Finally, observed GeV decay slope is
typically steeper than the predictions invoking a standard
adiabatic forward shock (e.g. 
Figs.\ref{080916C},\ref{090510},\ref{090902B},\ref{090926A},\ref{100414A}, 
see also Ghisellini et al. 2010). One needs to invoke a radiative
blastwave (Ghisellini et al. 2010) or a Klein-Nishina
cooling-dominated forward shock (Wang et al. 2010) to account for
the steepness of the decay slope.

The external shock model to interpret the entire GeV emission is 
challenged by the following two arguments. First, the GeV lightcurve
peak coincides the second peak of the GBM lightcurve for
GRB 080916C. This requires a fine-tuned bulk Lorentz factor of the
fireball to make the deceleration time coincide the epoch of
the second central engine activity. This is highly contrived.
Second, the external shock component should not have decayed 
steeply while the prompt emission is still on going. To examine 
this last point, we have applied the shell-blastwave code developed
by Maxham \& Zhang (2009) to model the
blastwave evolution of GRB 080916C using the observed data by
assuming that the outflow kinetic energy traces the observed
gamma-ray lightcurve (assuming a constant radiation
efficiency). The resulting LAT-band lightcurve always displays a
shallow decay phase caused by refreshing the forward shock by
materials ejected after the GeV lightcurve peak time even for a
radiative blastwave, in stark contrast to the data. This casts
doubts on the external shock origin of GeV emission during
the prompt phase (Maxham et al. 2011). 
We note that detailed modeling of GRB 090510 (He et al. 2010)
and GRB 090902B (Liu \& Wang 2011) with the external shock
model both suggests that the prompt GeV emission cannot be
interpreted as the emission from the external forward shock.

Collecting the observational evidence and the theoretical
arguments presented above, we suggest that at least during the prompt 
emission phase (when GBM-band emission is still on), the LAT-band
emission is not of external forward shock origin.

After the GBM-band prompt emission is over, the LAT-band emission
usually decays as a PL. We note that the long-term GeV lightcurve 
can be interpreted in more than one way.
(1) If one accepts that the prompt GeV emission is of internal origin, one may argue that the external shock 
component sets in before the end of the prompt emission
and thereafter dominates during the decay phase (Maxham et al. 
2011). This requires arguing for coincidence of the same 
decaying index for the early internal and the late 
external shock emission. Considering a possible superposition
effect (i.e. the observed flux during the transition epoch
includes the contributions from both the internal and external
shocks), this model is no more contrived
than the model that interprets prompt GeV emission as 
from external shocks, which requires 
coincidence of internal emission spectrum and the
external shock emission spectrum to mimic the same
Band spectrum in all time bins (Kumar \& Barniol Duran 2009).
(2) An alternative possibility is to
appeal to an internal origin of the entire GeV long-lasting 
afterglow, which reflects the gradual ``die-off'' of the central
engine activity. The difficulty of such a suggestion is that it must 
account for the different decaying behaviors between the
GBM-band emission and LAT-band emission in some (but not
all) GRBs (e.g. GRB 080916C). To differentiate between these
possibilities, one needs a bright GRB co-triggered by
{\em Fermi} LAT/GBM and {\em Swift} BAT, so that an early
{\em Swift} XRT lightcurve is available along with the
early GeV lightcurve. The external-shock-origin GeV
afterglow should be accompanied by a PL decaying early
X-ray lightcurve (Liang et al. 2009) instead of the canonical
steep-shallow-normal decaying pattern observed in most
{\em Swift} GRBs (Zhang et al. 2006; Nousek et al. 2006;
O'Brien et al. 2006). A violation of such a prediction would
suggest an internal origin of the GeV afterglow.

\section{Conclusions and discussion}

\begin{figure}
  \includegraphics[angle=0,scale=1.0]{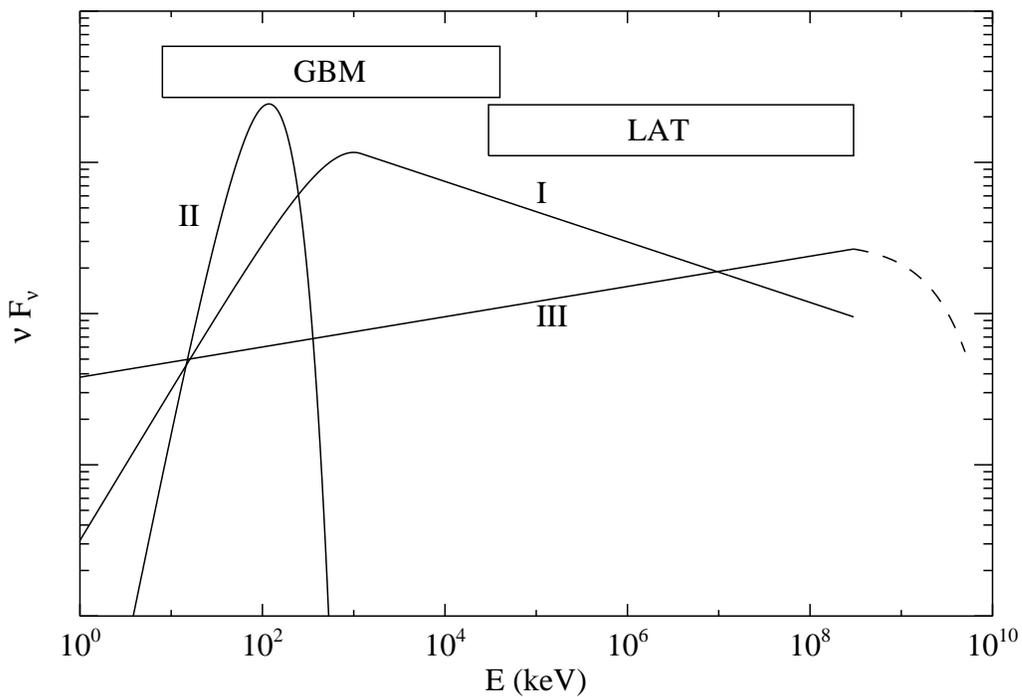}
\caption{A cartoon picture of three elemental spectral components
that shape GRB prompt emission spectra: (I) a Band-function
component that is likely of the non-thermal origin; (II) a
quasi-thermal component; and (III) an extra power-law component
that extends to high energy, which is expected to have a cut-off
near or above the high energy end of the LAT energy band.}
\label{Cartoon}
\end{figure}

\begin{figure}
 \includegraphics[angle=0,scale=.3]{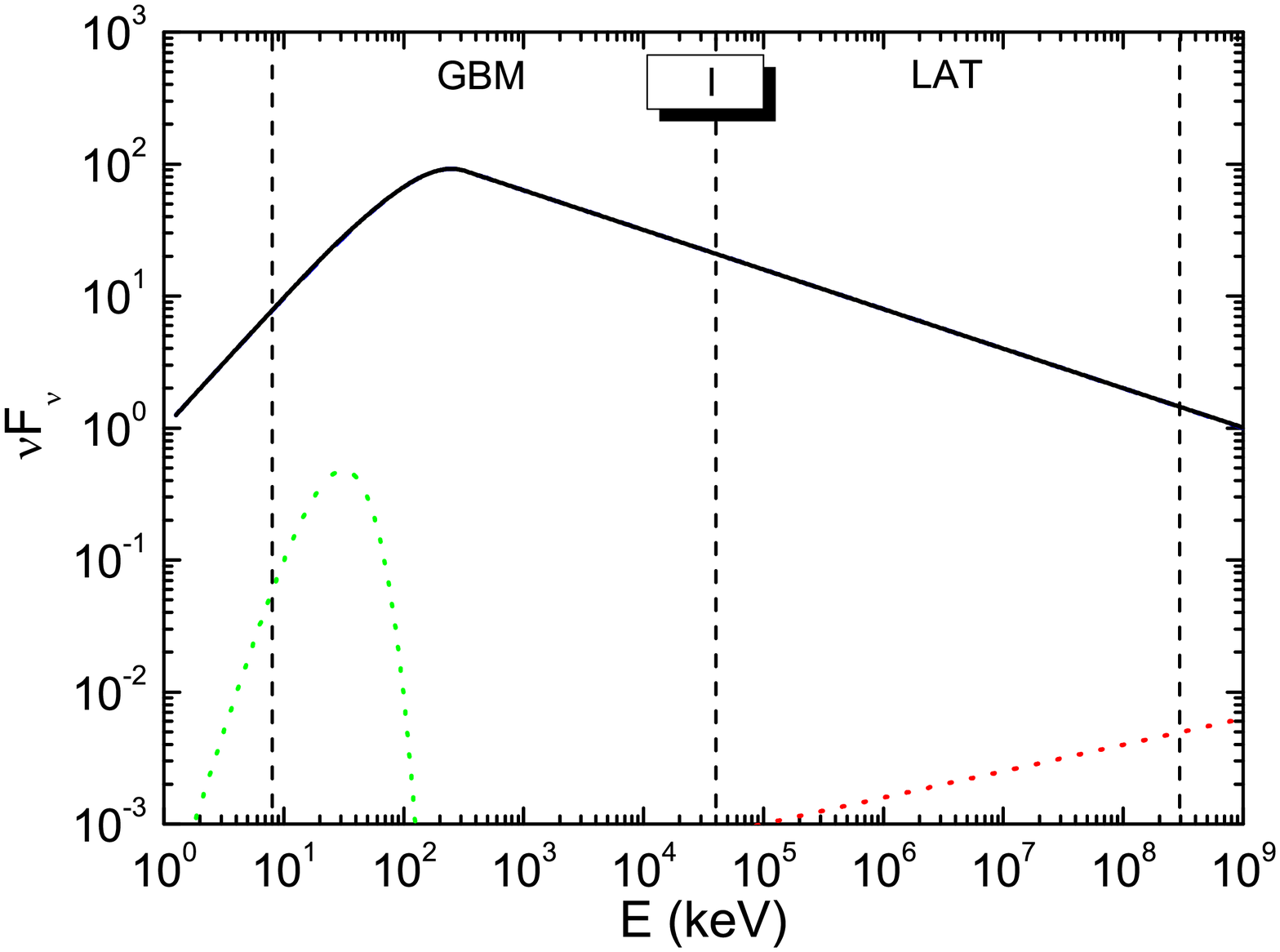}
  \includegraphics[angle=0,scale=.3]{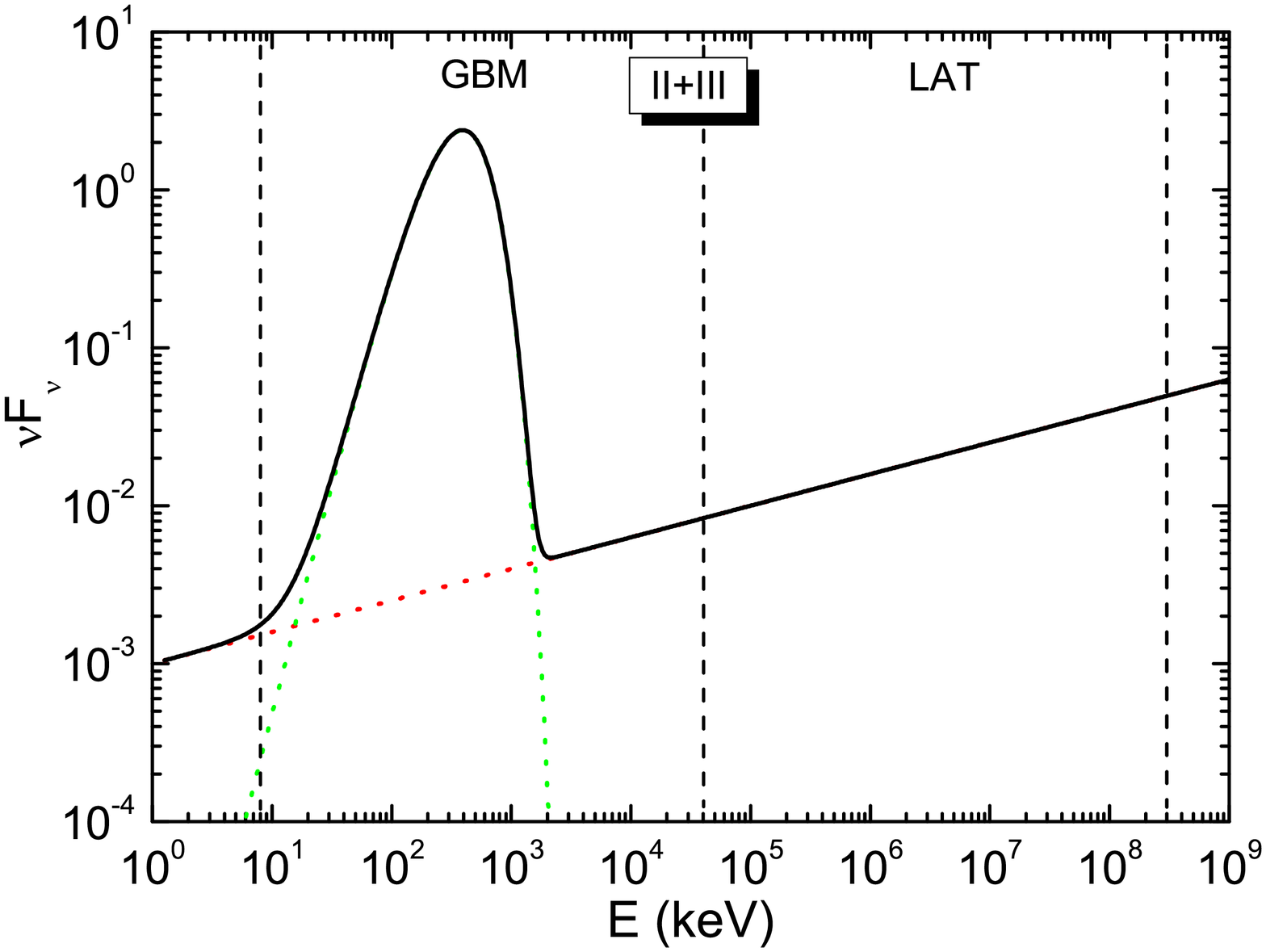}
  \includegraphics[angle=0,scale=.3]{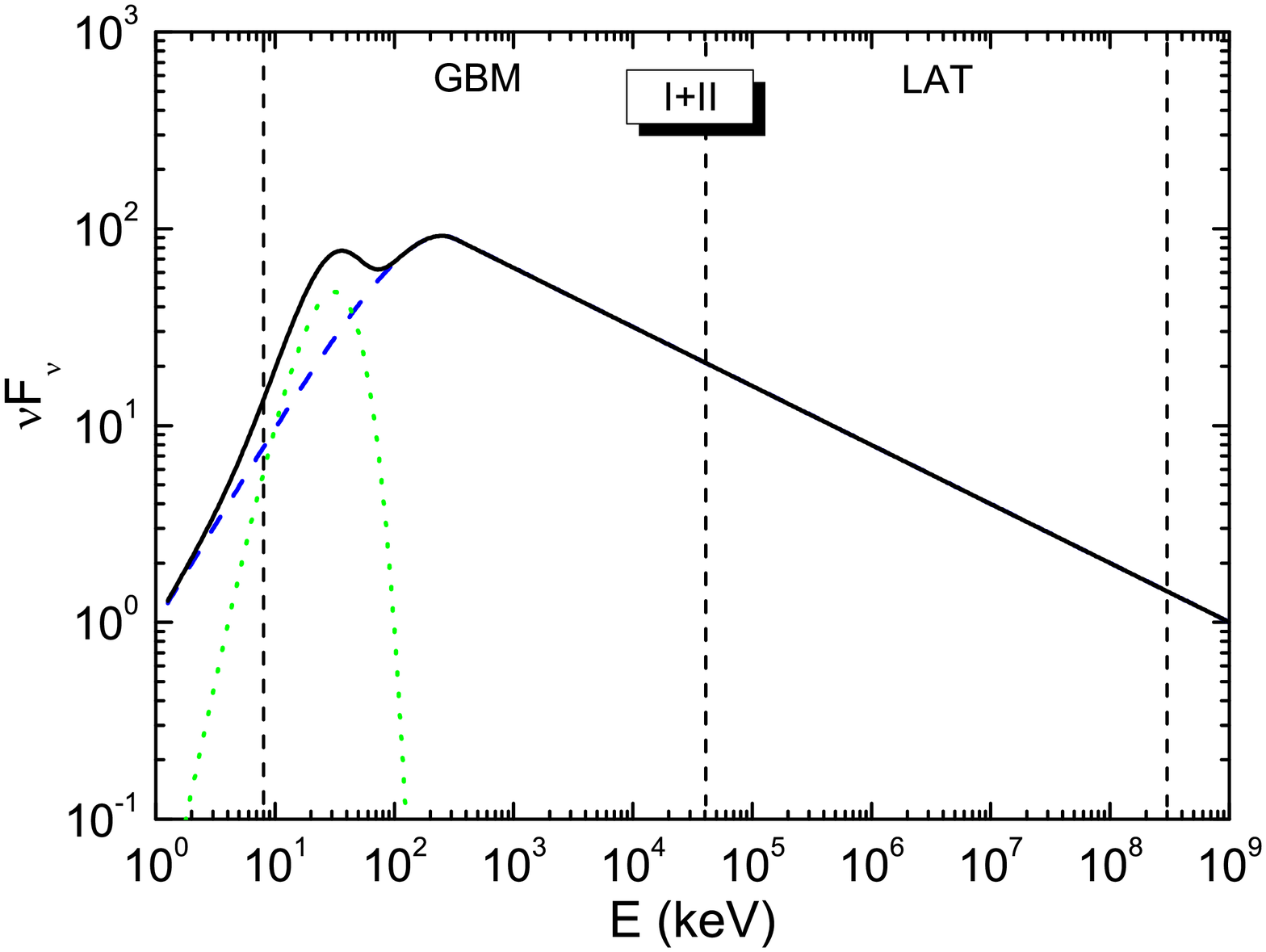}
  \includegraphics[angle=0,scale=.3]{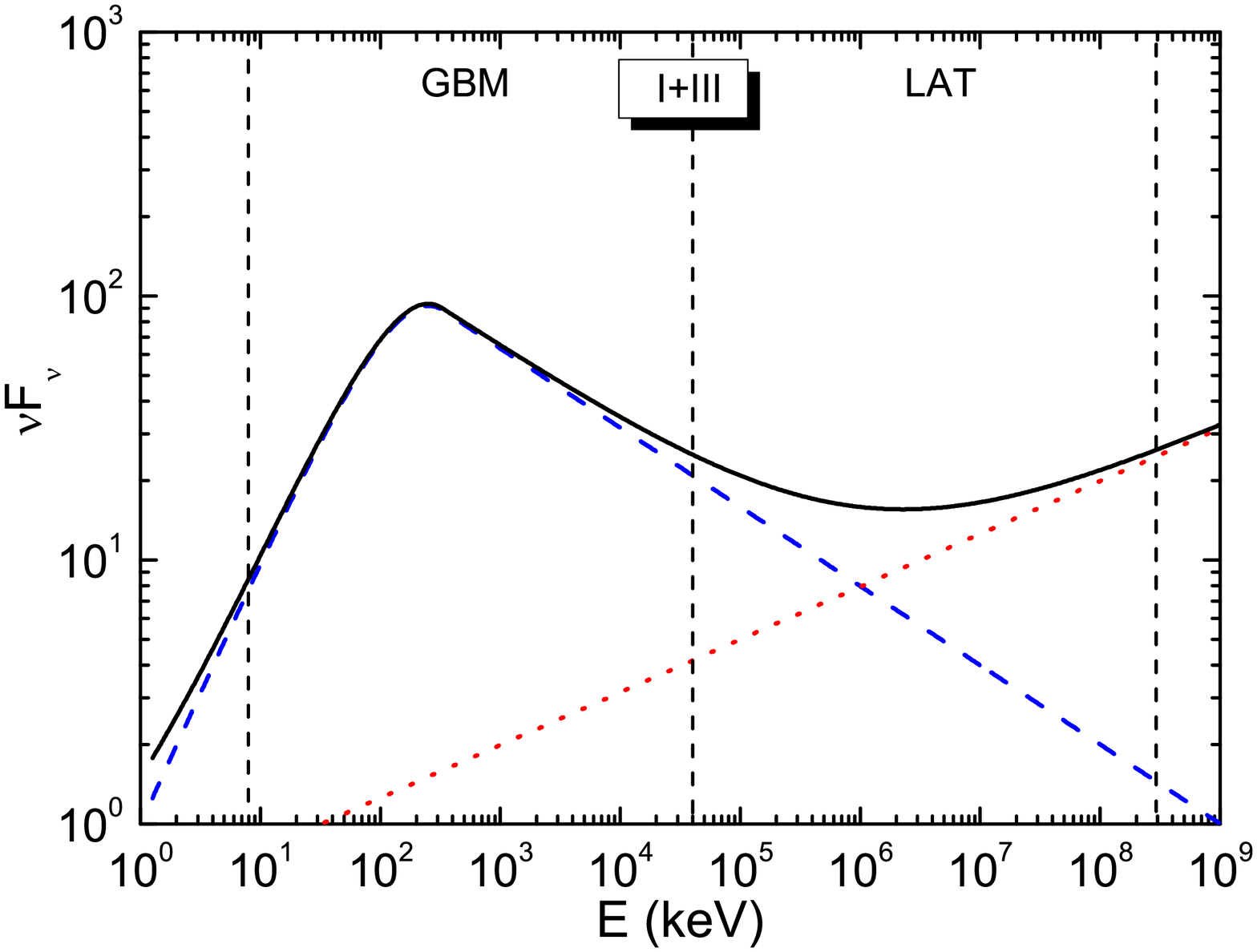}
  \includegraphics[angle=0,scale=.3]{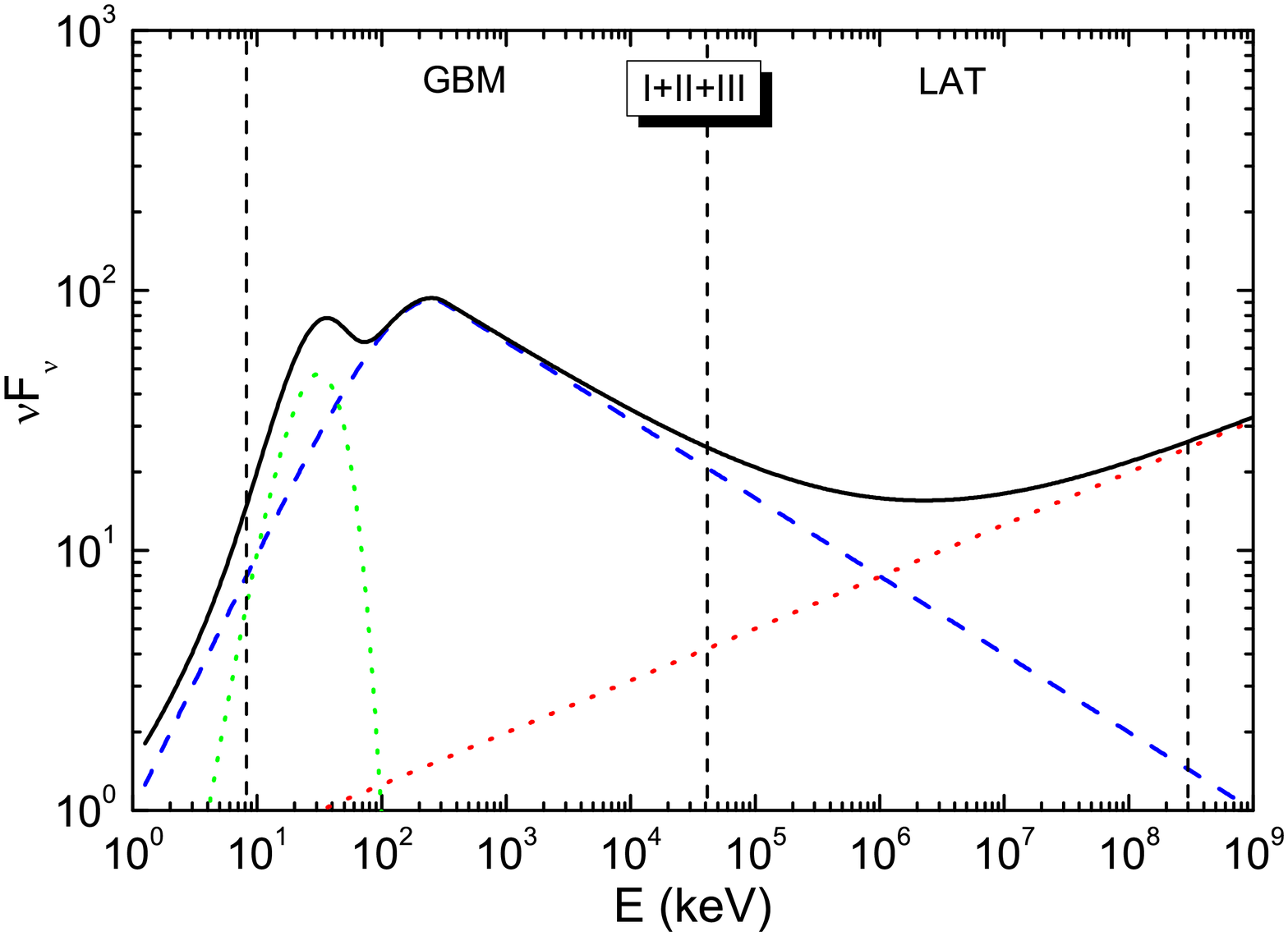}
\caption{Five possible spectral combinations with the three spectral components.}
\label{spectral-combinations}
\end{figure}

We have presented a comprehensive joint analysis of 17 GRBs 
co-detected by {\em Fermi} GBM and LAT. We carried out a time-resolved 
spectral analysis of all the bursts with the finest temporal resolution 
allowed by statistics, in order to { reduce} temporal smearing of different 
spectral components. Our data analysis results can be summarized as
the following:

\begin{itemize}
\item We found that the time-resolved spectra of 14 out of 17 GRBs 
are best modeled with the classical ``Band'' function over the entire 
{\em Fermi} spectral range, which may suggest a common origin for emissions 
detected by LAT and GBM. GRB 090902B and GRB 090510 are found to be
special in that the data require the superposition between
a MeV component and an extra power law component, and that
the MeV component has a sharp cutoff above $E_p$. More 
interestingly, the MeV component of GRB 090902B becomes 
progressively narrower as the time bin gets smaller, and can be 
fit with a Planck function as the time bin becomes small enough. 
This is in stark contrast to GRB 080916C, which shows no evidence
of ``narrowing'' with the reducing time bin. This suggests that
the Band-function component seen in GRB 080916C is 
physically different from the MeV component 
seen in GRB 090902B.
\item We tentatively propose that phenomenologically there can be
three elemental spectral components (Fig.\ref{Cartoon}), namely, 
(I): a Band-function component (Band) that extends to a 
wide spectral regime without ``narrowing'' with reduced time bins,
which is likely of non-thermal origin; 
(II): a quasi-thermal component (BB) that 
``narrows'' with reducing time bins and that can be reduced to
a blackbody (or multi-color blackbody) function; and (III):
a power-law component (PL) that has a positive slope in $\nu F_\nu$
space and extends to very high energy beyond the LAT energy band.
\item Component I (Band) is the most common spectral component,
which appears in 15 of 17 GRBs. Except GRB 090926A (which may
have Component III at late times), all these GRBs have a 
Band-only spectrum in the time-resolved spectral analysis.
\item Component II (BB) shows up in the time-resolved
spectral analysis of GRB 090902B and possibly also in GRB 090510.
The MeV component of these two GRBs can be fit with a power
law with exponential cutoff (CPL). Since data demand the 
superposition with an additional PL component (Component III), 
the uncertainty in the spectral index of the PL component 
makes it possible to have a range of low energy photon indices
for the CPL component. In particular, the MeV component of
GRB 090902B can be adjusted to be consistent with a blackbody
(Plank function).  This is not possible for GRB 090510, whose 
low energy photon index is softer. In any case, the MeV component 
of GRB 090510 may be consistent with a multi-color blackbody.

\item Component III (PL) shows up in both GRB 090902B and the 
short GRB 090510, and probably in the late epochs of GRB 090926A
as well. It has a positive slope in $\nu F_\nu$, which suggests 
that most energy in this component is released near or above the 
high energy end of the LAT energy band.

\item With the above three elemental emission components, one may
imagine 7 possible spectral combinations. Most ($\sim 80\%$) of 
GRBs in our sample have the Band-only spectra. GRB 090902B has the 
BB+PL spectra in the time resolved spectral analyses, and GRB 090510 
has a CPL + PL spectra. Both can be considered as the superposition 
between Components II and III. 
GRB 090926A may have the superposition between I and III at late
epochs. Other combinations are not identified yet
with the current analysis, but some combinations (e.g. I+II, 
I+II+III) may in principle exist.

\item LAT-band emission has a delayed onset with respect to GBM-band 
emission in some (but not all) GRBs and it usually lasts much longer.
In most cases (all except GRBs 090902B and 090510), however, the LAT
and GBM photons are consistent with belonging to the same spectral
component, suggesting a possible common origin. For bright bursts,
the LAT-band activities usually roughly track the GBM-band activities.
In the long-term, the LAT and GBM lightcurves sometimes (not always)
show different decaying behaviors. 
The LAT lightcurves continuously decay as a power-law up to
hundreds of seconds.

\item A statistical study of the spectral parameters in our sample
generally confirms the previously found correlations between $E_p$
and luminosity, both globally in the whole sample and individually
within each burst. We also discover preliminary rough correlations
between $\alpha$ and $\beta$ (negative correlation) and between 
flux and $\alpha$ (positive correlation). Both correlations need
confirmation from a larger sample.
\end{itemize}

From these results, we can draw the following physical implications
regarding the nature of GRBs.

The Band-only spectra are inconsistent with the simplest
fireball photosphere-internal-shock model. This is because if the
Band component is non-thermal emission from the internal shock,
the expected photosphere emission should be very bright. A natural
solution is to invoke a Poynting-flux-dominated flow. An alternative
possibility is to interpret the Band component as the photosphere
emission itself. However, the following results seem to disfavor
such a possibility. (1) In some cases (e.g. GRB 080916C), the
Band-only spectrum extends to energies as high as 10s of GeV;
(2) The low-energy photon indices in the time-resolved spectra
are typically $-1$, much softer than that expected in the
photosphere models; (3) There is no evidence that the Band
component is the temporal superposition of thermal-like emission 
components in the Band-only sample. We therefore suggest that 
GRB 080916C and probably all Band-only GRBs
may correspond to those GRBs whose jet composition is dominated
by a Poynting flux rather than a baryonic flux (Zhang \& Pe'er
2009; Zhang \& Yan 2011).

The existence of a bright photosphere component in GRB 
090902B (see also Ryde et al. 2010; Pe'er et al. 2010) suggests
that the composition of this GRB is likely a hot fireball without
strong magnetization. It is rare,
but its existence nonetheless suggests that GRB outflow composition
may be diverse. Its associated PL component is hard to interpret,
but it may be from the contributions of multiple non-thermal spectral
components (Pe'er et al. 2010). The case of GRB 090510 may be similar 
to GRB 090902B. The low-energy spectral index of the MeV component is
too shallow to be consistent with a blackbody, but the high-latitude
emission from an instantaneously ejected fireball (which is relevant
to short GRBs) would result in a multi-color blackbody due to the
angular superposition effect (Pe'er \& Ryde 2010).

The delayed onset of GeV emission may be simply due to one of the 
following two reasons: (1) The particle acceleration condition 
may be different throughout the burst. Initially, the electron 
spectral index may be steep initially (so that GeV emission is
too faint to be detected), but later it turns to a shallower value
so that GeV emission emerges above the detector sensitivity;
(2) Initially the ejecta may be more opaque so that there was
a pair-production spectral cutoff below the LAT band. This
cutoff energy later moves to higher energies to allow LAT photons
to be detected. Within this picture, the electron spectral index
is similar throughout the burst. There are other models discussed
in the literature to attribute GeV emission to a different origin
from the MeV component. This is reasonable for GRB 090510 and
GRB 090902B, but for most other GRBs this model is contrived
since the GeV emission appears as the natural extension of the
MeV Band-function to high energies.

The GeV emission during the prompt phase is very likely not of 
external forward shock origin. This is due to the following
facts: (1) In most GRBs the entire {\em Fermi}-band emission is well fit by
a single Band component. The GeV emission is consistent with being 
the extension of MeV to high energies. (2) During the prompt phase
and except for the delayed onset in some GRBs, the LAT-band
activities in bright GRBs generally track GBM-band activities.
The latter property is relevant even for GRB 090902B which shows
clearly two components in the spectra. (3) The peak of GeV lightcurve
coincides the second peak of GBM lightcurve for GRB 080916C.
A more reasonable possibility
is that the GeV emission during the prompt phase has an ``internal''
origin similar to its MeV counterpart.

The origin of the long lasting GeV afterglow after the prompt 
emission phase (end of the GBM-band emission) is unclear.
If it is from the external forward shock, one needs to introduce
abnormal shock parameters, and to argue for coincidence to
connect with the internal-origin early GeV emission to form a
simple PL decay lightcurve. Alternatively, the long lasting GeV
emission can be also of the internal origin. Future joint
{\em Fermi}/{\em Swift} observations of the early GeV/X-ray
afterglows of some bright GRBs will help to differentiate between
these possibilities.

The two tentative correlations ($\alpha-\beta$ and $\alpha$-flux)
proposed in this paper need to be confirmed with a larger data
sample, and their physical implications will be discussed then.

\begin{acknowledgements}
We thank Rob Preece for important instructions on {\em Fermi} data analysis.
This work is partially supported by NASA NNX09AT66G, NNX10AD48G, and
NSF AST-0908362 at UNLV. EWL, YZF, and XFW acknowledge National Basic
Research Program of China (973 Program 2009CB824800).
This work is partially supported by the National Natural Science Foundation of
China (grant 10873002 for EWL, and grants 10633040, 10921063 for XFW).
EWL is also supported by Guangxi Ten-Hundred-Thousand project (Grant 2007201),
Guangxi Science Foundation (2010GXNSFC013011), and the program for
100 Young and Middle-aged Disciplinary Leaders in Guangxi Higher
Education Institutions. YZF is also supported by a special grant from Purple Mountain Observatory and by the National Nature Science Foundation of China (grant 11073057). XFW is also supported by the Special Foundation for the
Authors of National Excellent Doctorial Dissertations of P. R. China by
Chinese Academy of Sciences. AP is supported by the Riccardo Giacconi
Fellowship award of the Space Telescope Science Institute.
\end{acknowledgements}



\begin{deluxetable}{llllllllllll}
\tabletypesize{\scriptsize}
\tablecaption{The GRBs co-detected by {\em Fermi} LAT and GBM since
{\em Fermi} science operation and until May, 2010}
\setlength{\tabcolsep}{0.035in}
 \startdata
\hline
\hline\noalign{\smallskip}
GRB  &  $z$   &  dur. [sec]  & $E_p$ [keV] &   $E_{\gamma,\rm iso}$ [erg] & Fluence ($1-10^4$ keV) & Spectral Type & Onset Delay& $E_{\rm max}$ \\

080825C &  - & $22$ & $192\pm15$ & - & $4.84_{-0.57}^{+0.59}\times 10^{-5}$ & BAND& Y &
$\sim 600$ MeV \\
080916C &  4.35 & $66$ & $1443_{-303}^{+433}$ & $5.7_{-0.41}^{+0.54}\times 10^{54} $
& $1.55_{-0.11}^{+0.15}\times 10^{-4}$ &  BAND & Y & $\sim 13.2$ GeV\\  
081024B &  - & $0.8$ & $1258_{-522}^{+2405}$ & -&  $(1,61\pm 3.8) \times 10^{-6}$ & BAND&Y
& $\sim 3$ GeV\\ 
081215A &  - & $7.7$ & $1014_{-123}^{+140}$ & - & $8.74_{-0.99}^{+1.21}\times
10^{-5}$ & BAND& - & -  \\  
090217 &  - & $32.8$ & $552_{-71}^{+85}$ & - & $4.48_{-0.56}^{+0.69}\times 10^{-5}$
& BAND& N & $\sim 1$ GeV\\ 
090323 &  3.57 & $150$ & $812_{-143}^{+181}$ & $>2.89_{-0.69}^{+6.56}\times 10^{54}$
& $>1.07_{-0.26}^{+0.24}\times 10^{-5}$ & BAND& N & $\sim 1$ GeV\\
090328 &  0.736 & $80$ & $756_{-72}^{+85}$ &  $1.02_{-0.083}^{+0.087}\times 10^{53}$
& $7.14_{-0.58}^{+0.61}\times 10^{-5}$ &  BAND& ? & $>100$ MeV\\
090510 &  0.903 & $0.3$ & $6010_{-1690}^{+2524}$ &  $4.47_{-3.77}^{+4.06}\times
10^{52}$ & $2.06_{-1.74}^{+1.88} \times 10^{-5}$ &  CPL+PL& Y& $\sim$ 31 GeV   \\
090626 &  -  & $70$ & $362_{-41}^{+47}$ & - &  $7.81_{-0.38}^{+0.44}\times 10^{-5}$
& BAND& ? & $\sim 30$ GeV \\
090902B &  1.822  & $21$ & $ 207\pm6$ [BB] & $(1.77\pm 0.01)\times
10^{52}$ & $(2.10\pm 0.02) \pm 10^{-4}$ &  BB+PL& Y & $33.4_{-3.5}^{+2.7}$ GeV\\
090926A & 2.1062 & $\sim 20$ & $ 412\pm 20$&  $2.10_{-0.08}^{+0.09}\times 10^{54}$ &
 $1.93_{-0.07}^{+0.08}\times 10^{-4}$  & BAND& Y & $\sim $20 GeV \\
091003 & 0.8969 & $21.1$ & $ 409_{-31}^{+34}$  &  $7.85_{-0.57}^{+0.73}\times
10^{52}$  &  $3.68_{-0.27}^{+0.34}\times 10^{-5}$& BAND& N &  $>150$ MeV \\
091031 & - & $\sim 40$ & $ 567_{-135}^{+197}$ & - &   $3.17_{-0.51}^{+0.64}\times
10^{-5}$ &  BAND&N  & $1.2$ GeV \\
100116A & - & $\sim 110$ & $ 1463_{-122}^{+163}$ & - &   $7.34_{-1.26}^{+1.42}\times
10^{-5}$  & BAND& N& $\sim 2.2 $ GeV \\
100225A & - & $13\pm 3 $ & $ 540_{-204}^{+381}$ & -  &   $1.21_{-0.57}^{+1.07}\times
10^{-5}$  &  BAND&Y& $\sim 300$ MeV \\
100325A & - & $8.3\pm 1.9 $ & $ 198_{-37}^{+44}$ & -&   $6.15_{-1.81}^{+2.85}\times
10^{-6}$  &  BAND&N& $\sim 800$ MeV \\
100414A & 1.368 & $26.4\pm 1.6 $ & $ 520_{-39}^{+42}$ & $5.88_{-0.65}^{+0.69}\times 10^{53}$ &   $1.20_{-0.10}^{+0.12}\times
10^{-5}$ &  BAND&N & $\sim 2.6$ GeV \\

\enddata
\tablecomments{References:
(1) GRB080825C: $z$, $T_{90}$ -- \cite{2008GCN..8141....1V};
(2) GRB090916C: $z$ -- \cite{2009A&A...498...89G}, $T_{90}$ -- \cite{2008GCN..8245....1G}, $E_{\gamma,\rm iso}$ -- \cite{2009Sci...323.1688A};
(3) GRB081024B: $T_{90}$ -- \cite{2010ApJ...712..558A};
(4) GRB081215A: $T_{90}$ -- \cite{2008GCN..8678....1P};
(5) GRB090328: $T_{90}$ -- \cite{2009GCN..9050....1G};
(6) GRB090510: $z$ -- \cite{2010A&A...516A..71M};
(7) GRB090902B: $z$ -- \cite{2009GCN..9873....1C};
(8) GRB090926A: $z$ -- \cite{2009GCN..9942....1M};
(9) GRB091003:
$z$ -- \cite{2009GCN.10031....1C};
(10) GRB100414A: $z$ -- \cite{2010GCN.10606....1C}}
\end{deluxetable}
\clearpage

\begin{deluxetable}{llllllllll}
\tabletypesize{\tiny}
\tablecaption{Time-resolved and time-integrated 
spectral fitting parameters of 17 {\em Fermi}/LAT GRBs.}
\setlength{\tabcolsep}{0.075in}

 \startdata
\hline
\hline

\ \\
\multicolumn{7}{c}{080825C Model : Band Function}\tabularnewline
\ \\
\hline\noalign{\smallskip}
 Seq &  Time &    $\alpha$ &   $\beta$ &   $E_{0}$  &   $K$  &  $\chi^2$ & $dof$ \\
 & s  & & & keV & $\frac{photons}{keV cm^{2}s}@100 keV$& & \\
\hline
$1$&0.00-6.75& $ -0.57_{-0.04}^{+0.05}$ &  $ -2.29\pm0.04$          &  $ 135_{-9}^{+10}$  &  $ 0.114_{-0.007}^{+0.008}$ &  $147.1$& $154$\\
$2$&6.75-18.1& $ -0.75\pm0.06$          &  $ -2.35_{-0.07}^{+0.09}$ &  $ 141_{-14}^{+16}$ &  $ 0.051_{-0.004}^{+0.005}$ &  $132.7$& $154$\\
$3$&18.1-25.0& $ -0.95_{-0.15}^{+0.17}$ &  $ -2.17_{-0.08}^{+0.17}$ &  $ 131_{-35}^{+56}$ &  $ 0.027_{-0.006}^{+0.009}$ &  $120.1$& $154$\\
Total&0.00-25.0& $ -0.73\pm0.03$          &  $ -2.33_{-0.03}^{+0.04}$ &  $ 148\pm9$         &  $ 0.058_{-0.003}^{+0.003}$ &  $265.6$& $154$\\
\hline
\ \\
\multicolumn{7}{c}{080916C Model : Band Function}\tabularnewline
\ \\
\hline\noalign{\smallskip}
 Seq &  Time &    $\alpha$ &   $\beta$ &   $E_{0}$  &   $K$  &  $\chi^2$ & $dof$ \\
 & s  & & & keV & $\frac{photons}{keV cm^{2}s}@100 keV$& & \\

\hline
$1$&0.00-3.70& $ -0.69_{-0.04}^{+0.05}$ &  $ -2.49_{-0.08}^{+0.13}$ &  $ 342_{-37}^{+43}$    &  $ 0.047_{-0.002}^{+0.003}$ &  $99.5$&  $124$\\
$2$&3.70-9.10& $ -1.14\pm0.03$          &  $ -2.32_{-0.05}^{+0.06}$ &  $ 1680_{-348}^{+500}$ &  $ 0.027\pm0.001$           &  $153.0$& $124$\\
$3$&9.10-17.0& $ -1.15_{-0.04}^{+0.05}$ &  $ -2.29_{-0.05}^{+0.07}$ &  $ 975_{-235}^{+361}$  &  $ 0.016\pm0.001$           &  $125.9$& $124$\\
$4$&17.0-25.0& $ -0.99\pm0.04$          &  $ -2.27_{-0.04}^{+0.06}$ &  $ 447_{-60}^{+75}$    &  $ 0.024\pm0.001$           &  $114.3$& $124$\\
$5$&25.0-41.0& $ -1.08\pm0.03$          &  $ -2.49_{-0.07}^{+0.10}$ &  $ 666_{-87}^{+111}$   &  $ 0.017\pm0.001$           &  $124.2$& $124$\\
$6$&41.0-66.0& $ -1.09\pm0.04$          &  $ -2.36_{-0.05}^{+0.06}$ &  $ 696_{-128}^{+186}$  &  $ 0.010\pm0.001$           &  $162.8$& $124$\\
Total&0.00-66.0& $ -1.05\pm0.02$          &  $ -2.30\pm0.02$          &  $ 664_{-46}^{+51}$    &  $ 0.018\pm0.001$           &  $427.5$& $124$\\
\hline
\ \\
\multicolumn{7}{c}{081024B Model : Band Function}\tabularnewline
\ \\
\hline\noalign{\smallskip}
 Seq &  Time &    $\alpha$ &   $\beta$ &   $E_{0}$  &   $K$  &  $\chi^2$ & $dof$ \\
 & s  & & & keV & $\frac{photons}{keV cm^{2}s}@100 keV$& & \\
\hline
$1$&-0.300-0.800& $ -1.15_{-0.16}^{+0.14}$ &  $ -2.20(fixed)$ &  $ 1478_{-551}^{+2810}$ &  $ 0.007\pm0.001$ &  $353.9$& $208$\\
\hline
\ \\
\multicolumn{7}{c}{081215A Model : Band Function}\tabularnewline
\ \\
\hline\noalign{\smallskip}
 Seq &  Time &    $\alpha$ &   $\beta$ &   $E_{0}$  &   $K$  &  $\chi^2$ & $dof$ \\
 & s  & & & keV & $\frac{photons}{keV cm^{2}s}@100 keV$& & \\
\hline
$1$&0.00-1.50& $ -0.65\pm0.05$          &  $ -2.27_{-0.11}^{+0.14}$ &  $ 753_{-88}^{+101}$ &  $ 0.059\pm0.002$           &  $80.0$ & $71$\\
$2$&1.50-2.28& $ -0.52_{-0.07}^{+0.08}$ &  $ -2.16_{-0.08}^{+0.10}$ &  $ 280_{-39}^{+43}$  &  $ 0.223_{-0.017}^{+0.020}$ &  $63.6$ & $61$\\
$3$&2.28-4.93& $ -0.60\pm0.06$          &  $ -2.34_{-0.08}^{+0.09}$ &  $ 178_{-17}^{+20}$  &  $ 0.156_{-0.012}^{+0.013}$ &  $66.1$ & $77$\\
$4$&4.93-5.59& $ -0.49_{-0.08}^{+0.09}$ &  $ -2.29_{-0.11}^{+0.15}$ &  $ 214_{-31}^{+36}$  &  $ 0.266_{-0.026}^{+0.032}$ &  $45.0$ & $54$\\
$5$&5.59-8.00& $ -0.72_{-0.14}^{+0.16}$ &  $ -2.19_{-0.10}^{+0.13}$ &  $ 102_{-22}^{+28}$  &  $ 0.093_{-0.019}^{+0.029}$ &  $47.5$ & $82$\\
Total&0.00-8.00& $ -0.71\pm0.03$          &  $ -2.16_{-0.03}^{+0.04}$ &  $ 289_{-21}^{+22}$  &  $ 0.110_{-0.004}^{+0.005}$ &  $179.9$& $86$\\
\hline
\tablebreak
\hline
\ \\
\multicolumn{7}{c}{090217 Model : Band Function}\tabularnewline
\ \\
\hline\noalign{\smallskip}
 Seq &  Time &    $\alpha$ &   $\beta$ &   $E_{0}$  &   $K$  &  $\chi^2$ & $dof$ \\
 & s  & & & keV & $\frac{photons}{keV cm^{2}s}@100 keV$& & \\
\hline
$1$&0.00-7.50& $ -0.59\pm0.04$          &  $ -2.56_{-0.07}^{+0.10}$ &  $ 365_{-30}^{+33}$ &  $ 0.027\pm0.001$           &  $165.1$& $156$\\
$2$&7.50-13.1& $ -0.83\pm0.05$          &  $ -2.66_{-0.14}^{+0.37}$ &  $ 470_{-58}^{+70}$ &  $ 0.021\pm0.001$           &  $135.5$& $156$\\
$3$&13.1-19.7& $ -0.96\pm0.09$          &  $ -2.38_{-0.10}^{+0.22}$ &  $ 257_{-51}^{+73}$ &  $ 0.015\pm0.002$           &  $131.1$& $156$\\
$4$&19.7-30.0& $ -0.52_{-0.25}^{+0.43}$ &  $ -2.22_{-0.09}^{+0.17}$ &  $ 118_{-52}^{+65}$ &  $ 0.008_{-0.003}^{+0.009}$ &  $175.4$& $156$\\
Total&0.00-30.0& $ -0.81\pm0.03$          &  $ -2.54_{-0.04}^{+0.06}$ &  $ 418_{-30}^{+33}$ &  $ 0.015\pm0.001$           &  $371.6$& $156$\\
\hline
\ \\
\multicolumn{7}{c}{090323 Model : Band Function}\tabularnewline
\ \\
\hline\noalign{\smallskip}
 Seq &  Time &    $\alpha$ &   $\beta$ &   $E_{0}$  &   $K$  &  $\chi^2$ & $dof$ \\
 & s  & & & keV & $\frac{photons}{keV cm^{2}s}@100 keV$& & \\
\hline
$1$&5.00-14.0& $ -0.97_{-0.04}^{+0.05}$ &  $ -2.58_{-0.13}^{+0.25}$ &  $ 792_{-136}^{+172}$ &  $ 0.016\pm0.001$           &  $98.4$ & $125$\\
$2$&14.0-25.0& $ -1.11\pm0.04$          &  $ -2.54_{-0.10}^{+0.18}$ &  $ 826_{-141}^{+198}$ &  $ 0.017\pm0.001$           &  $127.2$& $125$\\
$3$&35.0-50.0& $ -1.08\pm0.03$          &  $ -2.64_{-0.15}^{+0.39}$ &  $ 557_{-69}^{+84}$   &  $ 0.018\pm0.001$           &  $151.5$& $125$\\
$4$&50.0-60.0& $ -0.88\pm0.04$          &  $ -2.81_{-0.24}^{+1.13}$ &  $ 449_{-44}^{+52}$   &  $ 0.026\pm0.001$           &  $115.2$& $125$\\
$5$&60.0-135.& $ -1.31_{-0.01}^{+0.02}$ &  $ -2.62_{-0.07}^{+0.11}$ &  $ 987_{-116}^{+694}$ &  $ 0.010\pm0.001$           &  $496.7$& $125$\\
$6$&135.-145.& $ -1.30\pm0.06$          &  $ -2.34_{-0.12}^{+0.32}$ &  $ 294_{-57}^{+74}$   &  $ 0.017_{-0.001}^{+0.002}$ &  $208.3$& $125$\\
Total&0.00-150.& $ -1.22\pm0.01$          &  $ -2.68_{-0.04}^{+0.06}$ &  $ 880_{-50}^{+64}$   &  $ 0.012\pm0.001$           &  $857.3$& $125$\\
\hline
\ \\
\multicolumn{7}{c}{090328 Model : Band Function}\tabularnewline
\ \\
\hline\noalign{\smallskip}
 Seq &  Time &    $\alpha$ &   $\beta$ &   $E_{0}$  &   $K$  &  $\chi^2$ & $dof$ \\
 & s  & & & keV & $\frac{photons}{keV cm^{2}s}@100 keV$& & \\
\hline
$1$&3.00-8.00& $ -0.92_{-0.03}^{+0.04}$ &  $ -2.38_{-0.10}^{+0.16}$ &  $ 662_{-86}^{+99}$ &  $ 0.024\pm0.001$        &  $188.0$& $217$\\
$2$&12.0-20.0& $ -0.96\pm0.02$          &  $ -2.38_{-0.06}^{+0.09}$ &  $ 727_{-67}^{+80}$ &  $ 0.024\pm0.001$        &  $199.3$& $217$\\
$3$&20.0-30.0& $ -1.15\pm0.03$          &  $ -2.30_{-0.07}^{+0.09}$ &  $ 616_{-69}^{+81}$ &  $ 0.020\pm0.001$        &  $250.7$& $217$\\
Total&0.00-30.0& $ -1.05\pm0.01$          &  $ -2.44_{-0.04}^{+0.05}$ &  $ 791_{-50}^{+58}$ &  $ 0.018\pm0.001$        &  $472.5$& $217$\\
\hline
\ \\
\multicolumn{7}{c}{090510 Model : Cut-off Power-Law+Power Law}\tabularnewline
\ \\
\hline\noalign{\smallskip}
 Seq &  Time  &    $\Gamma_{\rm CPL}$ &   $E_0$  &   $K_{\rm CPL}$  &   $\Gamma_{\rm PL}$ &   $K_{\rm PL}$  &  $\chi^2$ & $dof$ \\
& s &  & keV & $\frac{photons}{keV cm^{2}s}@1 keV$  & & $\frac{photons}{keV cm^{2}s}@1 keV$  & & \\
\hline
$1$&0.450-0.600& $ -0.76\pm0.08$          &  $ 2688_{-765}^{+1360}$  &  $ 1.85_{-0.63}^{+0.85}$ &  $ ---$ &  $ ---$ &  $83.7$& $230$\\
$2$&0.600-0.800& $ -0.60_{-0.13}^{+0.14}$ &  $ 4286_{-1130}^{+1760}$ &  $ 0.47_{-0.26}^{+0.53}$ &  $ -1.73_{-0.07}^{+0.06}$ &  $ 23.2_{-12.3}^{+13.0}$ &  $154.9$& $251$\\
$3$&0.800-0.900& $ -0.75_{-0.31}^{+0.67}$ &  $ 777_{-464}^{+1900}$   &  $ 0.97_{-0.93}^{+3.41}$ &  $ -1.60_{-0.07}^{+0.11}$ &  $ 14.3_{-11.6}^{+17.9}$ &  $52.0$& $178$\\
$4$&0.900-1.00& $ ---$ &  $ ---$ &  $ ---$ &  $ -1.62\pm0.06$    &  $ 11.5_{-5.8}^{+7.4}$ &  $38.0$& $134$\\
Total&0.450-1.00& $ -0.76_{-0.07}^{+0.08}$ &  $ 3624_{-612}^{+759}$ &  $ 1.06_{-0.39}^{+0.54}$ &  $ -1.66_{-0.03}^{+0.05}$ &  $ 11.9_{-5.6}^{+6.2}$ &  $215.0$& $272$\\
\hline
\ \\
\multicolumn{7}{c}{090626 Model : Band Function}\tabularnewline
\ \\
\hline\noalign{\smallskip}
 Seq &  Time &    $\alpha$ &   $\beta$ &   $E_{0}$  &   $K$  &  $\chi^2$ & $dof$ \\
 & s  & & & keV & $\frac{photons}{keV cm^{2}s}@100 keV$& & \\
\hline
$1$&0.00-9.00& $ -0.99_{-0.02}^{+0.03}$ &  $ -2.47_{-0.03}^{+0.04}$ &  $ 193_{-11}^{+12}$ &  $ 0.079\pm0.003$ &  $340.3$& $186$\\
$2$&15.0-20.0& $ -1.42\pm0.03$          &  $ -2.47_{-0.08}^{+0.13}$ &  $ 391_{-50}^{+60}$ &  $ 0.040\pm0.002$ &  $155.6$& $186$\\
$3$&20.0-27.0& $ -1.28_{-0.02}^{+0.03}$ &  $ -2.58_{-0.08}^{+0.13}$ &  $ 504_{-54}^{+63}$ &  $ 0.034\pm0.001$ &  $136.5$& $186$\\
$4$&30.0-40.0& $ -1.30\pm0.03$          &  $ -2.49_{-0.06}^{+0.10}$ &  $ 444_{-50}^{+63}$ &  $ 0.025\pm0.001$ &  $211.7$& $186$\\
Total&0.00-60.0& $ -1.40\pm0.01$          &  $ -2.62_{-0.03}^{+0.04}$ &  $ 482_{-25}^{+27}$ &  $ 0.025\pm0.001$ &  $743.3$& $186$\\
\hline
\ \\
\multicolumn{7}{c}{090902B Model : Black Body+Power Law}\tabularnewline
\ \\
\hline\noalign{\smallskip}
 Seq &  Time   &    $kT$ (keV) &   $K_{\rm BB}$  &   $\Gamma_{\rm PL}$ &   $K_{\rm PL}$  &  $\chi^2$ & $dof$ \\
& s &  keV &  $\frac{L_{39}}{D_{10}^2}$& &$\frac{photons}{keV cm^{2}s}@1 keV$ \\
\hline
$1$&0.00-1.50 & $ 75.60_{-1.79}^{+1.86}$  & $ 38.84_{-1.03}^{+1.02}$   &  $ -1.88\pm0.02$          &  $ 43.0_{-3.8}^{+3.9}$    &  $330.6$& $264$\\
$2$&1.50-2.25 & $ 98.74_{-3.41}^{+3.57}$  & $ 57.13_{-2.19}^{+2.25}$   &  $ -1.84_{-0.04}^{+0.03}$ &  $ 31.1_{-4.3}^{+5.3}$    &  $226.3$& $237$\\
$3$&2.25-2.81 & $ 121.20_{-4.79}^{+5.00}$ & $ 84.54_{-3.72}^{+3.79}$   &  $ -1.81_{-0.04}^{+0.03}$ &  $ 27.5_{-4.3}^{+4.6}$    &  $217.5$& $238$\\
$4$&2.81-3.23 & $ 82.52_{-3.97}^{+4.32}$  & $ 58.00_{-2.88}^{+3.05}$   &  $ -1.80_{-0.04}^{+0.03}$ &  $ 33.6_{-5.3}^{+6.4}$    &  $199.0$& $217$\\
$5$&3.23-3.83 & $ 100.90_{-3.57}^{+3.76}$ & $ 69.22_{-2.71}^{+2.81}$   &  $ -1.83_{-0.04}^{+0.03}$ &  $ 34.7_{-4.8}^{+6.0}$    &  $190.7$& $240$\\
$6$&3.83-4.46 & $ 86.81_{-2.79}^{+2.92}$  & $ 60.01_{-2.14}^{+2.20}$   &  $ -1.83_{-0.04}^{+0.03}$ &  $ 33.4_{-4.7}^{+5.7}$    &  $218.3$& $236$\\
$7$&4.46-4.99 & $ 90.79_{-4.43}^{+4.78}$  & $ 47.82_{-2.52}^{+2.65}$   &  $ -1.83_{-0.04}^{+0.03}$ &  $ 38.6_{-5.2}^{+6.4}$    &  $207.4$& $225$\\
$8$&4.99-5.45 & $ 109.50_{-4.11}^{+4.32}$ & $ 88.50_{-3.68}^{+3.82}$   &  $ -1.82_{-0.05}^{+0.04}$ &  $ 31.5_{-5.2}^{+6.6}$    &  $185.3$& $228$\\
$9$&5.45-5.86 & $ 116.20_{-4.94}^{+5.20}$ & $ 85.70_{-4.13}^{+4.22}$   &  $ -1.82_{-0.05}^{+0.04}$ &  $ 34.6_{-5.9}^{+7.2}$    &  $180.5$& $227$\\
$10$&5.86-6.28& $ 132.60_{-4.21}^{+4.36}$ & $ 141.20_{-5.14}^{+5.27}$  &  $ -1.81_{-0.05}^{+0.04}$ &  $ 32.5_{-5.3}^{+6.5}$    &  $186.5$& $233$\\
$11$&6.28-6.61& $ 157.40_{-6.50}^{+6.74}$ & $ 155.60_{-7.36}^{+7.77}$  &  $ -1.81_{-0.06}^{+0.04}$ &  $ 38.0_{-6.0}^{+8.6}$    &  $186.2$& $228$\\
$12$&6.61-7.19& $ 171.10_{-4.85}^{+5.01}$ & $ 174.10_{-5.80}^{+5.97}$  &  $ -1.86_{-0.03}^{+0.02}$ &  $ 87.2_{-7.3}^{+8.6}$    &  $229.0$& $248$\\
$13$&7.19-7.65& $ 174.20_{-5.35}^{+5.55}$ & $ 207.90_{-7.37}^{+7.57}$  &  $ -1.87_{-0.03}^{+0.02}$ &  $ 124.3_{-10.3}^{+12.1}$ &  $231.3$& $244$\\
$14$&7.65-8.00& $ 217.80_{-7.29}^{+7.47}$ & $ 307.00_{-12.20}^{+12.50}$&  $ -1.87\pm0.02$          &  $ 203.5_{-13.2}^{+15.0}$ &  $223.0$& $243$\\
$15$&8.00-8.50& $ 204.80_{-5.48}^{+5.62}$ & $ 288.60_{-9.01}^{+9.22}$  &  $ -1.91\pm0.01$          &  $ 344.6_{-15.7}^{+17.3}$ &  $319.9$& $248$\\
$16$&8.50-9.00& $ 206.60_{-5.83}^{+5.97}$ & $ 281.00_{-9.16}^{+9.35}$  &  $ -1.93_{-0.02}^{+0.01}$ &  $ 375.7_{-19.3}^{+21.5}$ &  $260.2$& $249$\\
$17$&9.00-9.50& $ 206.20_{-5.83}^{+5.99}$ & $ 270.50_{-8.91}^{+9.11}$  &  $ -1.92\pm0.01$          &  $ 445.6_{-18.9}^{+20.5}$ &  $325.6$& $248$\\
$18$&9.50-10.0& $ 135.90_{-3.18}^{+3.26}$ & $ 209.90_{-5.45}^{+5.53}$  &  $ -1.96_{-0.02}^{+0.01}$ &  $ 553.2_{-26.0}^{+28.6}$ &  $271.2$& $244$\\
$19$&10.0-10.5& $ 168.80_{-4.47}^{+4.58}$ & $ 236.40_{-7.04}^{+7.18}$  &  $ -1.94\pm0.02$          &  $ 378.4_{-20.9}^{+23.8}$ &  $258.3$& $244$\\
$20$&10.5-11.0& $ 195.70_{-5.89}^{+6.03}$ & $ 246.60_{-8.50}^{+8.70}$  &  $ -1.90\pm0.01$          &  $ 352.5_{-16.0}^{+17.7}$ &  $348.6$& $247$\\
$21$&11.0-11.5& $ 145.20_{-4.34}^{+4.50}$ & $ 179.10_{-5.81}^{+5.98}$  &  $ -1.93\pm0.02$          &  $ 332.2_{-18.3}^{+20.8}$ &  $278.5$& $242$\\
$22$&11.5-12.0& $ 153.10_{-4.32}^{+4.43}$ & $ 169.30_{-5.56}^{+5.68}$  &  $ -1.92\pm0.02$          &  $ 253.5_{-16.2}^{+18.8}$ &  $241.9$& $241$\\
$23$&12.0-12.4& $ 61.07_{-2.90}^{+3.09}$  & $ 44.61_{-2.24}^{+2.31}$   &  $ -1.90\pm0.02$          &  $ 242.6_{-15.9}^{+18.4}$ &  $194.7$& $214$\\
$24$&12.4-13.2& $ 35.36_{-0.88}^{+0.92}$  & $ 31.80_{-0.90}^{+0.91}$   &  $ -1.92\pm0.01$          &  $ 271.2_{-11.9}^{+12.8}$ &  $324.6$& $231$\\
$25$&13.2-13.3& $ 42.30_{-1.59}^{+1.68}$  & $ 87.55_{-3.83}^{+3.92}$   &  $ -1.84\pm0.03$          &  $ 213.7_{-22.7}^{+27.0}$ &  $141.4$& $180$\\
$26$&13.3-13.6& $ 45.32_{-1.97}^{+2.10}$  & $ 57.60_{-2.72}^{+2.79}$   &  $ -1.87\pm0.02$          &  $ 276.6_{-20.6}^{+23.4}$ &  $175.3$& $192$\\
$27$&13.6-13.8& $ 53.27_{-1.94}^{+2.02}$  & $ 69.62_{-2.85}^{+2.90}$   &  $ -1.87_{-0.03}^{+0.02}$ &  $ 203.7_{-17.3}^{+20.6}$ &  $169.2$& $199$\\
$28$&13.8-14.1& $ 66.19_{-2.72}^{+2.92}$  & $ 89.79_{-3.80}^{+3.93}$   &  $ -1.84\pm0.02$          &  $ 187.8_{-13.8}^{+15.3}$ &  $275.3$& $206$\\
$29$&14.1-14.2& $ 105.70_{-4.91}^{+5.22}$ & $ 201.80_{-9.99}^{+10.2}$  &  $ -1.82\pm0.03$          &  $ 169.6_{-18.2}^{+20.2}$ &  $177.9$& $204$\\
$30$&14.2-14.4& $ 120.40_{-5.70}^{+5.93}$ & $ 199.60_{-10.00}^{+10.40}$&  $ -1.83_{-0.03}^{+0.02}$ &  $ 159.9_{-15.2}^{+18.7}$ &  $180.7$& $211$\\
$31$&14.4-14.6& $ 51.74_{-2.30}^{+2.45}$  & $ 57.16_{-2.79}^{+2.86}$   &  $ -1.86_{-0.03}^{+0.02}$ &  $ 186.8_{-16.2}^{+18.8}$ &  $164.6$& $194$\\
$32$&14.6-14.8& $ 99.11_{-4.00}^{+4.23}$  & $ 155.80_{-6.57}^{+6.88}$  &  $ -1.85\pm0.03$          &  $ 160.5_{-15.4}^{+19.3}$ &  $173.6$& $211$\\
$33$&14.8-15.0& $ 71.48_{-3.09}^{+3.30}$  & $ 115.90_{-5.38}^{+5.55}$  &  $ -1.82\pm0.03$          &  $ 149.0_{-15.9}^{+19.0}$ &  $165.7$& $196$\\
$34$&15.0-15.1& $ 102.20_{-5.26}^{+5.60}$ & $ 220.80_{-11.7}^{+12.2}$  &  $ -1.81\pm0.03$          &  $ 159.0_{-18.3}^{+21.9}$ &  $184.4$& $202$\\
$35$&15.1-15.2& $ 102.10_{-4.22}^{+4.40}$ & $ 233.10_{-10.1}^{+10.5}$  &  $ -1.81\pm0.03$          &  $ 144.6_{-15.4}^{+18.9}$ &  $212.1$& $199$\\
$36$&15.2-15.5& $ 127.0_{-3.73}^{+3.85}$ &  $ 223.0_{-7.18}^{+7.36}$ &  $ -1.85_{-0.0234}^{+0.0201}$ &  $ 160.7_{-12.5}^{+14.3}$ &  $216.60$& $215$\\
$37$&15.5-15.7& $ 150.70_{-5.99}^{+6.16}$ & $ 254.80_{-11.30}^{+11.80}$&  $ -1.83\pm0.03$          &  $ 120.5_{-12.4}^{+15.9}$ &  $168.4$& $221$\\
$38$&15.7-16.2& $ 59.42_{-1.74}^{+1.81}$  & $ 63.99_{-2.12}^{+2.15}$   &  $ -1.88\pm0.02$          &  $ 169.4_{-12.4}^{+14.3}$ &  $197.2$& $221$\\
$39$&16.2-16.3& $ 84.53_{-3.69}^{+3.95}$  & $ 132.10_{-6.08}^{+6.36}$  &  $ -1.84\pm0.03$          &  $ 168.9_{-16.5}^{+20.3}$ &  $190.3$& $203$\\
$40$&16.3-16.5& $ 90.82_{-3.47}^{+3.67}$  & $ 160.90_{-6.63}^{+6.85}$  &  $ -1.83\pm0.03$          &  $ 158.1_{-15.4}^{+18.3}$ &  $177.3$& $206$\\
$41$&16.5-16.7& $ 94.44_{-4.25}^{+4.55}$  & $ 143.00_{-6.81}^{+7.11}$  &  $ -1.84\pm0.03$          &  $ 160.6_{-15.8}^{+19.1}$ &  $169.6$& $210$\\
$42$&16.7-16.9& $ 78.69_{-4.10}^{+4.46}$  & $ 96.94_{-5.29}^{+5.55}$   &  $ -1.83_{-0.04}^{+0.03}$ &  $ 137.2_{-15.1}^{+18.4}$ &  $155.4$& $198$\\
$43$&16.9-17.1& $ 47.97_{-2.47}^{+2.65}$  & $ 40.30_{-2.26}^{+2.33}$   &  $ -1.84_{-0.03}^{+0.02}$ &  $ 138.7_{-13.1}^{+15.3}$ &  $144.2$& $191$\\
$44$&17.1-17.5& $ 63.52_{-2.19}^{+2.29}$  & $ 75.35_{-2.87}^{+2.93}$   &  $ -1.86_{-0.03}^{+0.02}$ &  $ 148.8_{-13.1}^{+15.6}$ &  $171.4$& $206$\\
$45$&17.5-17.8& $ 68.97_{-3.26}^{+3.46}$  & $ 54.62_{-2.76}^{+2.85}$   &  $ -1.85_{-0.03}^{+0.02}$ &  $ 113.7_{-10.6}^{+12.6}$ &  $191.9$& $209$\\
$46$&17.8-18.3& $ 46.21_{-1.50}^{+1.56}$  & $ 38.75_{-1.36}^{+1.39}$   &  $ -1.87\pm0.02$          &  $ 142.8_{-9.5}^{+10.4}$  &  $248.0$& $228$\\
$47$&18.3-18.9& $ 57.27_{-1.85}^{+1.95}$  & $ 52.36_{-1.75}^{+1.80}$   &  $ -1.88\pm0.02$          &  $ 166.4_{-9.8}^{+10.6}$  &  $334.0$& $233$\\
$48$&18.9-19.4& $ 57.29_{-1.87}^{+1.97}$  & $ 49.10_{-1.71}^{+1.75}$   &  $ -1.88\pm0.02$          &  $ 156.0_{-9.7}^{+10.7}$  &  $302.1$& $220$\\
$49$&19.4-19.6& $ 49.44_{-1.86}^{+1.96}$  & $ 81.63_{-3.39}^{+3.50}$   &  $ -1.83\pm0.03$          &  $ 147.1_{-15.7}^{+18.8}$ &  $167.7$& $189$\\
$50$&19.6-19.7& $ 54.68_{-2.14}^{+2.24}$  & $ 88.95_{-3.81}^{+3.88}$   &  $ -1.83_{-0.03}^{+0.02}$ &  $ 164.9_{-16.3}^{+18.9}$ &  $171.8$& $192$\\
$51$&19.7-19.9& $ 57.57_{-2.43}^{+2.54}$  & $ 94.89_{-4.21}^{+4.29}$   &  $ -1.83_{-0.03}^{+0.02}$ &  $ 178.0_{-16.0}^{+18.0}$ &  $202.2$& $194$\\
$52$&19.9-20.1& $ 72.81_{-3.90}^{+4.16}$  & $ 91.88_{-5.08}^{+5.28}$   &  $ -1.85_{-0.03}^{+0.02}$ &  $ 197.8_{-17.5}^{+20.5}$ &  $170.6$& $196$\\
$53$&20.1-20.3& $ 43.33_{-3.07}^{+3.37}$  & $ 42.35_{-2.88}^{+2.99}$   &  $ -1.82_{-0.03}^{+0.02}$ &  $ 136.6_{-14.9}^{+16.6}$ &  $165.1$& $189$\\
$54$&20.3-20.6& $ 50.94_{-2.41}^{+2.52}$  & $ 53.85_{-2.59}^{+2.64}$   &  $ -1.86\pm0.02$          &  $ 193.9_{-15.3}^{+17.2}$ &  $221.4$& $205$\\
$55$&20.6-20.9& $ 46.04_{-1.71}^{+1.79}$  & $ 51.23_{-2.12}^{+2.16}$   &  $ -1.87\pm0.02$          &  $ 192.5_{-14.8}^{+16.7}$ &  $192.6$& $196$\\
$56$&20.9-21.0& $ 42.49_{-2.04}^{+2.20}$  & $ 55.46_{-2.79}^{+2.90}$   &  $ -1.84\pm0.03$          &  $ 148.9_{-16.0}^{+18.6}$ &  $171.3$& $183$\\
$57$&21.0-21.3& $ 36.47_{-2.20}^{+2.44}$  & $ 23.88_{-1.53}^{+1.59}$   &  $ -1.87_{-0.03}^{+0.02}$ &  $ 152.9_{-14.5}^{+17.0}$ &  $143.5$& $189$\\
$58$&21.3-21.7& $ 42.84_{-1.19}^{+1.23}$  & $ 50.72_{-1.63}^{+1.67}$   &  $ -1.88_{-0.03}^{+0.02}$ &  $ 155.2_{-12.7}^{+14.8}$ &  $186.5$& $212$\\
$59$&21.7-21.9& $ 47.05_{-2.70}^{+2.89}$  & $ 46.19_{-2.80}^{+2.89}$   &  $ -1.84_{-0.03}^{+0.02}$ &  $ 161.9_{-15.5}^{+17.8}$ &  $152.6$& $195$\\
$60$&21.9-22.2& $ 49.53_{-3.13}^{+3.39}$  & $ 42.03_{-2.83}^{+2.94}$   &  $ -1.84_{-0.03}^{+0.02}$ &  $ 153.6_{-15.1}^{+17.5}$ &  $147.1$& $188$\\
$61$&22.2-23.0& $ 31.13_{-3.30}^{+4.08}$  & $ 5.72_{-0.60}^{+0.62}$    &  $ -1.90\pm0.02$          &  $ 126.0_{-9.4}^{+10.2}$  &  $187.3$& $233$\\
Total&0.00-30.0& $ 96.71_{-0.484}^{+0.461}$& $ 71.65_{-0.36}^{+0.34}$   &  $ -1.93\pm0.01$         &  $ 175.1_{-1.3}^{+1.2}$   &  $14732.0$& $276$\\
\hline

\multicolumn{7}{c}{ 090902B Model : Band Function + Power Law  }\tabularnewline

\hline
 &  Time  &    $\alpha$ &   $\beta$ &   $E_{0}$  &   $K$  &   $\Gamma_{\rm PL}$ &   $K_{\rm PL}$  &  $\chi^2$ & $dof$ \\
 & s  & & & keV & $\frac{photons}{keV cm^{2}s}@100 keV$& &  $\frac{photons}{keV cm^{2}s}@1 keV$\\

\hline

Total&0.00-23.0& $ -0.83\pm0.01$ &  $ -3.68_{-0.20}^{+0.12}$ &  $ 724_{-12}^{+13}$ &  $ 0.099\pm0.001$ &  $ -1.85_{-1.85}^{+1.85}$ &  $ 43.4\pm1.5$ &  $2024.3$& $275$\\
\ \\
\hline

\multicolumn{7}{c}{090926A Model : Band Function}\tabularnewline
\ \\
\hline\noalign{\smallskip}
 Seq &  Time &    $\alpha$ &   $\beta$ &   $E_{0}$  &   $K$  &  $\chi^2$ & $dof$ \\
 & s  & & & keV & $\frac{photons}{keV cm^{2}s}@100 keV$& & \\
\hline
$1$&0.00-2.81& $ -0.53_{-0.03}^{+0.04}$ &  $ -2.43_{-0.05}^{+0.06}$ &  $ 235_{-15}^{+16}$ &  $ 0.106\pm0.004$           &  $189.0$& $210$\\
$2$&2.81-3.75& $ -0.48\pm0.03$          &  $ -2.75_{-0.13}^{+0.21}$ &  $ 255_{-14}^{+15}$ &  $ 0.303_{-0.010}^{+0.011}$ &  $168.6$& $196$\\
$3$&3.75-5.62& $ -0.57\pm0.02$          &  $ -2.35\pm0.02$          &  $ 208\pm8$         &  $ 0.344\pm0.009$           &  $269.1$& $213$\\
$4$&5.62-7.50& $ -0.73\pm0.02$          &  $ -2.50_{-0.08}^{+0.13}$ &  $ 326\pm15$        &  $ 0.191\pm0.004$           &  $229.7$& $210$\\
$5$&7.50-9.38& $ -0.63\pm0.03$          &  $ -2.81_{-0.13}^{+0.17}$ &  $ 183_{-8}^{+9}$   &  $ 0.255_{-0.008}^{+0.009}$ &  $169.6$& $209$\\
$6$&9.38-11.2& $ -0.75\pm0.02$          &  $ -2.52_{-0.08}^{+0.10}$ &  $ 193_{-8}^{+9}$   &  $ 0.327_{-0.009}^{+0.010}$ &  $228.1$& $213$\\
$7$&11.2-13.1& $ -0.80\pm0.03$          &  $ -2.29_{-0.05}^{+0.06}$ &  $ 154_{-10}^{+11}$ &  $ 0.242_{-0.012}^{+0.014}$ &  $186.1$& $212$\\
$8$&13.1-15.9& $ -0.99\pm0.05$          &  $ -2.36_{-0.11}^{+0.22}$ &  $ 161_{-19}^{+22}$ &  $ 0.081_{-0.007}^{+0.008}$ &  $164.7$& $213$\\
$9$&15.9-20.0& $ -1.26\pm0.08$          &  $ -2.07_{-0.04}^{+0.07}$ &  $ 216_{-48}^{+68}$ &  $ 0.025_{-0.003}^{+0.004}$ &  $170.9$& $214$\\
Total&0.00-20.0& $ -0.74\pm0.01$         &  $ -2.34\pm0.01$          &  $ 226\pm4$         &  $ 0.165\pm0.002$           &  $777.1$& $216$\\
\hline
\ \\
\multicolumn{7}{c}{091003 Model : Band Function}\tabularnewline
\ \\
\hline\noalign{\smallskip}
 Seq &  Time &    $\alpha$ &   $\beta$ &   $E_{0}$  &   $K$  &  $\chi^2$ & $dof$ \\
 & s  & & & keV & $\frac{photons}{keV cm^{2}s}@100 keV$& & \\
\hline
$1$&7.00-15.0& $ -1.33\pm0.05$          &  $ -2.41_{-0.10}^{+0.20}$ &  $ 426_{-77}^{+101}$ &  $ 0.012\pm0.001$ &  $234.5$& $246$\\
$2$&15.0-18.0& $ -1.01\pm0.04$          &  $ -2.52_{-0.10}^{+0.19}$ &  $ 337_{-38}^{+43}$  &  $ 0.040\pm0.002$ &  $152.4$& $243$\\
$3$&18.0-20.0& $ -0.85\pm0.03$          &  $ -2.55_{-0.07}^{+0.10}$ &  $ 357_{-26}^{+28}$  &  $ 0.094\pm0.003$ &  $218.9$& $242$\\
$4$&20.0-26.0& $ -1.36_{-0.05}^{+0.06}$ &  $ -2.35_{-0.08}^{+0.15}$ &  $ 429_{-97}^{+143}$ &  $ 0.014\pm0.001$ &  $189.2$& $246$\\
Total&0.00-26.0& $ -1.09_{-0.01}^{+0.02}$ &  $ -2.58_{-0.04}^{+0.05}$ &  $ 474_{-25}^{+27}$  &  $ 0.024\pm0.001$ &  $446.2$& $246$\\
\hline
\ \\
\multicolumn{7}{c}{091031 Model : Band Function}\tabularnewline
\ \\
\hline\noalign{\smallskip}
 Seq &  Time &    $\alpha$ &   $\beta$ &   $E_{0}$  &   $K$  &  $\chi^2$ & $dof$ \\
 & s  & & & keV & $\frac{photons}{keV cm^{2}s}@100 keV$& & \\
\hline
$1$&0.00-8.00& $ -0.89\pm0.06$          &  $ -2.44_{-0.07}^{+0.09}$ &  $ 496_{-84}^{+111}$  &  $ 0.013\pm0.001$ &  $177.1$& $186$\\
$2$&8.00-15.0& $ -0.86_{-0.05}^{+0.06}$ &  $ -2.50_{-0.08}^{+0.13}$ &  $ 357_{-47}^{+55}$   &  $ 0.020\pm0.001$ &  $173.3$& $186$\\
$3$&15.0-25.0& $ -0.78_{-0.10}^{+0.11}$ &  $ -2.55_{-0.12}^{+0.26}$ &  $ 467_{-104}^{+157}$ &  $ 0.006\pm0.001$ &  $187.1$& $186$\\
Total&0.00-25.0& $ -0.87_{-0.03}^{+0.04}$ &  $ -2.55_{-0.05}^{+0.06}$ &  $ 458_{-33}^{+51}$   &  $ 0.012\pm0.001$ &  $347.2$& $186$\\
\hline
\ \\
\multicolumn{7}{c}{100116A Model : Band Function}\tabularnewline
\ \\
\hline\noalign{\smallskip}
 Seq &  Time &    $\alpha$ &   $\beta$ &   $E_{0}$  &   $K$  &  $\chi^2$ & $dof$ \\
 & s  & & & keV & $\frac{photons}{keV cm^{2}s}@100 keV$& & \\
\hline
$1$&-2.00-5.00& $ -1.03_{-0.11}^{+0.13}$ &  $ -2.54_{-0.24}^{+2.54}$ &  $ 384_{-124}^{+201}$  &  $ 0.006\pm0.001$ &  $104.8$& $155$\\
$2$&80.0-90.0 & $ -1.03_{-0.04}^{+0.05}$ &  $ -2.80_{-0.21}^{+0.97}$ &  $ 791_{-142}^{+192}$  &  $ 0.010\pm0.001$ &  $127.8$& $155$\\
$3$&90.0-95.0 & $ -1.00\pm0.01$          &  $ -3.22_{-0.25}^{+1.51}$ &  $ 1459_{-121}^{+161}$ &  $ 0.033\pm0.001$ &  $156.9$& $155$\\
$4$&95.0-110. & $ -1.03\pm0.05$          &  $ -2.63_{-0.11}^{+0.23}$ &  $ 677_{-120}^{+169}$  &  $ 0.009\pm0.001$ &  $127.0$& $155$\\
Total&0.00-110. & $ -1.11_{-0.02}^{+0.01}$ &  $ -3.13_{-0.09}^{+0.11}$ &  $ 2867_{-283}^{+430}$ &  $ 0.004\pm0.001$ &  $415.6$& $155$\\
\hline
\ \\
\multicolumn{7}{c}{100225A Model : Band Function}\tabularnewline
\ \\
\hline\noalign{\smallskip}
 Seq &  Time &    $\alpha$ &   $\beta$ &   $E_{0}$  &   $K$  &  $\chi^2$ & $dof$ \\
 & s  & & & keV & $\frac{photons}{keV cm^{2}s}@100 keV$& & \\
\hline
$1$&0.00-6.00& $ -0.53_{-0.19}^{+0.22}$ &  $ -2.43_{-0.19}^{+0.87}$ &  $ 263_{-74}^{+120}$  &  $ 0.010\pm0.002$           &  $51.8$& $94$\\
$2$&6.00-12.0& $ -0.93_{-0.13}^{+0.15}$ &  $ -2.30_{-0.12}^{+0.26}$ &  $ 507_{-181}^{+351}$ &  $ 0.009_{-0.001}^{+0.002}$ &  $40.3$& $93$\\
Total&0.00-12.0& $ -0.77_{-0.11}^{+0.12}$ &  $ -2.37_{-0.10}^{+0.18}$ &  $ 375_{-86}^{+129}$  &  $ 0.010\pm0.001$           &  $64.5$& $94$\\
\hline
\ \\
\multicolumn{7}{c}{100325A Model : Band Function}\tabularnewline
\ \\
\hline\noalign{\smallskip}
 Seq &  Time &    $\alpha$ &   $\beta$ &   $E_{0}$  &   $K$  &  $\chi^2$ & $dof$ \\
 & s  & & & keV & $\frac{photons}{keV cm^{2}s}@100 keV$& & \\
\hline
$1$&-3.00-10.0& $ -0.72_{-0.10}^{+0.11}$ &  $ -2.60_{-0.21}^{+1.89}$ &  $ 155_{-26}^{+32}$ &  $ 0.014\pm0.002$ &  $151.6$& $125$\\
\hline
\ \\
\multicolumn{7}{c}{100414A Model : Band Function}\tabularnewline
\ \\
\hline\noalign{\smallskip}
 Seq &  Time &    $\alpha$ &   $\beta$ &   $E_{0}$  &   $K$  &  $\chi^2$ & $dof$ \\
 & s  & & & keV & $\frac{photons}{keV cm^{2}s}@100 keV$& & \\
\hline
$1$&1.00-7.25& $ -0.19_{-0.05}^{+0.06}$ &  $ -2.54_{-0.10}^{+0.16}$ &  $ 256_{-20}^{+22}$ &  $ 0.036\pm0.002$           &  $124.3$& $156$\\
$2$&7.25-14.3& $ -0.25_{-0.04}^{+0.05}$ &  $ -2.89_{-0.24}^{+0.51}$ &  $ 281_{-20}^{+19}$ &  $ 0.040_{-0.001}^{+0.002}$ &  $124.5$& $156$\\
$3$&14.3-19.6& $ -0.56_{-0.03}^{+0.04}$ &  $ -2.53_{-0.10}^{+0.16}$ &  $ 361_{-26}^{+28}$ &  $ 0.047\pm0.002$           &  $135.1$& $156$\\
$4$&19.6-25.5& $ -0.76\pm0.03$          &  $ -2.45_{-0.07}^{+0.11}$ &  $ 386_{-28}^{+30}$ &  $ 0.052\pm0.002$           &  $131.9$& $156$\\
Total&1.00-26.0& $ -0.52\pm0.02$          &  $ -2.62_{-0.05}^{+0.07}$ &  $ 344_{-12}^{+12}$ &  $ 0.042\pm0.001$           &  $281.7$& $156$\\

\hline
\hline
\enddata
\tablecomments{Spectral models used in this paper, i.e, Band Function (BAND), 
Black-Body (BB), Cut-off Power Law (CPL), and Power-Law (PL), correspond to 
grbm, bbody, cutoffpl and powerlaw in XSPEC package, respectively. Details 
of the formulae of these models can be found at 
\url{http://heasarc.gsfc.nasa.gov/docs/xanadu/xspec/manual/XspecModels.html}. }

\end{deluxetable}


\clearpage

\begin{deluxetable}{lll}
\tabletypesize{\scriptsize}
\tablecaption{Temporal decay slopes and the time integrated photon indices of 
the long-term LAT count rate lightcurves}

 \startdata
\hline
Name & $\alpha_{\rm LAT}$  & ${\bar\Gamma_{\rm LAT}}$\\
\hline

080825C  &   $-0.47    \pm  0.74$  &$  -1.71$\\
080916C  &   $-1.33     \pm  0.08$& $ -1.77$\\
081024B   &  $-1.37     \pm  0.41$ & $  -1.98$\\
081215A   &  - & -\\
090217    & $-0.81  \pm    0.23$&$  -1.97$ \\
090323    & $-0.52  \pm    0.67$ & $  -1.75$\\
090328    & $-0.96    \pm  0.44$& $  -1.82$\\
090510    & $-1.70     \pm  0.08$& $  -1.94 $\\
090626    &- & $  -1.53$\\
090902B   & $-1.40     \pm  0.06$ & $  -1.76$\\
090926A   & $-2.05    \pm   0.14$ &$  -2.03$\\
091003    & $< -0.93    $ & $  -1.74$\\
091031    & $-0.57     \pm  0.28$ & $  -1.73$\\  
100116A   & - &$  -1.68$\\ 
100225A   & - & $  -1.77$\\ 
100325A   & $<-1.04     $ & $   -1.53$\\ 
100414A   & $-1.64      \pm 0.89$&  $  -1.85 $\\ 
\hline

\enddata
\end{deluxetable}

\clearpage

\newpage
\clearpage

\begin{figure}
 \includegraphics[angle=0,scale=.8]{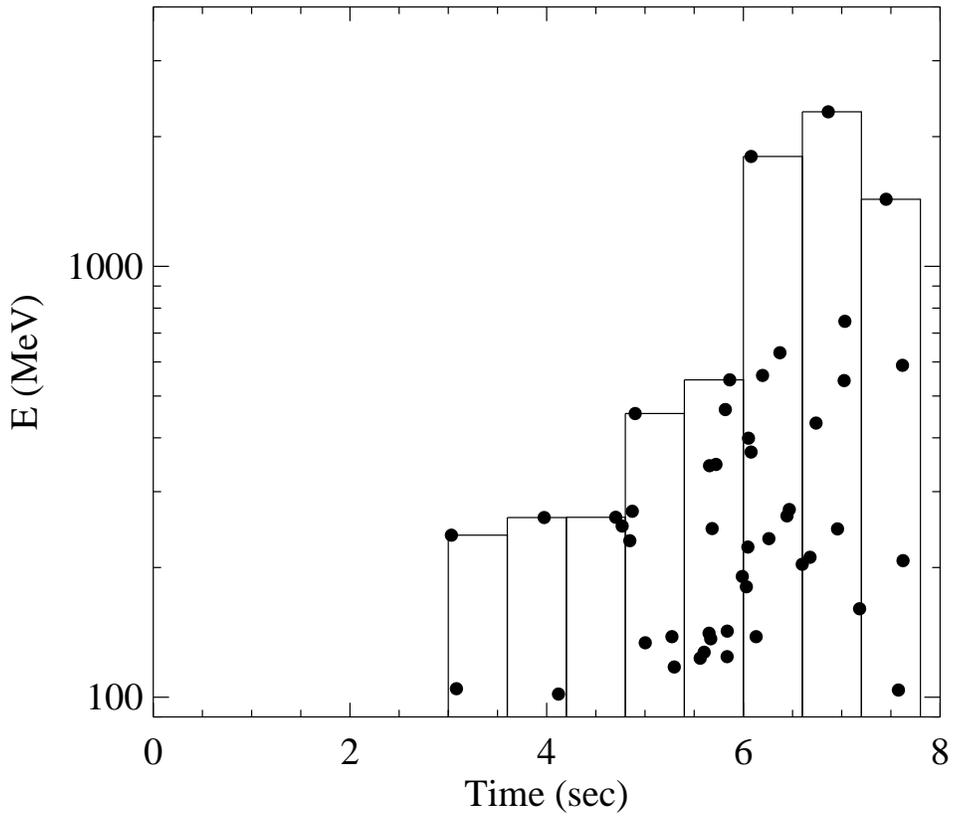}

\caption{LAT photon arrival time distribution for GRB 080916C. A rough trend of gradual increase of the maximum photon energy with time is seen.}
\end{figure}

\end{document}